\documentclass[aps,prb,twocolumn,superscriptaddress]{revtex4-2}
\usepackage{graphicx}
\usepackage{amsmath}
\usepackage{amssymb}
\usepackage{mathtools}
\usepackage{xcolor}
\usepackage[colorlinks=true, citecolor=blue, urlcolor=blue, linkcolor=blue,bookmarks=false,hypertexnames=true]{hyperref} 

\newcommand{\ket}[1]{\left| #1 \right\rangle}

\usepackage{array}
\usepackage[column=O]{cellspace}
\newcolumntype{P}[1]{>{\centering\arraybackslash}p{#1}}

\begin{document}
\graphicspath{}

\title{Theory of Glide Symmetry Protected Helical Edge States in WTe$_{2}$ Monolayer}

\author{Maciej Bieniek}
\affiliation{Institut f\"ur Theoretische Physik und Astrophysik, Universit\"at W\"urzburg, 97074 W\"urzburg, Germany}
\affiliation{Department of Theoretical Physics, Wroc\l aw University of Science and Technology, Wybrze\.ze Wyspia\'nskiego 27, 50-370 Wroc\l aw, Poland}
\affiliation{Department of Physics, University of Ottawa, Ottawa, Ontario, Canada K1N 6N5}
\date{\today}

\author{Jukka I. V\"ayrynen}
\affiliation{Department of Physics and Astronomy, Purdue University, West Lafayette, Indiana 47907 USA}

\author{Gang Li}
\affiliation{School of Physical Science and Technology, ShanghaiTech University, Shanghai 201210, China}

\author{Titus Neupert}
\affiliation{Department of Physics, University of Zurich, Winterthurerstrasse 190, 8057 Zurich, Switzerland}

\author{Ronny Thomale}
\affiliation{Institut f\"ur Theoretische Physik und Astrophysik, Universit\"at W\"urzburg, 97074 W\"urzburg, Germany}

\begin{abstract}
Helical edge states in quantum spin Hall (QSH) materials are central building blocks of topological matter design and engineering. Despite their principal topological protection against elastic backscattering, the level of operational stability depends on manifold parameters such as the band gap of the given semiconductor system in the ``inverted'' regime, temperature, disorder, and crystal orientation. We theoretically investigate electronic and transport properties of QSH edge states in large gap 1-T' WTe$_{2}$ monolayers. We explore the impact of edge termination, disorder, temperature, and interactions on experimentally addressable edge state observables, such as local density of states and conductance. We show that conductance quantization can remain surprisingly robust even for heavily disordered samples because of an anomalously small edge state decay length and additional protection related to the large direct gap allowed by glide symmetry. From the simulation of temperature-dependent resistance, we find that moderate disorder enhances the stability of conductance by localizing bulk states. 
We evaluate the edge state velocity and Luttinger liquid parameter as functions of the  chemical potential, finding prospects for physics beyond 
linear helical Luttinger liquids in samples with ultra-clean and well-defined edges. 
\end{abstract}

\maketitle

\section{Introduction}
Quantum spin Hall (QSH) insulators are a pillar of topological matter in which helical edge states offer a novel route towards dissipationless transport and quantum computation \cite{Hasan_Kane_2010, Qi_Zhang_2011, Alicea_2012, Ren_Niu_2016, Culcer_Tkachov_2020}. The first proposals in graphene \cite{Kane_Mele_2005a, Kane_Mele_2005b} quickly turned out to be insufficient for the observation of QSH effect due to small spin-orbit coupling~\cite{Yao_Fang_2007} and disadvantageous orbital composition~\cite{Li_Thomale_2018}, where both shortcomings could be overcome in Bismuthene as the realization of a Kane-Mele type QSH system at room temperature~\cite{Reis_Claessen_2017,Li_Thomale_2018,Stuhler_Claessen_2020}. It has, however, proven difficult to perform transport experiments in Bismuthene because of the challenging synthesis of sufficiently large homogeneous samples, a shortcoming which might be overcome in other Xene monolayer/substrate compounds\cite{Deng_Hou_2018} or in up to now less understood classes of materials such as the jacutingaite family \cite{Marrazzo_Marzari_2018, Kandrai_Nemes-Incze_2020, Wu_DiSante_2019}. 

Quantum well heterostructures accomplish the QSH insulating regime from band inversion through the reduction of the inherent point group symmetry of the core semiconductor. By construction, these setups are suitable for performing transport experiments where as local spectroscopy is nearly impossible due to their composite layer nature. While quantum wells have provided the first observation of QSH effect~\cite{Konig_Zhang_2007, Roth_Zhang_2009,Knez_Sullivan_2011}, and much subsequent progress has been made to enhance their measurability and operability \cite{Spanton_Moler_2014,Pribiag_Kouwenhoven_2015, Du_Du_2015,Li_Du_2015,Du_Du_2017, Du_Du_2017b,Bendias_Molenkamp_2018, Lunczer_Molenkamp_2019, Xiao_Hu_2019, Han_Du_2019,Piatrusha_Khrapai_2019,Strunz_Molenkamp_2020, Shamim_Molenkamp_2020, Dartiailh_Bocquillon_2020, Shamim_Molenkamp_2021}, their large penetration depth of edge states (in order of tens of nanometers) prevents an ideal QSH setting. This manifests in the lack of topological protection due to, e.g., coupling to charge puddles \cite{Vayrynen_Glazman_2013, Vayrynen_Glazman_2014}, impurities of non-magnetic \cite{Novelli_Polini_2019} or Kondo \cite{Maciejko_Zhang_2009, Tanaka_Matveev_2011, Altshuler_Yudson_2013} type, incoherent electromagnetic noise \cite{Vayrynen_Alicea_2018}, interaction-mediated localization \cite{Wu_Zhang_2006, Xu_Moore_2006}, nuclear spins \cite{Maestro_Rosenow_2013, Hsu_Loss_2017, Hsu_Loss_2018}, or axial spin symmetry breaking by Rashba effect \cite{Schmidt_Glazman_2012, Chou_Foster_2015}. It is thus desirable to identify systems with smaller edge state penetration depths, larger band gaps and better thermal stability which are accessible through transport experiments~\cite{Schleder_Fazzio_2021}.

The transition metal dichalcogenide WTe$_{2}$ is an intriguing platform which provides a promising realization of QSH edge states accessible through both transport experiments and local spectroscopy. In its 3D form, it is a type-II Weyl semimetal \cite{DiSante_Panaccione_2017, Das_Panaccione_2019}, while when thinned down to a monolayer  becomes a QSH insulator \cite{Qian_Li_2014,Fei_Cobden_2017,Tang_Shen_2017,Wu_Jarillo-Herrero_2018,Li_Tang_2020}. 
The signal associated with QSH edge states has been demonstrated in transport studies \cite{Fei_Cobden_2017, Wu_Jarillo-Herrero_2018}. Scanning tunneling microscopy/spectroscopy also revealed the existence of states on the boundaries \cite{Tang_Shen_2017, Jia_Li_2017, Peng_Fu_2017, Maximenko_Madhavan_2022}, consistent with a QSH edge state scenario. 
Furthermore, various spectral  features of a bulk gap have been detected such as the Coulomb gap at the Fermi energy \cite{Song_Li_2018}, NbSe$_{2}$ proximity-induced superconducting gap \cite{Lupke_Hunt_2020, Tao_Weber_2022}, strain-induced gap \cite{Zhao_Jia_2020}, and CrI$_{3}$ antiferromagnet exchange-field gap \cite{Zhao_Cobden_2020}. As the electron concentration is increased, transitions to metallic and superconducting bulk states were observed \cite{Fatemi_Jarillo-Herrero_2018, Sajadi_Cobden_2018}. It is further plausible that doping might have an influence on the nature of the many-body ground state in WTe$_{2}$ even beyond the mere metallic or superconducting character, as it was recently claimed for an excitonic insulator phase in WTe$_{2}$~\cite{Jia_Wu_2022, Sun_Cobden_2022}. Given the evidence of electron correlation effects in the bulk, it is plausible that the QSH edge states may also experience strong electron-electron interactions. While unambiguous Luttinger liquid behaviour of the edge channels still need to be explored and detected, the material's propensity towards such 1D channels is hinted at in initial reports for a twisted WTe$_{2}$ bilayer geometry \cite{Wang_Wu_2022}. 

The plethora of experimental results stimulated a significant body of theoretical work related to analyzing the electronic, spin, and many-body properties of WTe$_{2}$. Various intertwined methods have been used to address the principal electronic structure in this material, including density functional theory~\cite{Qian_Li_2014, Lv_Sun_2015, Zheng_Feng_2016, Xiang_Liu_2016, Lin_Ni_2017, Tang_Shen_2017, Peng_Fu_2017, Hu_Liu_2018, Jelver_Jacobsen_2019, Ok_Neupert_2019, Lau_Akhmerov_2019, Muechler_Car_2020, Zhang_Li_2020, Zhao_Jia_2020, Yang_Zhang_2020, Maximenko_Madhavan_2022, Lu_Sushko_2021}, tight-binding approaches~\cite{Meuchler_Carr_2016, Ok_Neupert_2019, Lau_Akhmerov_2019, Hsu_Sau_2020, Copenhaver_Vayrynen_2021, Hu_Liu_2021}, and low-energy $k\cdot p$ models~\cite{Xu_Jarillo-Herrero_2018, Shi_Song_2019, Xie_Law_2020, Garcia_Roche_2020, Nandy_Pesin_2021, Jia_Wu_2022, Sun_Cobden_2022, Hu_Liu_2021}. Remarkably, theoretical models predict a small (if at all) bulk gap in WTe$_{2}$ from first principles which is superficially at odds with the rather stable QSH behaviour observed experimentally.  Since the first identification of several 1T' TMD's as topological insulators \cite{Qian_Li_2014},  understanding the nature of the d-d band inversion process \cite{Choe_Chang_2016}, general phase diagram \cite{Meuchler_Carr_2016}, the role of glide symmetry in strong localization of edge states \cite{Ok_Neupert_2019} and the role of edge termination \cite{Ok_Neupert_2019,Lau_Akhmerov_2019, Zhang_Li_2020, Lu_Sushko_2021} has persisted as a highly challenging task. This similarly applies to the impact of edge roughness on conductance \cite{Meuchler_Carr_2016}, spin dynamics, and anomalous Hall conductivity \cite{Nandy_Pesin_2021}, the role of disorder on edge spin transport \cite{Copenhaver_Vayrynen_2021}, the edge magnetoresistance due to orbital moments \cite{Arora_Song_2020}, the possibility of spontaneous magnetization of edge states \cite{Jelver_Jacobsen_2019} and gate-activated canted spin texture~\cite{Shi_Song_2019, Garcia_Roche_2020}, the  possible pairing mechanisms and symmetries of the superconducting state~\cite{ Xie_Law_2020, Hsu_Sau_2020, Yang_Zhang_2020, Lee_Son_2021, Crepel_Fu_2021}, and the nature of the excitonic insulator~\cite{Varsano_Rontani_2020,Lee_2021, He_Lee_2021,Kwan_Parameswaran_2021}. 

In our article, we center the theoretical analysis around the assertion that the standalone microscopic features of WTe$_2$ are rooted in its glide symmetry~\cite{Ok_Neupert_2019}. It implies that the Dirac cones are not pinned in momentum space and are hence allowed to shift, which leads to large direct gaps experienced by the QSH edge states despite a small bulk gap. While the large direct gaps in principle naturally explain the small penetration depth of QSH edge states in WTe$_2$, it also renders the specific sample boundary termination of WTe$_2$ pivotal to accurately describe the QSH profile, see Fig.~\ref{fig1}. We will further assume that different microscopic sources of disorder, e.g. Te defect states \cite{Muechler_Car_2020} or edge inhomogeneity \cite{Lau_Akhmerov_2019}, can be on average modelled by Anderson-type disorder. Combined, we embark on a theoretical analysis of local spectroscopy and transport in WTe$_2$ related to scanning tunneling microscopy (STM) studies of the edge (Fig.~\ref{fig2}), longitudinal resistance measurements (Fig.~\ref{fig3}), and the temperature dependence of conductance (Fig.~\ref{fig4}). While we will leave a detailed analysis of the correlated bulk nature of WTe$_2$ for future studies, we will estimate the strength of electron-electron interactions on the edge states, and calculate the Luttinger liquid parameter as a function of the chemical potential, see Fig.~\ref{fig5}. Our calculation reveals a strong dependence on edge termination in the interaction strengths and a potential for tuning it by changing the edge electron density.

\section{Electronic structure}

\begin{figure*}\
\includegraphics[scale=0.6]{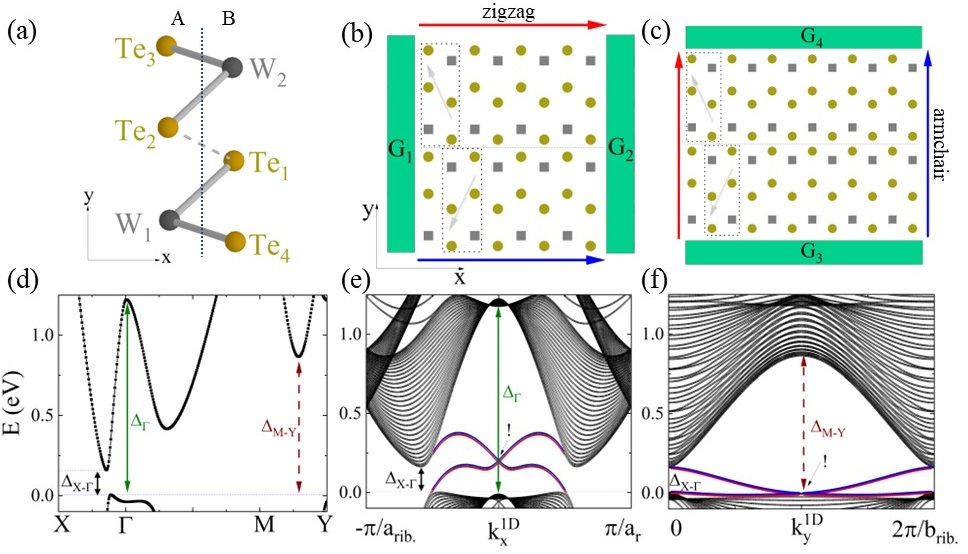}\
\caption{Structural and electronic properties of WTe$_{2}$ for two types of edge termination. (a) Schematic arrangement of atoms inside the unit cell in the xy plane. Top-view of WTe$_{2}$ ribbon with (b) zigzag and (c) armchair edge type. Rectangles and circles denote W and Te atoms, respectively. Contacts $G_{1-4}$ define the two-terminal setup. Edge states are schematically shown as blue and red arrows. The dashed line and grey arrows help to visualize glide reflection. (d) Band structure of infinite WTe$_{2}$ along the $X-\Gamma-M-Y$ line in the rectangular Brilloiun zone. Note 3 gaps $\Delta_{X-\Gamma}$ (fundamental), $\Delta_{\Gamma}$ and $\Delta_{M-Y}$. (e-f) Band structures of (e) the zigzag and (f) the armchair of 20 nm wide ribbons, with edge and bulk states colored as in (a). Exclamation mark '!' denotes the position of 1D Dirac points.}\
\label{fig1}
\end{figure*}\

To set the stage, let us begin with the discussion of structural and electronic properties of WTe$_{2}$ bulk and nanoribbons, focusing on two types of edge terminations. In Fig. \ref{fig1} (a) the top-view of two tungsten and four tellurium atoms inside the unit cell is shown. The lattice constants are calculated from the ab initio DFT@PBE level giving $a=3.50 $ \AA \ and $b=6.33 $ \AA. The real space primitive unit cell vectors are $\vec{a}_{1} = a(1,0)$ and $\vec{a}_{2} = b(0,1)$. Those give reciprocal lattice vectors $\vec{G}_{1}=2\pi/a(1,0)$ and $\vec{G}_{2}=2\pi/b(0,1)$ which define a rectangular Brillouin zone with $X$, $Y$ and $M$ points defined by $X=(\pi/a,0)$, $Y=(0,\pi/b)$ and $M=(\pi/a,\pi/b)$. Further details of geometry are discussed in \hyperref[app:A]{Appendix A}. Using terminology introduced in an earlier paper \cite{Ok_Neupert_2019} by some of us, we define two ribbon geometries with 'zigzag' and 'armchair' edge terminations, shown in Fig. \ref{fig1} (b) and (c), respectively.  We note that in the zigzag case edge states move parallel to the glide symmetry line, while for an armchair they move perpendicular to it. We will see below that the combination of larger glide-symmetry-enabled direct gap, smaller penetration depth and different A/B sublattice localization makes the zigzag edge states more robust to disorder.

Throughout this work we use ab initio based tight-binding model developed in Ref. [\onlinecite{Muechler_Car_2016}] for generic tilted Dirac-fermion and extended for massive tilted Dirac fermion in WTe$_{2}$ in Ref. [\onlinecite{Ok_Neupert_2019}], where the mass is related to spin-orbit coupling and Fock exchange-controlled band gap opening. The effective Hamiltonian of the system takes into account 4 orbitals, two $d_{x^2-y^2}$ localized on W atoms and two $p_{x}$ localized on Te atoms. Choosing the basis ordering for A/B sublattice and d/p orbitals as $\ket{A,d}, \ket{A,p}, \ket{B,d}, \ket{B,p}$ the 2D system Hamiltonian $\hat{H}_{0}(\vec{k})$ with its non-zero matrix elements is,
\begin{equation}
\hat{H}_{0}(\vec{k})=
\begin{bmatrix}
  \begin{matrix} H^{A}_{d} &  \\  & H^{A}_{p} \end{matrix} & \begin{matrix} H^{AB}_{dd} &H^{AB}_{dp} \\ H^{AB}_{pd} & H^{AB}_{pp} \end{matrix} \\
    $h.c.$ & \begin{matrix} H^{B}_{d} &  \\  & H^{B}_{p} \end{matrix} \\
\end{bmatrix}.
\label{eq1}
\end{equation}
Precise form of the elements and parameters are listed in \hyperref[app:B]{Appendix B}. The total spinful Hamiltonian $\hat{H}_{tot.}$ is given by
\begin{equation} 
\hat{H}_{tot.}(\vec{k}) = \hat{\sigma_{0}}\otimes H_{0}(\vec{k}) + V\hat{\sigma_{2}}\otimes \hat{\rho_{3}} \otimes \hat{\tau_{2}},
\end{equation}
 where $\hat{\sigma}, \hat{\rho}, \hat{\tau}$ are Pauli matrices acting on spin, sublattice A/B and orbital d/p degrees of freedom. Spin-orbit coupling strength is defined by the parameter $V=0.115$~eV. The Fermi level is set at the top of the valence band. This model has an indirect band gap of 165 meV exactly on $\Gamma - X$ line in the rectangular Brillouin zone, as shown in Fig. 1 (d). The glide symmetry allows for massive tilted Dirac fermions to be localized in k-space away from high-symmetry points. This has an interesting implications for relative gaps for an edge state dispersion in both zigzag and armchair geometries, as discussed below. To better understand direct band gaps for 1D Dirac cones, in addition to fundamental gap $\Delta_{X-\Gamma}$, we also mark gap at the $\Gamma$ point ($\Delta_{\Gamma}$) and the gap along $M-Y$ line ($\Delta_{M-Y}$), both calculated from the top of the valence band. 

Now we discuss the dispersion properties of the edge states in clean ribbons, as shown in Fig. \ref{fig1} (e)-(f). In ribbon geometry the width of the system is fixed to 20 nm, order of magnitude larger than the largest edge state penetration depth considered in a clean system. Overall, one can clearly distinguish between two types of edge by the position of the Dirac cone on 1D BZ. In the zigzag ribbon it is located within the conduction bulk states, as shown in Fig. \ref{fig1} (e). In armchair ribbon, the 1D Dirac cone overlaps with the top of the valence band, 8 meV below the band edge, as shown in Fig. \ref{fig1} (f). We note that in both cases the direct gap for bulk bands, between which edge states exist, is much larger than the fundamental one, $\Delta_{X-\Gamma} = 165$ meV. For the Dirac point k-space position in zigzag  $\Delta_\Gamma = 1.22$ eV and for the armchair $\Delta_{M-Y}=0.87$ eV. Those large gaps are responsible for anomalously small edge state penetration depths, as already discussed in Ref. [\onlinecite{Ok_Neupert_2019}]. Focusing on energies where the bulk is insulating and only a single helical pair exists,  their edge-localization can be related to the direct gap at the given wave number  $k_{1D}$. 
However, the simplest models in which the penetration depth $\lambda$ is proportional to velocity over the gap $\lambda \sim (\partial E/\partial k) / \Delta$, fail to properly capture quantitative behavior, which is in contrast to HgTe quantum wells described by the Bernevig-Hughes-Zhang model \cite{Bernevig_Zhang_2006}. We discuss further details of localization properties in \hyperref[app:C]{Appendix C}.

\section{Local density of states near the edge \label{sec3} }
\begin{figure*}\
\includegraphics[scale=0.6]{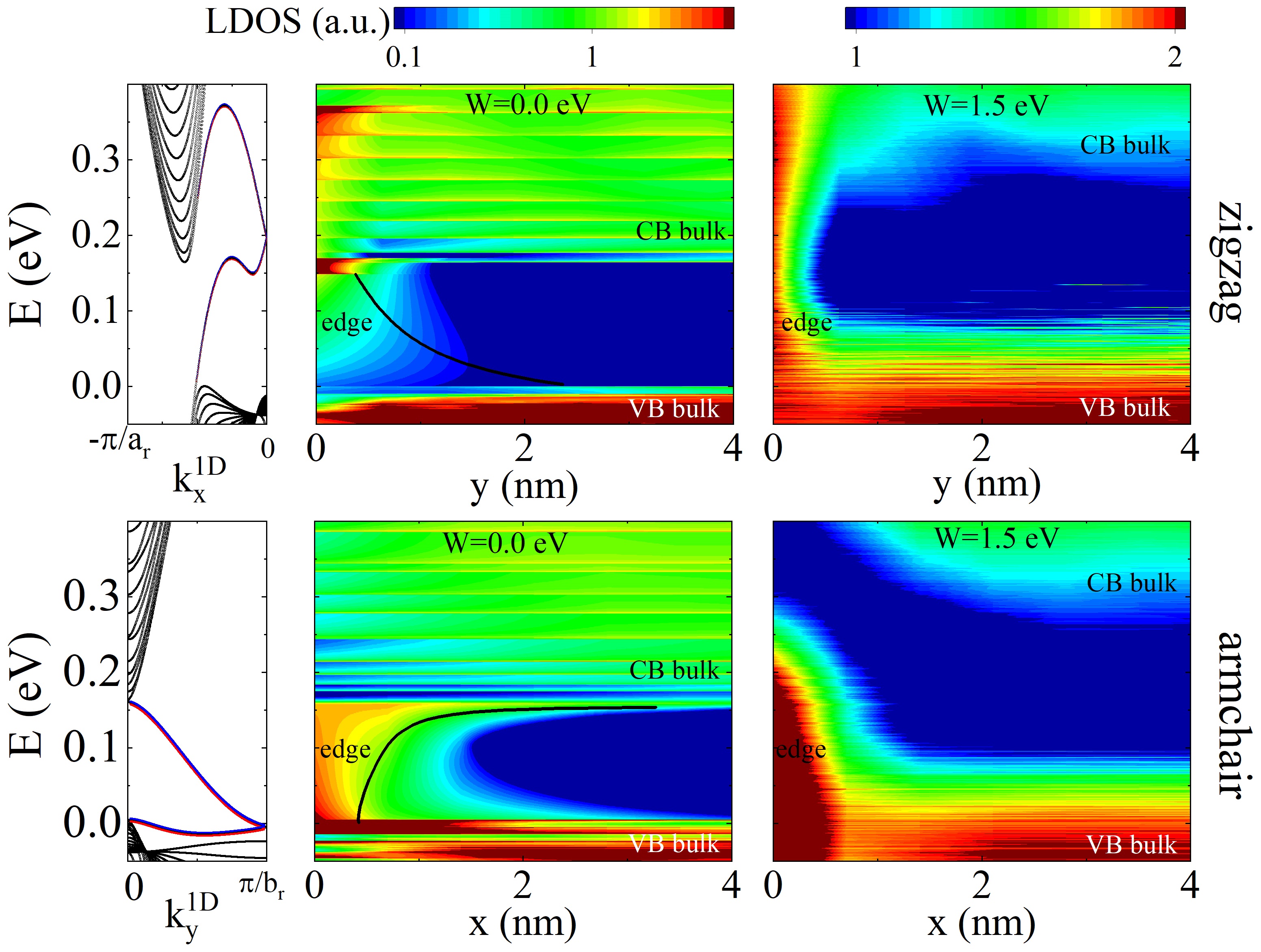}\
\caption{The energy-position resolved local density of states near one edge of the ribbon. Top panels are for zigzag and bottom panels are for armchair type ribbon, respectively. The left panel shows a zoom into the dispersion of edge states in a clean system. The vertical axis of the LDOS maps (middle and right panels) begins at the edge of the system: $y=0$ for zigzag, $x=0$ for armchair, with coordinates consistent with those in Fig 1 (b-c). The middle panel presents LDOS for a clean system (W=0.0 eV). The solid black line represents edge state penetration depth calculated using wavefunctions for the clean systems. The right panel shows the corresponding maps for the disordered case (W=1.5 eV). Note that the color scale encoding LDOS is logarithmic and changes between clean and disordered maps.}\
\label{fig2}
\end{figure*}\
The energetic position of 1D Dirac cones greatly influences the local density of states in both clean and disordered systems. Penetration depths are calculated using ribbon wavefunctions and local density of states (LDOS), given by
\begin{equation}
A(\vec{r},E)=-(1/\pi) \textrm{Im}\sum_{\alpha}G^{r}(\vec{r},\alpha,E),
\end{equation} 
computed from the retarded Green's function $G^{r}(\vec{r},\alpha,E)$ for both clean and disordered cases. Summation over $\alpha$ is performed over the spin, orbital, and atoms inside the ”block”, as described in \hyperref[app:A]{Appendix A} and \hyperref[app:D]{Appendix D}.  The disorder effects are introduced using Anderson model with disorder strength $W$ defining on each atom the random potential chosen from uniform distribution $[-W/2, W/2]$. 

Fig. \ref{fig2} presents the evolution of the local density of states as a function of the energy $E$, the distance from one of the edges (x/y) and the disorder strength $W$. We find that due to the different positions of the Dirac cones and localization properties of the bulk states in two types of ribbons, the LDOS maps can clearly distinguish in the experimentally accessible energy window ($\Delta E=300$ meV) between the zigzag and the armchair type of edge. This is also true for heavily disordered samples (e.g. value $W=1.5$ eV on the right panels of Fig. \ref{fig2}), especially when the energy is tuned into the conduction band. These results are consistent with a recent experiment reported in Ref. [\onlinecite{Tang_Shen_2017}], suggesting armchair type of the edge there. 
We note that in our simulation we average over disorder realizations and many ``scans'' of edge over different places on the sample to extract universal features of LDOS. More details of our averaging procedure and further studies of evolution of those maps as a function of disorder are discussed in \hyperref[app:D]{Appendix D}. 

\section{Transport in disordered systems}
\begin{figure*}\
\includegraphics[scale=0.6]{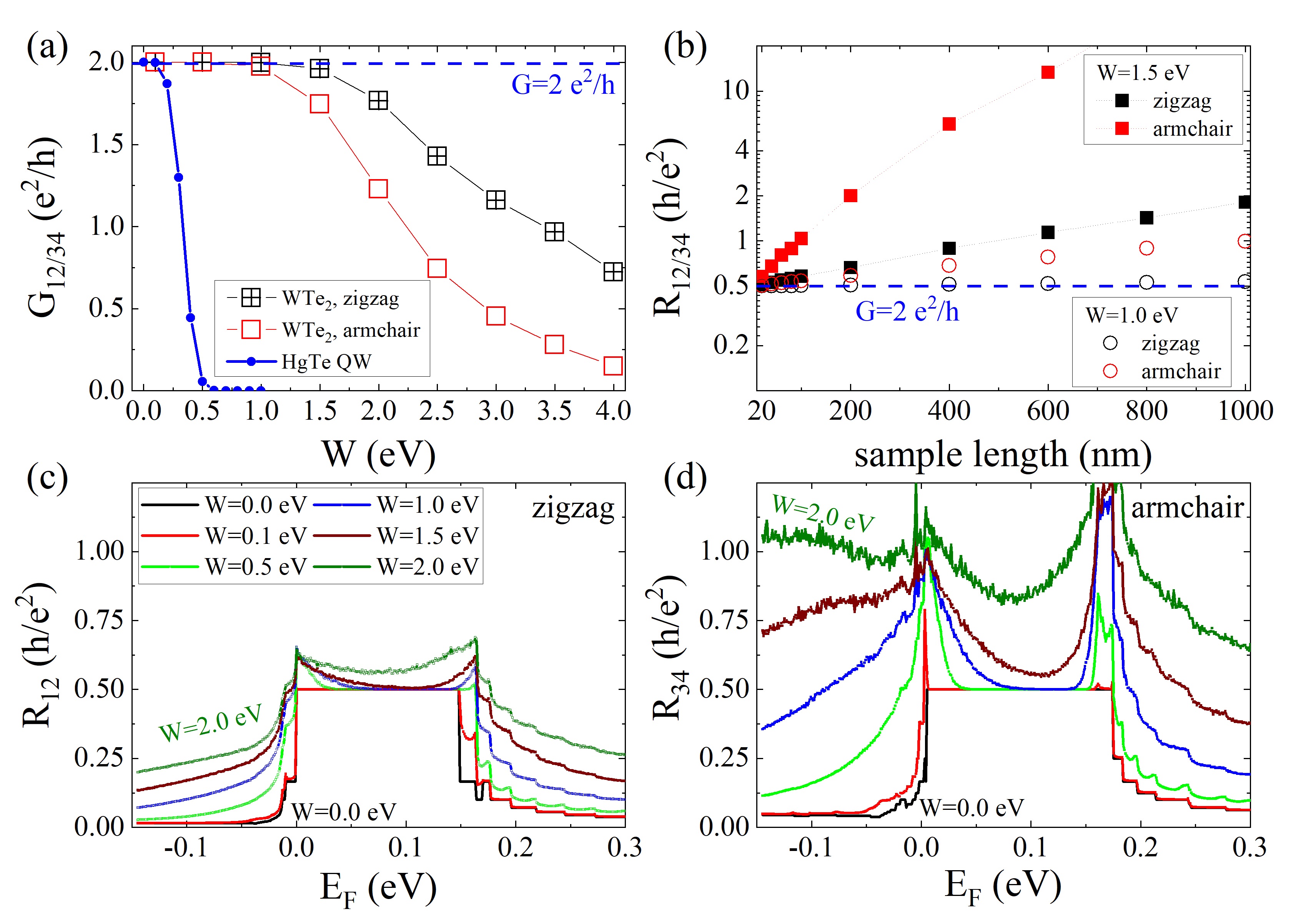}\
\caption{Transport in disordered ribbons. (a) Two-terminal conductance $G$ versus disorder strength $W$ for zigzag (black squares) and armchair (red squares) ribbons. The system size here is 20 x 20 nm$^2$ and $E_F = 82.5 $ meV. Contact configurations are shown in Fig. 1. The blue curve shows the corresponding G(W) characteristic for HgTe QW, as described in the text. (b) Scaling of longitudinal resistance $R=1/G$ with increasing sample length up to $1 \mu m$ for two values of disorder strength for both types of ribbon edges. Note the logarithmic scale on the y-axis. (c-d) Evolution of resistance with respect to increasing disorder strength $W$ as a function of the Fermi energy tuned between the valence and the conduction bands for (c) zigzag and (d) armchair terminations.}\
\label{fig3}
\end{figure*}\
Transport properties are calculated using standard Landauer-B\"uttiker formalism \cite{Landauer_1987, Datta_1997} where the two-terminal conductance $G = \frac{e^2}{h} \mathcal{T}$ is written in terms of the  transmission coefficient
\begin{equation}
\mathcal{T} = \textrm{Tr}\left[\Gamma_{L}G^{r}_{1,N}\Gamma_{R}(G^{r}_{1,N})^{\dagger}\right],
\end{equation} 
which is obtained using recursive Green's function scheme \cite{Thouless_Kirkpatrick_1981, Caroli_Saint-James_1971, Lee_Fisher_1981} (see \hyperref[app:E]{Appendix E} for more technical details). This method allows us to simulate edge length $\sim 1\mu$m, beyond previously-studied relatively short edge  lengths~\cite{Copenhaver_Vayrynen_2021}. 
The effects of disorder are included as a random on-site potential discussed above in Sec.~\ref{sec3}, as pioneered in studies of QSHE in quantum wells \cite{Li_Shen_2009, Groth_Beenakker_2009}. Unless stated otherwise, averaging over $10^3$ disorder realizations is performed (the results in Fig.~\ref{fig3}a were averaged up to $10^5$ realizations). 

We begin a discussion of transport by comparing clean and disordered relatively short ribbons. The results of two-terminal conductance as a function of disorder strength $W$ for the Fermi energy exactly in the middle of the bulk band gap, $E_F = 82.5$ meV, are analyzed in Fig. \ref{fig3}(a). 
As expected, strong enough disorder will lower the conductance, localizing the edge states. We note however a drastic difference between armchair and zigzag edge terminations. 
For the same scattering region size (20 $\times$ 20 nm$^2$) the conductance remains more robustly quantized to $G=2$ e$^{2}$/h for the zigzag type ribbon. 
This behavior can be rationalized by analysing how edge states localize on the A/B sublattices of WTe$_{2}$ \cite{Muechler_Car_2016}, in relation to glide symmetry. In case of zigzag edge disorder first has to induce A-B sublattice mixing on given edge and only then increasing disorder can on average induce penetration depth increase. This is in contrast to armchair termination, where even in a clean case the sublattices are already mixed, contributing to a less robust edge state. 

We also  note that due to the anomalously small penetration depths of the edge states in WTe$_{2}$, the quantized conductance $G= 2$ e$^{2}$/h survives up to much larger values of the disorder strength $W$ as compared to the HgTe quantum well (QW). In Fig. \ref{fig3}(a) the blue line shows $G(W)$ dependence for HgTe QW in the topological regime \cite{Qi_Zhang_2011} (well width $d=70$ \AA , Dirac mass parameter $M=-10$ meV). The size of the scattering region is chosen in this case to be 500 $\times$ 500 nm. We note that in principle we should compare systems with the same size, however we were not yet able to perform a 500 $\times$ 500 nm WTe$_{2}$ calculation because it is too large for atomistic calculation. On the other side, in the HgTe system smaller than 500 $\times$ 500 nm, the edge states become gapped due to size quantization, as discussed in Ref.~\cite{Zhou_Niu_2008}. Since penetration depth of HgTe is approximately 50 nm and less than 2 nm in WTe$_{2}$, the ratio of width to penetration depth is similar in our comparison. Although the precise value of $W$ for which ``Anderson localization'' of the edge state begins to set in depends on the size of the system, the edge state dispersion details and the Fermi energy \cite{Bieniek_Potasz_2017}, in general we can observe that for HgTe with the gap 20 meV the critical value of $W$ for which deviations from $G = 2$ e$^{2}$/h are observed is approximately $W_{crit.}= 0.2 $ eV, while for WTe$_{2}$ (gap 165 meV) it is as high as $W_{crit.} = 1.0$ eV. 

Next, we address the transition from short to long channel behavior in WTe$_{2}$ samples, in analogy to the experiment reported in Ref. \cite{Wu_Jarillo-Herrero_2018}. 
In Fig.~\ref{fig3} (b) we show that the edge resistance $R = 1/G$ grows approximately exponentially with the edge length, signifying Anderson localization.  
This transition is analogous to the Anderson transition of QSH states coupling two edge states by a sufficiently strong disorder in sufficiently long devices. A detailed study of $G(W, L_{y})$ is shown in \hyperref[app:F]{Appendix F}. 
By studying ribbons with 20 nm width and lengths up to 1 $\mu$m we show in Fig. \ref{fig3} (b) that in general resistance of long armchair samples should be much larger than zigzag samples, with differences increasing for more disordered samples. 
Focusing on two cases of disorder $W=1.0$ eV and $W=1.5$ eV we predict that the response of two types of edge should be clearly distinguishable, at least in a device of 20 nm width (realistic ones are much wider $>1$ $\mu$m, however those system sizes are currently not reachable using the atomistic approach). Next, we extract the localization lengths of the edge states defined as $G(L)=G_{0}\exp(-L/\xi)$. For $W=1.5$ eV in the zigzag and armchair we obtain $\xi \approx 750 \pm 10$ nm and $\xi \approx 150 \pm 10$ nm, respectively. For smaller value of disorder one can expect large enhancement of those values, which, for example, for $W = 1.0$ eV are for zigzag $\xi \approx 15200  \pm 200$ nm and for armchair $\xi \approx 1380 \pm 20$ nm.

In realistic experiments, the disordered sample should be described by one average disorder strength parameter $W$. In Fig. \ref{fig3} (c-d) we show general trends for resistance vs $E_F$ when the disorder strength is increased. For longitudinal resistance $R$ one can distinguish between zigzag and armchair edge response in experiment in which Fermi energy is tuned by the top gate. Notably, in both cases for the Fermi energy in the middle of the gap $R\approx 0.5 $ h/e$^{2}$ up to $W\approx 1$ eV. However, when $E_{F}$ is tuned away, the armchair edge signal exhibits resistance peaks  (close to the band edges) and a comparably larger asymmetry between the bulk valence and the conduction band response compared to zigzag ribbon orientation. We conclude that two sets of perpendicular contacts (shown in Fig. \ref{fig1} (b-c)) should in principle be able to discriminate between edge types for a realistic small sample of WTe$_{2}$. We note that the relation between Fermi energy in Fig. \ref{fig2} and gate voltage in experiment \cite{Fei_Cobden_2017} is difficult to estimate and in general might be strongly sample-dependent. 

\section{Temperature dependence of conductance on a disordered edge}
\begin{figure}\
\includegraphics[scale=0.33]{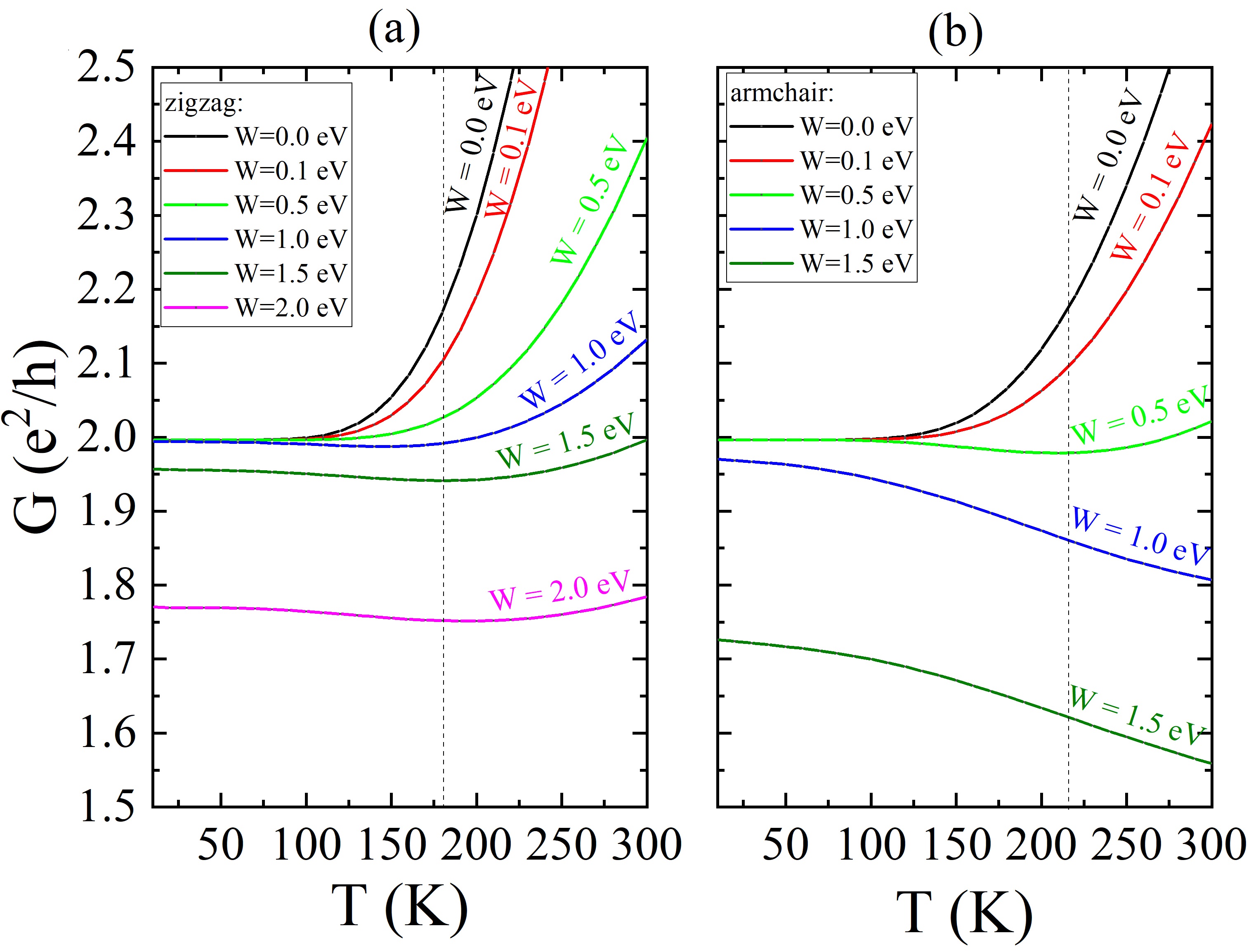}\
\caption{Temperature dependence of conductance for clean and disordered samples. Different colors of $G(T,W)$ for zigzag (a) and armchair (b) denote different values of disorder strength.}\
\label{fig4}
\end{figure}\
We now consider a disordered quantum spin Hall edge in the presence of a non-zero temperature. 
The effect of increased resistance (decreased conductance) close to the band edges (see Fig.~\ref{fig3}) is important in understanding the temperature stability of the approximately quantized conductance. 
In our simulation of free electrons, temperature will broaden the distribution of occupied electron states. 
To model this, we calculate for both clean and disordered 20 $\times$ 20 nm$^2$ system the temperature-dependent conductance, given by the integral,  
\begin{equation}
G(T)=\frac{e^{2}}{4hk_{B}T}\int_{-\infty}^{+\infty}\mathcal{T}(E)\cosh^{-2}\left( \frac{E-E_{F}}{2k_{B}T}\right)dE,
\end{equation}
using 1 meV energy $E$ discretization and again setting the Fermi level to middle of the bulk gap,  $E_F = 82.5$ meV. 
The dependence of the integration kernel on energy and disorder is studied in \hyperref[app:G]{Appendix G}. We conclude that due to the reduced conductance of the bulk states close to the edge of the band, the temperature stability of $G=2$ e$^{2}$/h conductance quantization is enhanced in realistic samples with moderate disorder, as shown e.g. for the response of $G(T)$ in Fig.~\ref{fig4} for $W=0.5$ eV compared to $W=0$. 
When the disorder strength is further increased, the zero-temperature conductance is no longer quantized but falls below $2$ e$^{2}$/h and the minimum conductance moves to a non-zero temperature, see the largest values of $W$ in Fig.~\ref{fig4}. 
Interestingly, the density of bulk states near the band edge can be also reduced by nanostructuring, i.e. by decreasing the width of the ribbon. Scaling of $G(T)$ with respect to ribbon width (and increasing number of bulk states)  in clean systems is shown in \hyperref[app:G]{Appendix G}, Fig.~\ref{figA7}. Linear extrapolation of our results for large ribbon widths leads to close match with recently measured values \cite{Wu_Jarillo-Herrero_2018}, suggesting low to moderate level of disorder in those experiments. 

\section{Helical 1D liquid properties}

\begin{figure}[t]\
\includegraphics[scale=0.45]{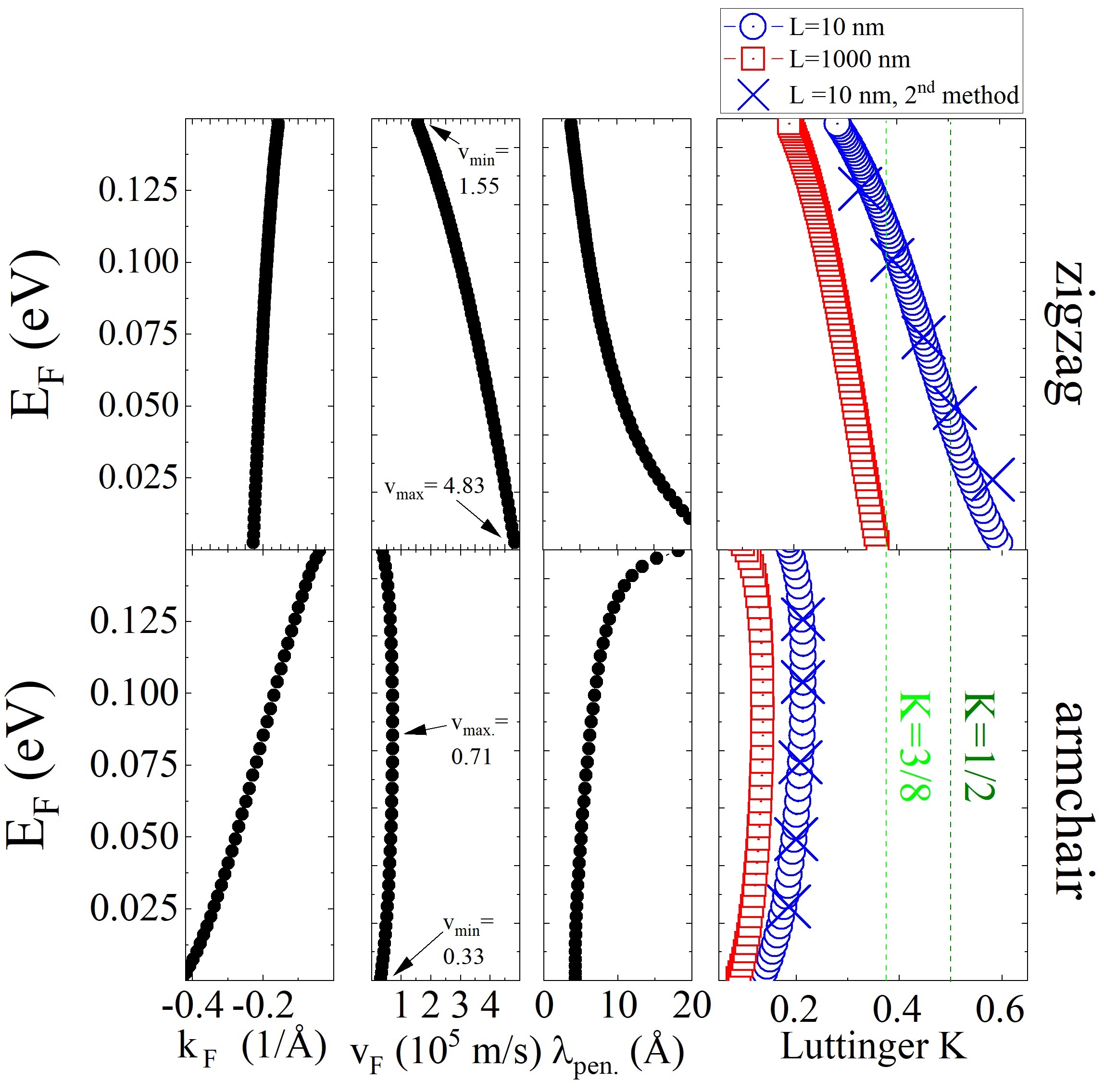}\
\caption{Luttinger liquid parameter $K$ estimation in ribbons. Top panels correspond to zigzag termination and bottom ones to armchair. From left to right: $E_{F}$ vs $k_{F}$ dispersion, corresponding Fermi velocity $v_{F}$, penetration depth $\lambda_{pen.}$ and $K$ parameter is presented for different Fermi energies $E_{F}$ on the y-axis. Two different lengths of channels used for $K$ estimation are given by red and blue curves on the rightmost panels. Blue crosses represent $K$ estimation using microscopic ribbon wave functions, as described in the text. }\
\label{fig5}
\end{figure}\
In the last part we discuss the role of strong electron-electron interactions. Because of the strong 1D localization, one may expect effects related to the helical Tomonaga-Luttinger liquid (TLL). It is well established that in two-dimensional semiconductors Coulomb interactions screening is reduced as compared to 3D. 
This may lead to strong electron-electron interactions in quasi-1D channels in such systems, as recently reviewed in Ref.~[\onlinecite{Song_Gao_2021}]. 
In TLL theory, the short-range interaction strength is characterized by the dimensionless parameter $K$ which equals 1 in the absence of interactions while $K<1$ for repulsive interactions. 
Although in general it is difficult to predict the value of $K$ theoretically  \cite{Muller_Kashuba_2017}, several attempts are available in literature  \cite{Glazman_Shklovskii_1992, Teo_Kane_2009, Maciejko_Zhang_2009}. 
Following Ref. \cite{Stuhler_Claessen_2020} we estimated a value of $K$ for edge states in both zigzag and armchair ribbons by using the formula 
\begin{equation}
\begin{split}
K=&\bigg[1-\frac{e^2}{\pi^2 \hbar v_{F} \varepsilon_{0} \left(\varepsilon_r + 1\right)}\times\\
&\textrm{ln}\left(e^{-\frac{1}{2}}2^{-\varepsilon_r}\frac{w}{L}+e^{1-\gamma}2^{-1}\frac{\lambda_{pen.}}{L}\right) \bigg]^{-\frac{1}{2}}  \,,
\label{eq6}
\end{split} 
\end{equation}
where for hBN encapsulated WTe$_{2}$ 
the relative static dielectric constant of hBN is $\varepsilon_r=4.5$, the effective thickness of the system $w=0.7\cdot 10^{-9}$ m.
The Fermi velocity $v_F$ and the edge state penetration depth $\lambda_{pen.}$ both depend on the Fermi energy and are calculated in \hyperref[app:C]{Appendix C} and shown in Fig. \ref{fig5}. 
We compare two lengths of the channel, $L=10^{-8}$ m and $L=10^{-6}$ m, and plot 
the Fermi energy dependent $K$ for the zigzag and armchair in-gap edge states in the rightmost panels of Fig.~\ref{fig5}. 
Contrary to usual case \cite{Wu_Zhang_2006}, $K$ varies strongly as a function of Fermi energy, stemming from the energy dependent Dirac velocity and penetration depth of the edge states. For zigzag termination monotonic decrease of $K$ ($0.6\rightarrow 0.3$) with  increasing $E_F$ is predicted, while for armchair maximal value of $K\approx 0.2$ is obtained in the middle of bulk band gap  and decreases slightly when $E_F$ is tuned towards either band edge. 

We confirm the trends obtained from Eq.~(\ref{eq6}) by calculating $K$ using microscopic ribbon  wavefunctions, following $K$ estimation in nanowires and nanotubes  \cite{Li_DasSarma_1991, Ando_2010}, where effective 1D channel radius is introduced. When radius  is chosen to reproduce one value of $K$ in the middle of bulk band gap, the trends predicted by Eq.~\ref{eq6} are nicely reproduced, as shown by blue crosses in Fig.~\ref{fig5}. Further details of this method are presented in \hyperref[app:H]{Appendix H}. 

Our prediction of values of $K$ (which can be $K<3/8$ in both types of edge termination) suggests that when disorder is present, 
insulating behavior may be expected due to Anderson localization of the edge state~\cite{PhysRevB.90.075118}. 
Because in armchair $K<1/4$ one may expect that even a single magnetic impurity will destabilize the edge state~\cite{Wu_Zhang_2006}. 
These effects should be observable in transport experiments as interaction-induced localization of the edge state. 
However, we note that when different dielectric environment  is used we expect that $K$ may increase due to additional screening of  electron-electron interactions. 
Observation of edge state transport together with zero-bias anomaly and   characteristic scaling \cite{Stuhler_Claessen_2020} of DOS near Fermi level would support the scenario of a delocalized TLL in WTe$_2$. 

Interestingly, non-linear contribution to edge dispersion, complicating significantly the Luttinger liquid picture \cite{Imambekov_Glazman_2012}, should in principle be taken into account in both edge terminations (see Fig.~\ref{fig5}), at least for well-defined, perfectly clean edges.  
Moreover, as expected from the Fermi energy dependence of the Luttinger parameter $K$, 
the microscopic two-particle interaction amplitudes' dependence on the Fermi energy and momentum transfer has to be  accounted for in realistic studies of edge physics~\cite{Markhof_Meden_2016}. 
Further studies combining Fermi energy dependence of the velocity and interaction parameters 
together with disorder and temperature effects are necessary to establish phase diagram and response functions in such realistic 1D quantum liquid \cite{Maciejko_Zhang_2009, Imambekov_Glazman_2012, Daviet_Dupuis_2020}. 

{\it{Note added:}} During the preparation of this manuscript, we became aware of the experimental results of the group of B. Weber \cite{Weber_2022} that confirm different $K$ parameters for different edges of WTe$_2$ and support scenario of strong interactions ($K<0.5$) in the studied system. 

\section{Conclusions}
We have analyzed the response of disordered WTe$_{2}$ ribbons with a focus on possible experimental differences due to two types of edge terminations. Our study suggests that careful sample preparation with respect to edge termination can serve as an additional tuning knob to optimize properties of quantum spin Hall edge states in WTe$_2$ and related compounds. We have theoretically rationalized that even for heavily disordered samples, WTe$_2$ states are identifiable in STM and transport measurements, due to their short penetration depth into the bulk. From the modelling of several types of observables such as edge state tunneling spectra, gate tunable longitudinal resistance, length dependence of channel resistance, and temperature dependence of conductance, we conclude that the edge termination  crucially determines the robustness of topological protection (with zigzag edge termination being more robust) and also impacts a possible helical Luttinger liquid description of the QSH edge modes. 

\section*{Acknowledgments}
The authors thank B. Weber, D. Pesin, L. Muechler, T. Helbig, and T. Schwemmer for discussions. We acknowledge support from the Deutsche Forschungsgemeinschaft
(DFG, German Research Foundation) through QUAST
FOR 5249-449872909 (Project P3).
The work in W\"urzburg is further supported by the Deutsche Forschungsgemeinschaft (DFG, German Research Foundation) through Project-ID 258499086-SFB 1170 and the Würzburg-Dresden Cluster of Excellence on Complexity and Topology in Quantum Matter – ct.qmat Project-ID 390858490-EXC 2147.
M. B. further acknowledges financial support from the Polish National Agency for Academic Exchange (NAWA), Poland, grant PPI/APM/2019/1/00085/U/00001. Computing resources from Compute Canada and the Wroc\l aw Center for Networking and Supercomputing are gratefully acknowledged.
J.I.V. acknowledges support by the US Department of Energy (DOE) Office of Science through the Quantum Science Center (QSC, a National Quantum Information Science Research Center). 

\section*{Appendix A: Details of geometry}
\label{app:A}

\begin{figure}[t]
	\centering
	\includegraphics[width=0.48\textwidth]{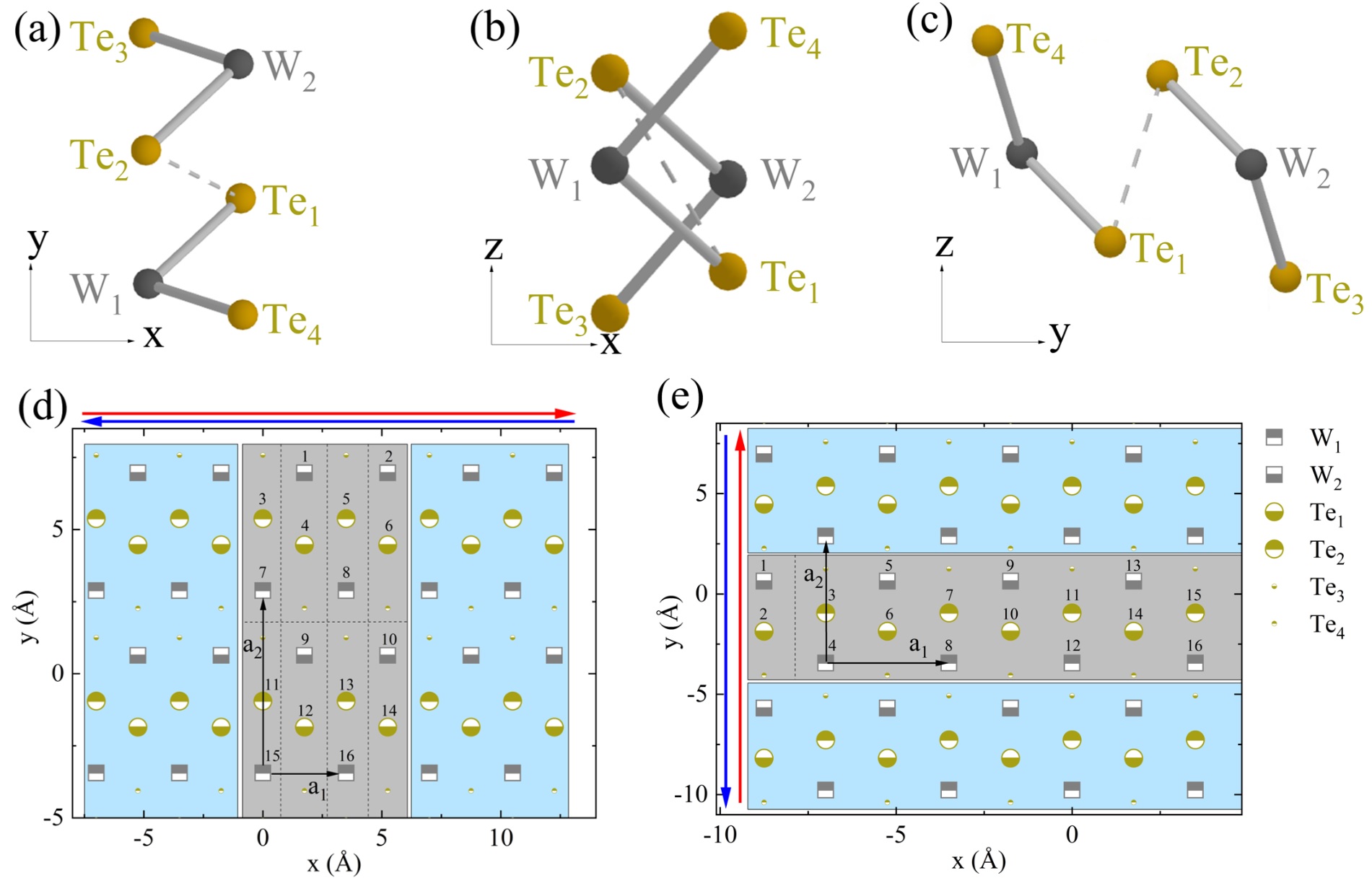}
	\caption{Atom arrangement in 1T' unit cell of the WTe$_{2}$ monolayer for (a) xy, (b) xz, and (c) yz projections. (d) Latice of atoms in a W-terminated "zigzag" ribbon; the gray region shows one "stripe" of the ribbon. (e) Corresponding arrangement of atoms and stripes in a "armchair"-type ribbon. The red and blue arrows on (d-e) denote counterpropagating quantum spin Hall states along the edges.}
	\label{figA1}
\end{figure}
Now we elaborate further on the geometric properties of WTe$_{2}$. Six relevant atom positions inside the unit cell of W$_2$Te$_4$ are (z=0 plane chosen between W atoms) are given in Table \ref{tab:SM1}. In Fig. \ref{figA1} (a-c) we show three projections of those atoms inside the unit cell.
\begin{table}[h]
\caption{\label{tab:SM1}Positions of atoms inside the unit cell.}
\begin{tabular}{l|l|l|l|}
 & x (\AA) & y(\AA) & z(\AA)\\
 \hline
W$_1$  & 1.754 &	4.447 & -1.479 \\
W$_2$  & 0.000 &	5.376 &	 1.479 \\
Te$_1$ & 1.754 &	0.633 & -0.102 \\
Te$_2$ & 0.000 &	2.878 &	 0.102 \\
Te$_3$ & 1.754 &	2.267 &	 2.105 \\
Te$_4$ & 0.000 &	1.244 & -2.105 
\end{tabular}
\end{table}

We stress that in effective tight-binding theory where only four orbitals are used. Ribbons dubbed "zigzag" (W-W or Te-Te chains along the edge, analogous to A-A or B-B carbon chains in graphene ribbons) and "armchair" (effectively W-Te chains, analogous to A-B chains in graphene armchair ribbons) are constructed, as shown in Fig. \ref{figA1} (d) and (e), respectively. We opt for choosing "uniform" basic building "block" for both consisting of eight atoms. For the clean (translationally invariant) case, the zigzag ribbon is chosen to be periodic along the x direction with periodicity $2a_{1}$, while the armchair ribbon is periodic along the y direction with periodicity $a_{2}$. 

\section*{Appendix B: Tight-binding model and ribbon electronic structure}
\label{app:B}
\begin{figure}[t]
	\centering
	\includegraphics[width=0.48\textwidth]{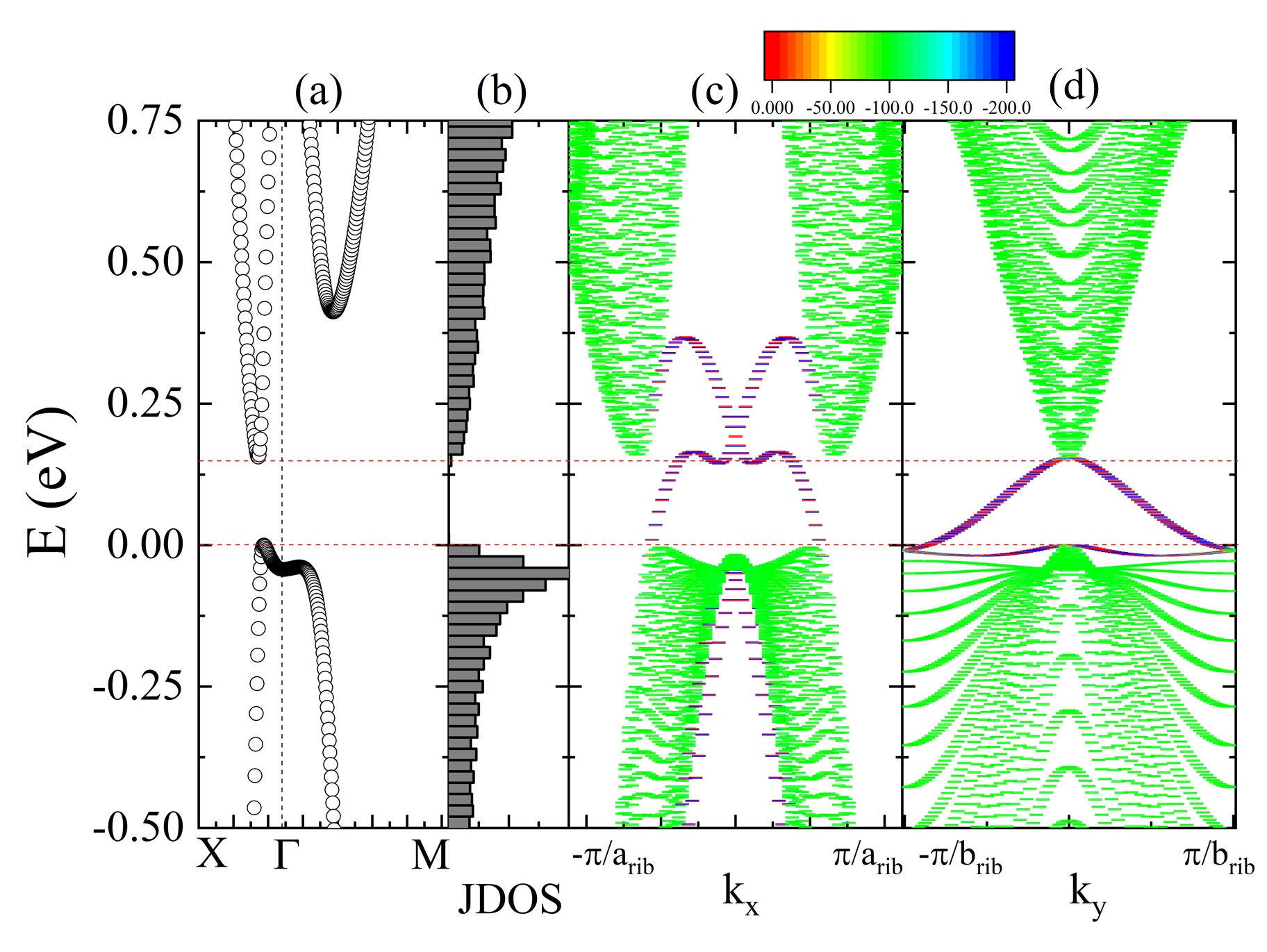}
	\caption{Band structure of bulk and nanoribbons. (a) Band structure near the Fermi level of WTe$_{2}$ obtained from a tight-binding Hamiltonian. (b) Joint energy-resolved density of states in a periodic system (without edge states). Band structures of (c) zigzag and (d) armchair ribbons. The color bar on (c-d) describes where along dimension perpendicular to periodicity (from one edge to the other) the wavefunction density is localized (green - bulk states, red/blue  = edge states on opposite edges).  }
	\label{figA2}
\end{figure}
Now we discuss further details of the tight-binding model. We recall the low-energy effective Wannier orbital model derived in Refs. [\onlinecite{Muechler_Car_2016, Ok_Neupert_2019}] takes into account only 4 orbitals, two $d_{x^2-y^2}$ localized on W$_{1}$ and W$_{2}$ atoms and two $p_{x}$ localized on atoms Te$_{1}$ and Te$_{2}$, see Fig. \ref{figA1} (a-c). Non-zero matrix elements of the 2D system Hamiltonian $\hat{H}_{0}(\vec{k}) $ given by Eq. (\ref{eq1}) are 
\begin{equation}
\begin{split}
&H^{A}_{d} = H^{B}_{d}=E_{d}+2t_{d}\cos(k_{x}a)+2t_{d}^{'}\cos(2k_{x}a), \\
&H^{A}_{p} = H^{B}_{p}=E_{p}+2t_{p}\cos(k_{x}a)+2t_{p}^{'}\cos(2k_{x}a),\\
&H^{AB}_{dd} = t^{AB}_{d}\exp(i\vec{k}\cdot \vec{R}_{W-W})(1+\exp(-ik_{x}a))\exp(-ik_{y}b),\\
&H^{AB}_{dp} = t^{AB}_{d-p}\exp(i\vec{k}\vec{R}_{W-Te}) (1-\exp(-ik_{x}a)) ,\\
&H^{AB}_{pd} = -t^{AB}_{d-p}\exp(i\vec{k}\vec{R}_{W-Te}(1-\exp(-ik_{x}a))),\\
&H^{AB}_{pp} = t^{AB}_{p}\exp(i\vec{k}\vec{R}_{Te-Te}) (1+\exp(-ik_{x}a)).
\end{split}
\end{equation}
The vectors defining atoms inside the unit cell are $\vec{R}_{W-W}=(1.750, 4.081)$ \AA, $\vec{R}_{W-Te}=(1.750, 1.588)$ \AA,$\vec{R}_{Te-Te}=(1.750, -0.905) $ \AA. The parameters (in eV) of this model are 
$E_{d} = 1.3265, E_{p} = -0.4935, t_{d}=-0.28, t_{d}^{'}=0.075, t_{p}=0.93, t_{p}^{'}=0.075, t^{AB}_{d}=0.52, t^{AB}_{p}=0.4, t^{AB}_{d-p}=1.02, V = 0.115 $ eV. Note that we did not include Rashba SOC in this work. The bandstructure of the Hamiltonian $\hat{H}_{tot.}$ along the high-symmetry $X-\Gamma-M$ line is plotted in Fig. \ref{figA2} (a). We checked that in the system without edges (periodic 2D system) there are no in-gap states, as shown by density of states in Fig. \ref{figA2} (b). 

The band structures of the quasi-1D ribbon (width 200 \AA) for  the zigzag and armchair are given in Fig. \ref{figA2} (c) and (d), respectively. The color bar denotes the position of the center of density of the wavefunction for a given $(k_{1d}, E)$ point, showing bulk states in green and top/bottom (left/right) localized edge states for zigzag (armchair) by red/blue.

\section*{Appendix C: Localization of edge states in clean system}
\label{app:C}
\begin{figure}[t]
	\centering
	\includegraphics[width=0.48\textwidth]{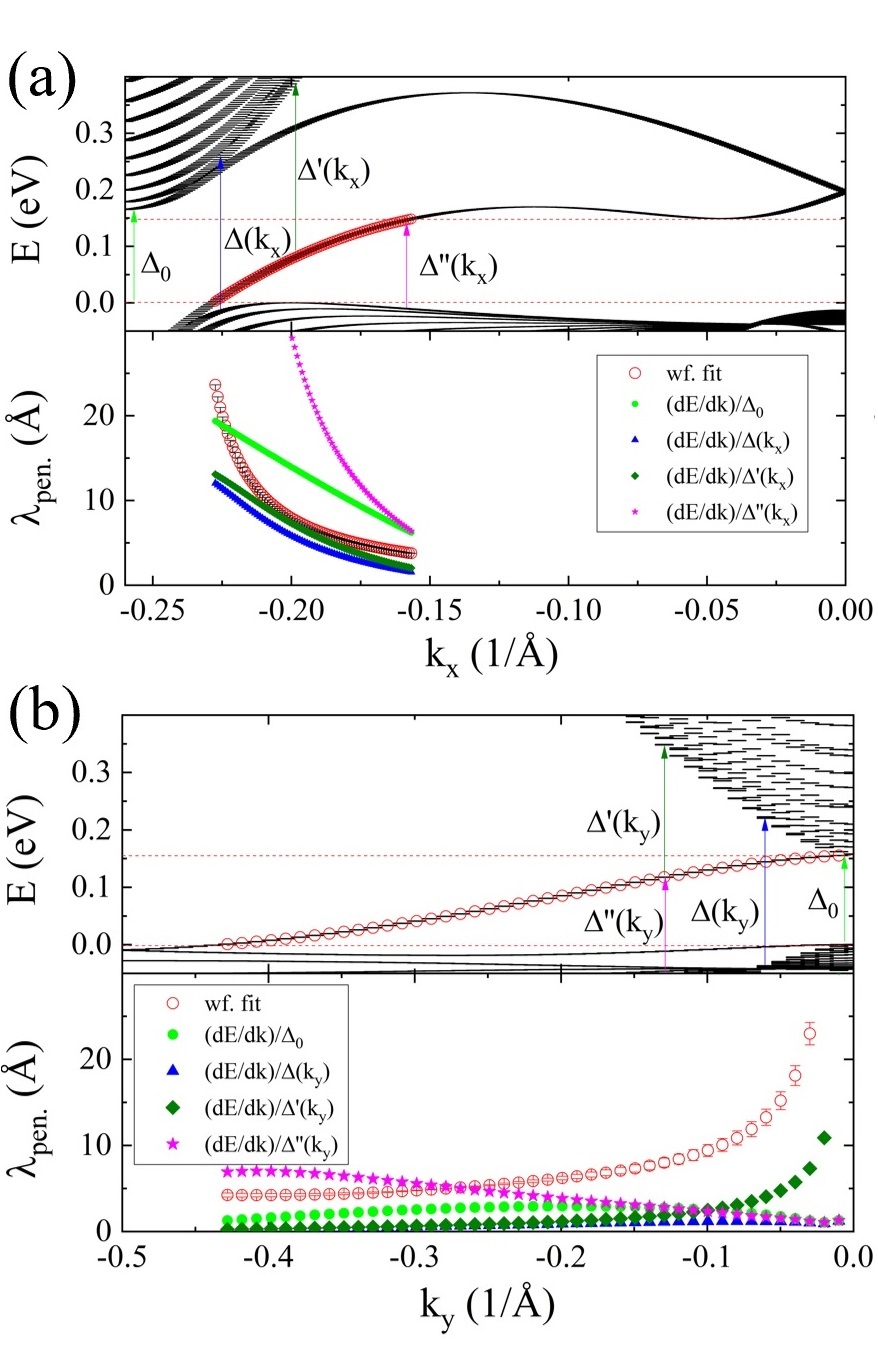}
	\caption{Localization properties of edge states. Red circles in top panels of (a) and (b) represent choice of purely in-gap, non-overlapping edge states. The lower panels show the corresponding numerically obtained penetration depths compared with different models $\lambda_{pen.}$ for (a) zigzag and (b) armchair ribbons.}
	\label{figA3}
\end{figure}
In the next step, the quantum spin Hall edge states shown in Fig. \ref{figA3} (a-b, top panels) are studied. We first specify states of interest as non-overlapping with either ribbon bulk states or themselves, a situation that occurs close to bulk conduction band in zigzag ribbon for 1D wavevector $k\approx -0.15 [1/\textrm{\AA}]$. The rationale behind such a choice is that we want to address the most straightforward situation when edge states are well protected in the topological sense and it is possible to compare numerically obtained penetration depths with theoretical model estimation. First, we calculate the depths $\lambda_{pen.}$ of the states in the ribbon defined for zigzag as $|\Psi|^{2}=A_{0}\exp\left(-y/\lambda_{pen.}\right)$ and for the armchair as $|\Psi|^{2}=A_{0}\exp\left(-x/\lambda_{pen.}\right)$. Before straightforward numerical fitting of those functions to the density of wavefunctions, we average over eight-atom blocks described in \hyperref[app:A]{Appendix A}, assuming average density in each at the center of the "block". We do this to avoid rapid oscillation of density inside those "blocks" which is unavoidable in such a model when atom-projected density is considered. The penetration depth $\lambda_{pen}$  values obtained using this procedure are shown in the bottom panels of Fig. \ref{figA3} and are fully consistent with the results of Ref. \cite{Ok_Neupert_2019}. Analyzing this result, we first note that these depths are related to the direct gap between the ribbon bulk states at a given 1D wavevector k. However, our attempt to the model penetration depth as $\lambda_{pen.}\sim (dE/dk)/\Delta(k)$, using different choices of gaps does not yield satisfactory quantitative values in the full region of interest in the k-space. We analyze $\lambda_{pen.}$ depending on: $\Delta_0$ - global indirect bandgap; $\Delta(k)$ - direct gap at given k between bulk conduction and valence band; $\Delta^{'}(k)$ - direct gap from edge state to bulk conduction band;  $\Delta^{''}(k)$ - similar to former one but from valence band to edge state. This is in contrast to the BHZ model for HgTe QW in which penetration depth could be easily described using state velocity and the gap between bulk states.

\section*{Appendix D: Local density of edge states}
\label{app:D}
\begin{figure}[ht]
	\centering
	\includegraphics[width=0.45\textwidth]{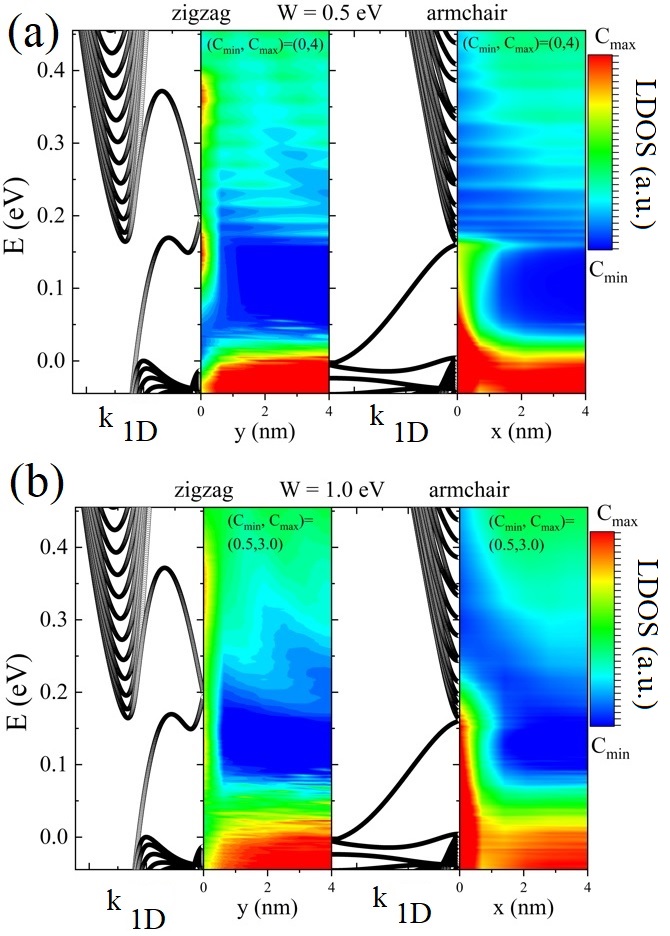}
	\caption{Energy - position LDOS maps for two different values of disorder, (a) $W=0.5$ eV and (b) $W=1.0$ eV. From left to right the panels show the corresponding bandstructure of the zigzag ribbon, the energy-position resolved LDOS map for the zigzag, then the armchair bandstructure and the LDOS map for the armchair. The LDOS scale is given by a pair of $(C_{min}, C_{max})$, $(C_{min}=0, C_{max}=4)$ for (a) and  $(C_{min}=0.5, C_{max}=3.0)$ for (b).   }
	\label{figA4}
\end{figure}

The ability to obtain retarded Green's function $G^{r}$ of our 20 nm x 20 nm scattering region allows us to study the local density of states $A(i,E)=-(1/\pi) \textrm{Im}\sum_{\alpha}G^{r}(i,i,\alpha,E)$ where i denotes respective eight-site "block" of ribbon. We note that in both cases we use Green's functions that are "collapsed" on eight - atom "blocks" due to the complicated structure of those quantities when atom-projection is considered.  Summation over $\alpha$ is therefore over the spin, orbital, and atoms inside the "block". When Anderson disorder is present, we first average over $10^{3}$ disorder realizations and subsequently over different stripes of the ribbons, which corresponds to the experiment in which edge LDOS is summed over many lines scanned perpendicular to the sample edge. We motivate this procedure by interest in general features of LDOS in position-energy maps (e.g., the role of the Dirac cone position).

In Fig. \ref{figA4} we compare energy-position LDOS maps for two different values of disorder strength W for both zigzag and armchair ribbons. The corresponding band structure of clean ribbons is shown. We note also that on x-axis we show a zoom to one edge of the ribbon (0-4 nm) which still has width 20 nm; therefore most of the density of bulk ribbon states, especially in CB, is localized near center of the map (=10 nm) and not visible in our plots. We stress that each map has a different color scale. This is because disorder broadening of LDOS introduces a "background" signal which can be subtracted (and probably is in realistic STM experiments) for clarity. The first striking observation is that when disorder is included, edge states become much more visible in LDOS. The second intriguing feature is the behavior of LDOS in the bulk conduction band. For zigzag, because the edge state forming Dirac cone inside the CB there is overlap of edge and bulk states, resulting in a strong signal from the edge. This feature is not present in the armchair due to the significantly different position of the 1D Dirac cone, which overlaps with the valence band. Interestingly, this effect survives even in strongly disordered samples ($W = 1.5$ eV, shown in the main text) and disappears for values of $W > 2.0$ eV. 

\section*{Appendix E: Conductance as a function of the Fermi energy}
\label{app:E}
\begin{figure}[t]
	\centering
	\includegraphics[width=0.45\textwidth]{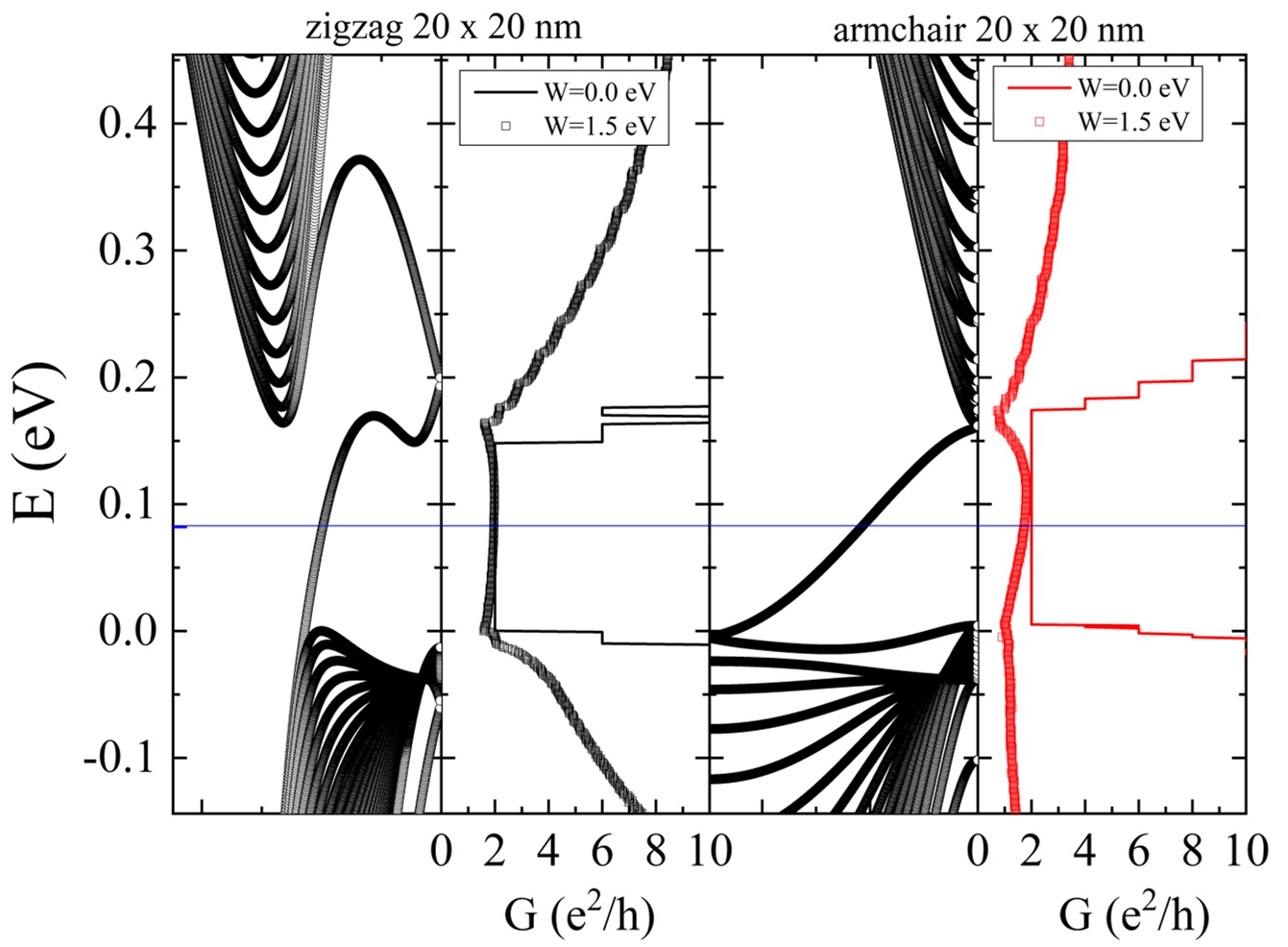}
	\caption{Comparison of conductance $G$ in zigzag (black) and armchair (red) ribbons for the clean (solid line) and disordered (W=1.5 eV, rectangles) case. The left panels show the corresponding band structure of clean ribbons.}
	\label{figA5}
\end{figure}
Next, we study the Fermi energy dependence of the conductance G. First, we calculate G in clean samples. The Landauer formula for the differential conductance is given by $G=\frac{e^{2}}{h}\mathcal{T}$, where $\mathcal{T}$ is a transmission coefficient between left and right contacts, calculated using the recursive Green's functions method. Remembering that we divided our scattering region to "slices" (shown, e.g., as grey regions in Fig. \ref{figA1}) enumerated from $1$ to $N$, recursion in this case is efficient due to the fact that only $(1,N)$ part of total Green's function of the system is needed. The inversion of the full matrix can be avoided and only inversions of "slices" are necessary. 
The coupling between the slices is performed using the Dyson equation. Transmission $\mathcal{T}$ is calculated from so called Caroli formula $\mathcal{T} = \textrm{Tr}\left[\Gamma_{L}G^{r}_{1,N}\Gamma_{R}(G^{r}_{1,N})^{\dagger}\right]$, where $G^{r}_{1,N}$ is a matrix representing the retarded Green's function between the first and the N-th slice. $\Gamma_{L(R)} $ is defined as a difference of semi-infinite lead self-energies ${(\Gamma_{L(R)}=\Sigma_{L(R)}-\Sigma_{L(R)}^{\dagger})}$, where electron self-energies are calculated using the Sancho-Rubio iterative algorithm. Those calculations are performed for the non-interacting case and at $T=0$. Semi-infinite leads, attached to the edges of the system, are considered to be made from the same material as the studied system to avoid the contact resistance effect. Disorder can be introduced only in the scattering region.

Then, for the Fermi energy window 1 meV we calculate average over $10^{3}$ disorder realizations for $W=1.5$ eV, producing the $G(E)$  that is plotted in Fig.~\ref{figA5} and used in subsequent finite-temperature calculations in Sec.~\ref{app:G}. Already from the $G(E)$ function we can observe that when the Fermi level is exactly in the middle of the bulk gap, the QSH edge state is protected even in the presence of strong disorder. However, when we tune away from such Fermi level, $G$  is no longer quantized to $2$ e$^{2}$/h. The deviation from this exact quantization is more apparent for the armchair ribbon. In both cases, the deviation is larger when $E_{F}$ is closer to the bulk edge. Then, when the bulk states began to contribute to the conductance G begins to increase. These two effects result in the appearance of drops in G, which, however, are suspected to become smaller in wider samples, as can be deduced from the analysis in Ref. [\onlinecite{Li_Shen_2009}]. These drops are directly responsible  for the appearance of "cusps" in $R$ shown in Fig. \ref{fig2} (b) in the main text.

\section*{Appendix F: Scaling in disordered samples}
\label{app:F}
\begin{figure}[t]
	\centering
	\includegraphics[width=0.45\textwidth]{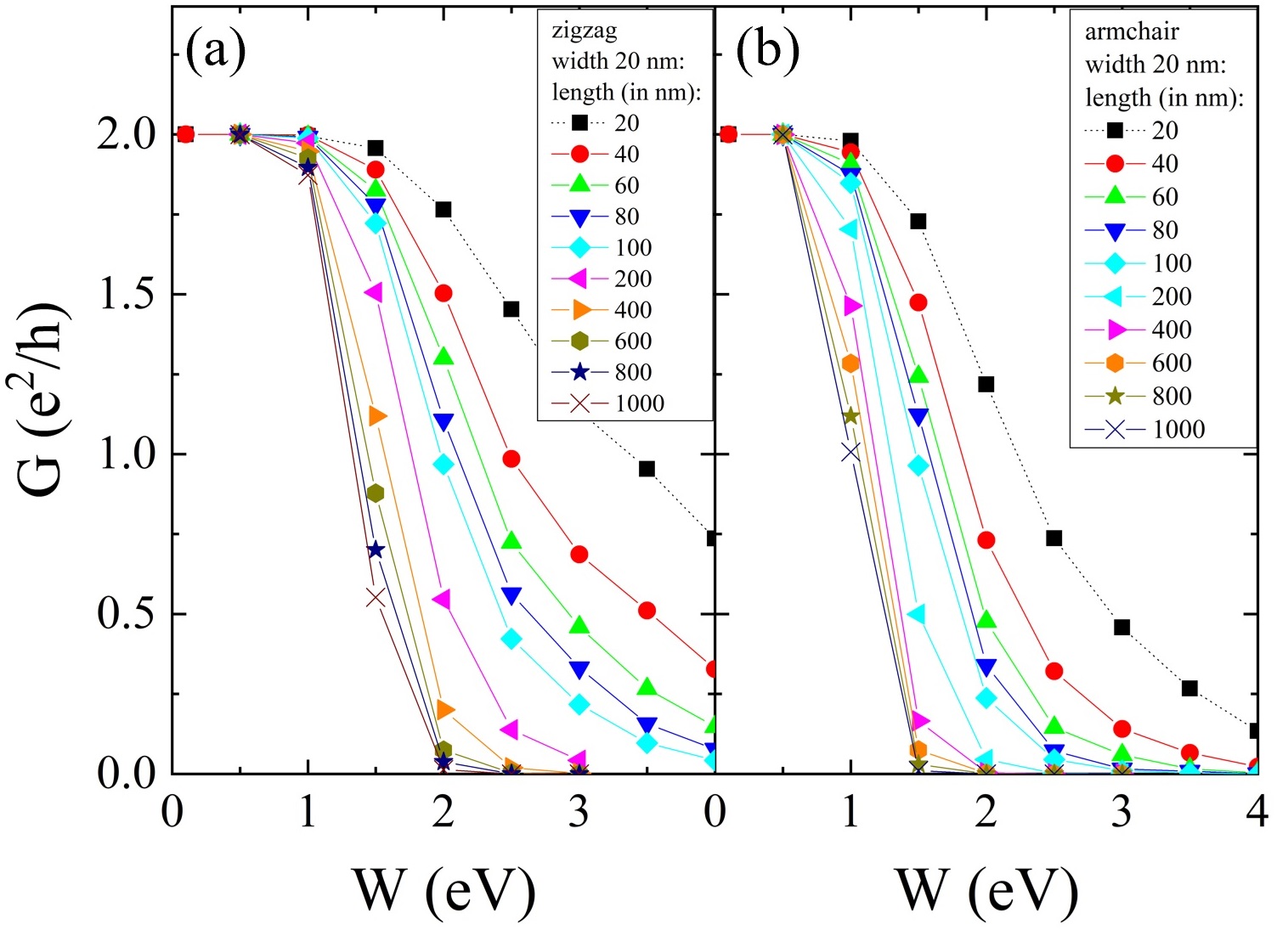}
	\caption{Scaling of G(W) curves as a function of the length of the samples (20 - 1000 nm, constant width = 20 nm) for (a) zigzag and (b) armchair type of the edge. The vertical cut for some disorder strength W gives values of G for different lengths used to calculate longitudinal resistance in Fig. \ref{fig3} (b) in the main text.}
	\label{figA8}
\end{figure}
Now we explain the rationale behind the short-to-long channel transition, as shown in Fig. \ref{fig4} in the main text. Assuming system width to be constant (20 nm), we study the conductance dependence on the disorder strength $G(W)$ as a function of the system length, changing it from 20 nm to 1000 nm. We note that this is in contrast to the experimental setup in Ref. \cite{Wu_Jarillo-Herrero_2018} in which the system width is greater than 1000 nm (and lengths vary between 50 and 1000 nm). In our calculation presented in Fig. \ref{figA8} we first note the general behavior that above some "critical" value of disorder strength (here $\approx 500$ meV) we observe decrease of conductance for longer samples.  This result can be understood  semiclassically as an edge state that has more and more possibilities to percolate to the other side of the sample and backscatter into counterpropagating edge state with the same spin. This drop in conductance means that, for a given sample (with sufficiently uniform disorder with strength $W$), the longitudinal resistance will grow with increasing length. Interestingly, due to the different robustness of transport for two types of ribbon termination, e.g., in the heavily disordered case ($W=1.5$ eV) it should be possible to distinguish between zigzag and armchair edge transport for generalized terminal geometry proposed in Fig. \ref{fig2} in the main text in which such a short-to-long channel transition can be measured along perpendicular edges. We also note that precise estimation of disorder strength in realistic samples is rather difficult, because the "critical" value of $W$ for which short and long channel behavior can be distinguished depends significantly on the width of the sample, moving the value of W for which $G$ begins to deviate from $G=2$ e$^2$/h to larger values, as well as making the function $G(W)$ more steeply vanishing. 
Precise scaling studies of this effect require rather massive computational capabilities and are beyond the scope of this work.

\section*{Appendix G: Temperature effect}
\label{app:G}
\begin{figure*}
	\centering
	\includegraphics[width=0.85\textwidth]{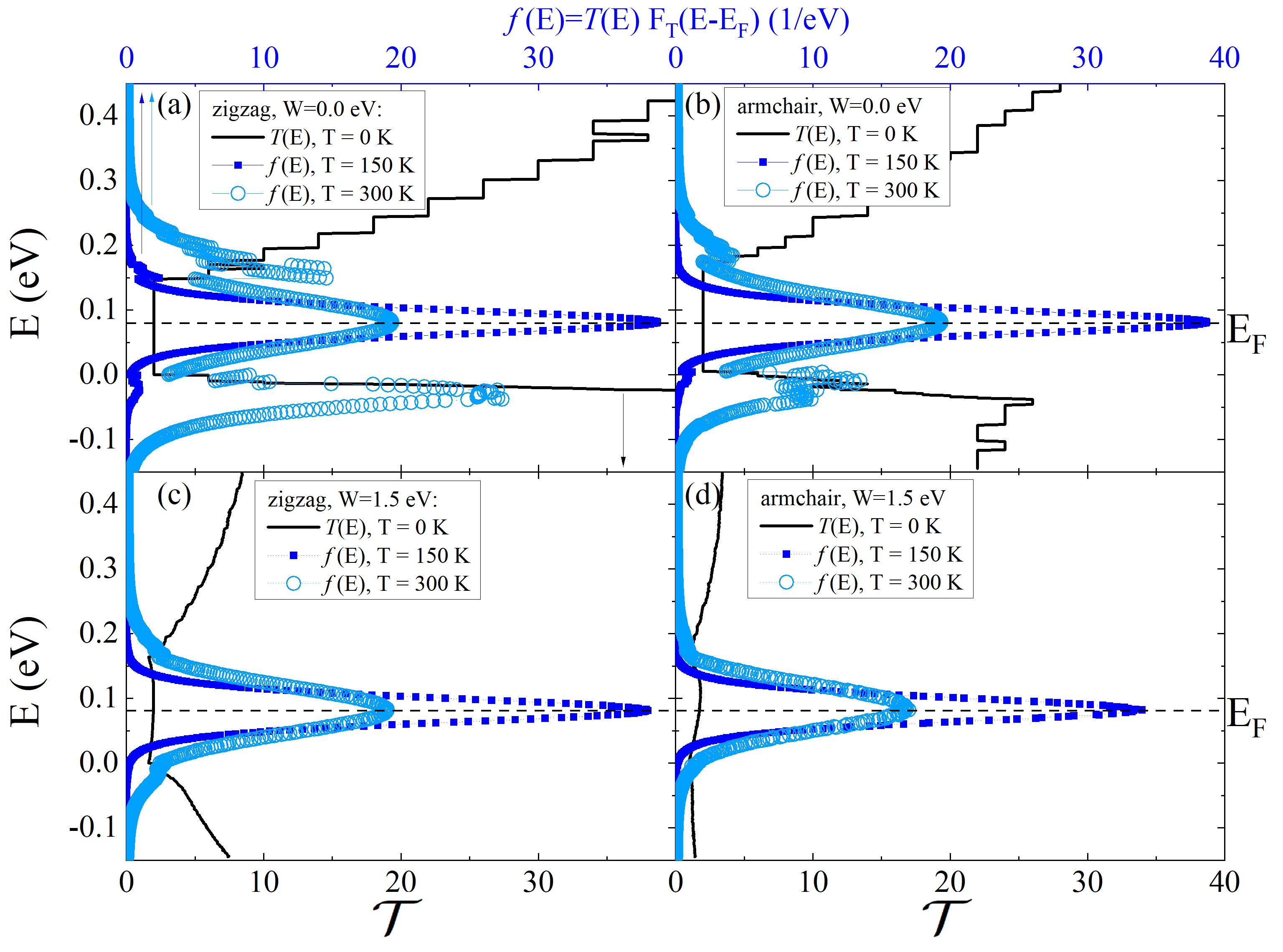}
	\caption{ Transmission $\mathcal{T}$ (black solid lines) of a clean sample at T=0 K and $\it{f}(E)$ function describing thermal broadening in (a) clean zigzag, (b) clean armchair, (c) disordered zigzag, and (d) disordered armchair cases. In all cases we compare the broadening for two temperatures: T=150 K (rectangles) and T=300 K (circles).  }
	\label{figA6}
\end{figure*}

In the next part, we focus on temperature dependence of conductance $G(T)$. We include the temperature using the standard \cite{Datta_1997} thermal broadening function
\begin{equation}
G(T) = \frac{e^{2}}{h}\int\mathcal{T}(E)F_{T}(E-E_{F})dE\,,
\end{equation}
where
\begin{equation}
\begin{split}
F_{T}(E) =& -\frac{\partial}{\partial E}\left(\frac{1}{\exp(E/k_{B}T)+1} \right) \\
& = \frac{1}{4k_{B}T}\textrm{cosh}^{-2}\left(\frac{E}{2k_{B}T} \right)
\end{split}\,,
\end{equation}
and the transmission $\mathcal{T}(E)$ is averaged over $10^{3}$ disorder realizations for each energy $E$. 
The integral is calculated numerically in 600 meV window around center of bulk gap with 1 meV discretization. In Fig. \ref{figA6} we compare the averaged transmission with the kernel of the above integral when the Fermi energy $E_{F}$ is set in the middle of the gap. The left and right panels correspond to zigzag (a,c) and armchair (b,d), while the top (a,b) and bottom (c,d) correspond to clean and disordered systems, respectively. In each panel, we compare the function $\it{f}(E)=\mathcal{T}(E)F_{T}(E-E_{F})$ for two temperatures, $T=150$ K (blue symbols) and $T=300$ K (red symbols). Immediately one can note that for temperature 150 K coupling of thermally broadened function to bulk ribbon states is small. On the other hand, such coupling (red points on top panels) becomes significant at 300 K, which is an especially strong effect in a clean sample. This is related to the large change in conductance when bulk states in a clean ribbon begin to contribute to conductance. On the other hand, in the disordered case, because bulk states become localized and their overall conductance sharply decrease, the effect of thermal broadening is visibly smaller. This explains why paradoxically, when moderate disorder is present, the quantized plateau $G=2$ e$^{2}$/h  is more robust against finite temperature, simply by suppression of the conductance of the bulk states to which the edge states are thermally coupled. 

\begin{figure}[ht]
	\centering
	\includegraphics[width=0.45\textwidth]{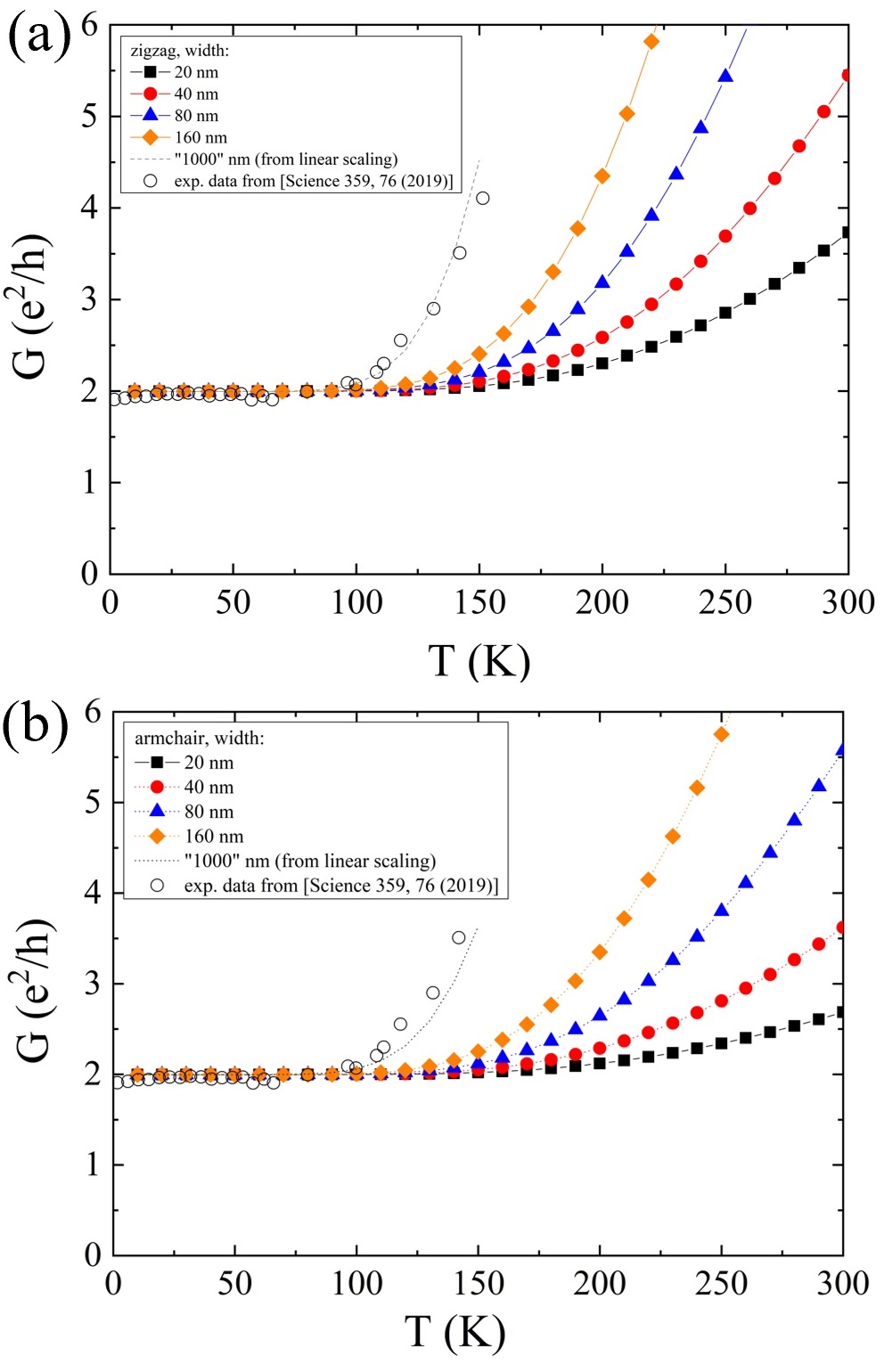}
	\caption{Conductance dependence on temperature in the range 10-300 K for (a) zigzag and (b) armchair ribbons with different widths (20-160 nm). The dotted line shows our linear interpolation to the system width equal to 1000 nm. The open circles on both graphs are experimental values extracted from Ref. [\onlinecite{Wu_Jarillo-Herrero_2018}]. }
	\label{figA7}
\end{figure}
As described above, one can expect that the density of the states of the bulk bands will determine the temperature response of the edge state. Due to the computationally demanding  nature of the problem we were able to calculate the temperature dependence of $G(T)$ only for wider ribbons in the clean (therefore translationally invariant) case, in which we do not need to average over disorder realizations. As expected, the number of bulk states in both CB and VB becomes larger for wider ribbons in both zigzag and armchair cases. After calculating $G(T)$ curves, see Fig. \ref{figA7} for system widths 20, 40, 80 and 160 nm we try to extrapolate this data, which seem to be well described by linear relation. In our extrapolation, we reach the sizes investigated experimentally ($\approx 1000$ nm). Using extrapolated $G(T)$ we observe good match with experimental values, although we cannot claim that this theory can distinguish between zigzag and armchair ribbon type temperature response due to only approximate nature of linear scaling and lack of comparison of this scaling with low/high disordered strength case.

\section*{Appendix H: On Tomonoga-Luttinger liquid theory in WTe$_{2}$}
\label{app:H}

The TLL Hamiltonian for 1D helical edge states has to take into account the energy-dependence of both the  velocity and the coupling constants. For a single edge the Hamiltonian is given by, 
\begin{equation}
\begin{split}
\hat{H} = & \sum_{k\in(-k_{F}-k_{0}, -k_{F}+k_{0})} \hbar v_{F}(k_F)(k+k_{F})\hat{c}^{\dagger}_{k,R\uparrow}\hat{c}_{k,R\uparrow}\\
&+ \sum_{k\in(k_{F}-k_{0},k_{F}+k_{0})} \hbar v_{F}(k_F)(-k+k_{F})\hat{c}^{\dagger}_{k,L\downarrow}\hat{c}_{k,L\downarrow}\\
&+\sum_{k_1 k_2 p}\lambda_{2}(k_1,k_2,p)\hat{c}^{\dagger}_{k_1,L\downarrow}\hat{c}^{\dagger}_{k_2,R\uparrow}\hat{c}_{(k_2 + p)R\uparrow}\hat{c}_{(k_1 - p)L\downarrow}\\
&+\sum_{k_1 k_2 p}\frac{\lambda_{4}(k_1,k_2,p)}{2}\big(\hat{c}^{\dagger}_{k_1,L\downarrow}\hat{c}^{\dagger}_{k_2,L\downarrow}\hat{c}_{(k_2 + p)L\downarrow}\hat{c}_{(k_1 - p)L\downarrow}\\
&+\hat{c}^{\dagger}_{k_1,R\uparrow}\hat{c}^{\dagger}_{k_2,R\uparrow}\hat{c}_{(k_2 + p)R\uparrow}\hat{c}_{(k_1 - p)R\uparrow}\big),
\end{split}
\label{eq:2}
\end{equation}
where for example $\hat{c}_{k,R\uparrow}$ are fermion annihilation operator for right mover with spin up.  
Forward scattering interaction coupling strength is $\lambda_{2} = g_{2\perp}$ is in standard 'g-ology' notation \cite{Solyom_1979}. Chiral interaction is parametrized by $\lambda_{4}=g_{4\parallel}$. The energy and momentum cut-offs, $E_{0}$ and $k_{0}$, determine the applicability of the Hamiltonian Eq.~(\ref{eq:2}). 
Those values correspond to energy/momentum windows for which edge states are non-overlapping with each other and with bulk ribbon states. Such edge states are denoted by, e.g., red circles in Fig. \ref{figA3} on ribbon bandstructure plots. We take $2E_0=150$ meV in zigzag and $2E_0=\Delta_{0}$ armchair and $2k_0=0.071$ 1/\AA \  for the zigzag and $2k_0=0.358$ 1/\AA \  for the armchair edge. As is well known, interactions renormalize the Fermi velocity as
\begin{equation}
v_{F}^{*}=v_{F}\sqrt{\left(1+\frac{\lambda_{4}}{2\pi v_{F}}\right)^{2}-\left(\frac{\lambda_{2}}{2\pi v_{F}}\right)^{2}}\,,
\end{equation}
while the Luttinger liquid parameter is given as
\begin{equation}
K=\sqrt{\frac{ 2\pi v_{F} + \lambda_{4} -\lambda_{2} }{2\pi v_{F} + \lambda_{4} + \lambda_{2}}} \,,
\end{equation}
where $\lambda_2$ and $\lambda_4$ are calculated for $k_1 =k_2 =k_F$ and $p=0$. 

We note that we do not take into account umklapp scattering (2-particle backscattering) which in the clean case becomes imporant only for very specific filling at $k_F=\pi/(2a)$, although for both zigzag and armchair it is in principle possible for the in-gap state in both zigzag (close to the valence band) and armchair (close to the middle of the gap) terminations. Because uniform umklapp scattering is relevant when $K<1/2$, it might be important for both terminations in long and ultra-clean samples. Then, one may expect a gap opening \cite{Wu_Zhang_2006} of the order of $\Delta = \lambda_{u}^{1/2-4K}/a$, where $a$ is ribbon periodicity constant and $\lambda_{u}=g_{3\parallel}$ parametrizes the strength of the umklapp processes
\begin{equation}
\begin{split}
H_{um}=&\sum_{k_1 k_2 p}\frac{\lambda_{u}(k_1,k_2,p)}{2}\\
&\big(\hat{c}^{\dagger}_{k_1,L\downarrow}\hat{c}^{\dagger}_{k_2,L\downarrow}\hat{c}_{(k_2 + p-2k_{F})R\uparrow}\hat{c}_{(k_1 - p-2k_{F})R\uparrow}  \\
&+ \hat{c}^{\dagger}_{k_1,R\uparrow}\hat{c}^{\dagger}_{k_2,R\uparrow}\hat{c}_{(k_2 + p+2k_{F})L\downarrow}\hat{c}_{(k_1 - p+2k_{F})L\downarrow} \big).
\end{split}
\label{eq:2b}
\end{equation}
Finally, we note that we have also ignored the 1-particle backscattering interaction~\cite{Schmidt_Glazman_2012, PhysRevB.90.075118} which in the clean case becomes only important near the edge Dirac point. 

Next, we discuss in more detail how the coupling constant $\lambda_{2}$, defined in Eq. (\ref{eq:2}), depends on Bloch wavefunctions of an infinite ribbon. The simplest Fermi wavevector $k_F$ and momentum transfer $p$ dependent matrix element is given by
\begin{equation}
\begin{split}
 &\lambda_{2}\left(k_{1}=+k_{F},k_{2}=-k_{F},p\right)=\\
&\iint_{R^{3}}d^3rd^3r' V^{3D}\left(\left|\vec{r}-\vec{r}'\right|\right) \Psi^{*}_{L\downarrow}\left(k_{F},\vec{r}\right) \Psi^{*}_{R\uparrow}\left(-k_{F},\vec{r}'\right) \times \\
&\Psi_{R\uparrow}\left(-k_{F}+p,\vec{r}'\right) \Psi_{L\downarrow}\left(k_{F}-p,\vec{r}\right).
		\end{split}
\end{equation}
Ribbon wavefunctions in Bloch form can be written as 
\begin{equation}
\begin{split}
&\Psi_{n}\left(k,\vec{r}\right)=\\
&\frac{1}{\sqrt{N_{\textrm{UC}}}}\sum_{i=1}^{N_{\textrm{UC}}}\sum_{\alpha=1}^{N_{y}}e^{ik(U_{i}+\tau_{\alpha})}\nu_{\alpha}^{n}(k)\varphi_{\alpha}\left(\vec{r}-U_{i}-\vec{\tau}_{\alpha}\right).
		\end{split}
\end{equation}
In the equation above, $k$ is a 1D wavevector, $n$ describes a left/right mover with spin up/down, $N_{UC}$ is the number of unit cells (understood as stripes described in \hyperref[app:A]{Appendix A}) that formally go to infinity and $N_{y}$ is the number of atoms in a given slice of ribbon. The two dimensional vector $\vec{\tau}$ describes the position of a given localized orbital inside the unit cell. The ribbon wavefunction $\nu$ and are obtained from the numerical diagonalization of the Hamiltonian $H(k)$. Localized orbitals are denoted by $\varphi$.

\begin{figure}[ht]
	\centering
	\includegraphics[width=0.45\textwidth]{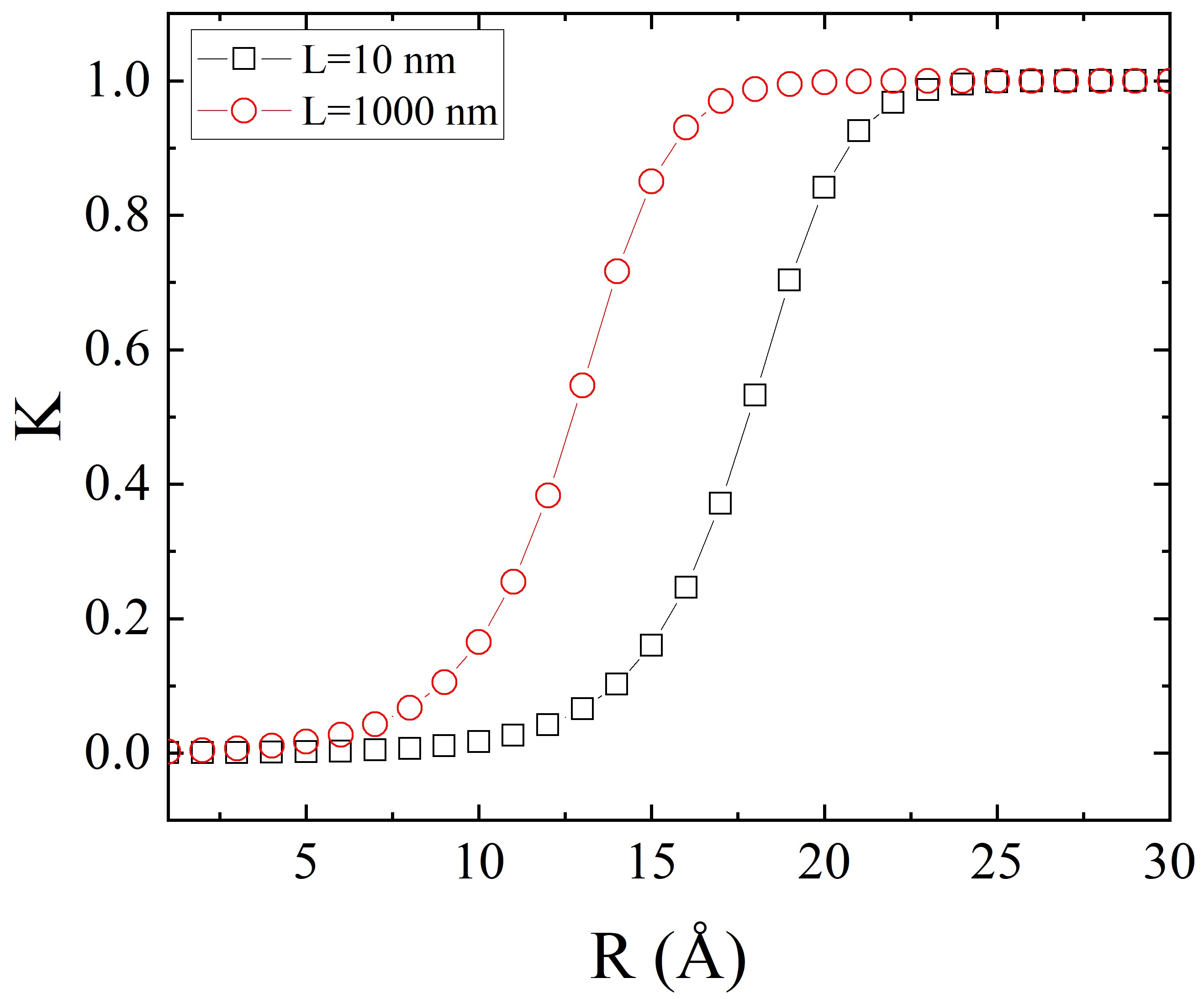}
	\caption{Dependence of the K parameter on the channel radius R and the length of the channel L for the zigzag edge.}
	\label{figA10}
\end{figure}

Next, conceptually following the procedure known from exciton physics where interactions need to be calculated including microscopic wavefunctions \cite{Bieniek_Hawrylak_2020},  we perform 1D Fourier transform of 3D Coulomb interaction, expand pairs of Bloch wavefunctions for the same coordinate in Fourier series and assume structureless delta-like localized orbitals. Then we regularize the short-range Coulomb interaction by introducing the effective Luttinger liquid channel radius R. Long-range regularization can be performed as in the jellium model by excluding $G=0$ from summation over reciprocal 1D. 
The final expression for the coupling constant is
\begin{equation}
\begin{split}
 &\lambda_{2}\left(k_{1}=+k_{F},k_{2}=-k_{F},p\right)\approx\\
&\frac{e^2}{2\pi\varepsilon_{0}L}\sum_{\alpha=1}^{Ny}\sum_{\alpha'=1}^{Ny} \nu^{*L\downarrow}_{\alpha}(k_{F})  \nu^{L\downarrow}_{\alpha}(k_{F}-p) \times\\ 
&\nu^{*R\uparrow}_{\alpha'}(-k_{F}) \nu^{R\uparrow}_{\alpha'}(-k_{F}+p)e^{ip(-\tau_{x,\alpha}+\tau_{x,\alpha})} \times \\
 &\sum_{G\ne 0}\frac{1}{\varepsilon_{r}(G,p,\tau,R)}\times \\
&K_{0}\left[(-G+p)\sqrt{(\tau_{y\alpha}-\tau_{y'\alpha'})^2 + (\tau_{z\alpha}-\tau_{z'\alpha'})^2 + R^2}\right] \,,
    \end{split}
\end{equation}
where $K_{0}$ are modified Bessel functions of the second kind. The Luttinger $K$ parameter is then calculated (using $\lambda_{2}=\lambda_{4}$) as $K=1/(1+\lambda_{2}/(\pi\hbar v_{F}))$. We note that $K$ calculated in this way depends on the effective 1D channel "radius" R, as shown in Fig. \ref{figA10}.

To address the Fermi energy dependence of K in this method we first choose the channel radius reproducing value obtained from the model in Ref. \cite{Stuhler_Claessen_2020}. For the energy in the middle of the energy window ($E_F=0.075$ eV) considered we choose R=17.52 \AA \ and R=15.81 \AA \ for zigzag and armchair for models to match the values of K. As shown in the rightmost panel of Fig. \ref{fig5} (blue crosses), the presented method gives a very similar  dependence of $K$ on the Fermi energy. 

\begin{figure}[ht]
	\centering
	\includegraphics[width=0.45\textwidth]{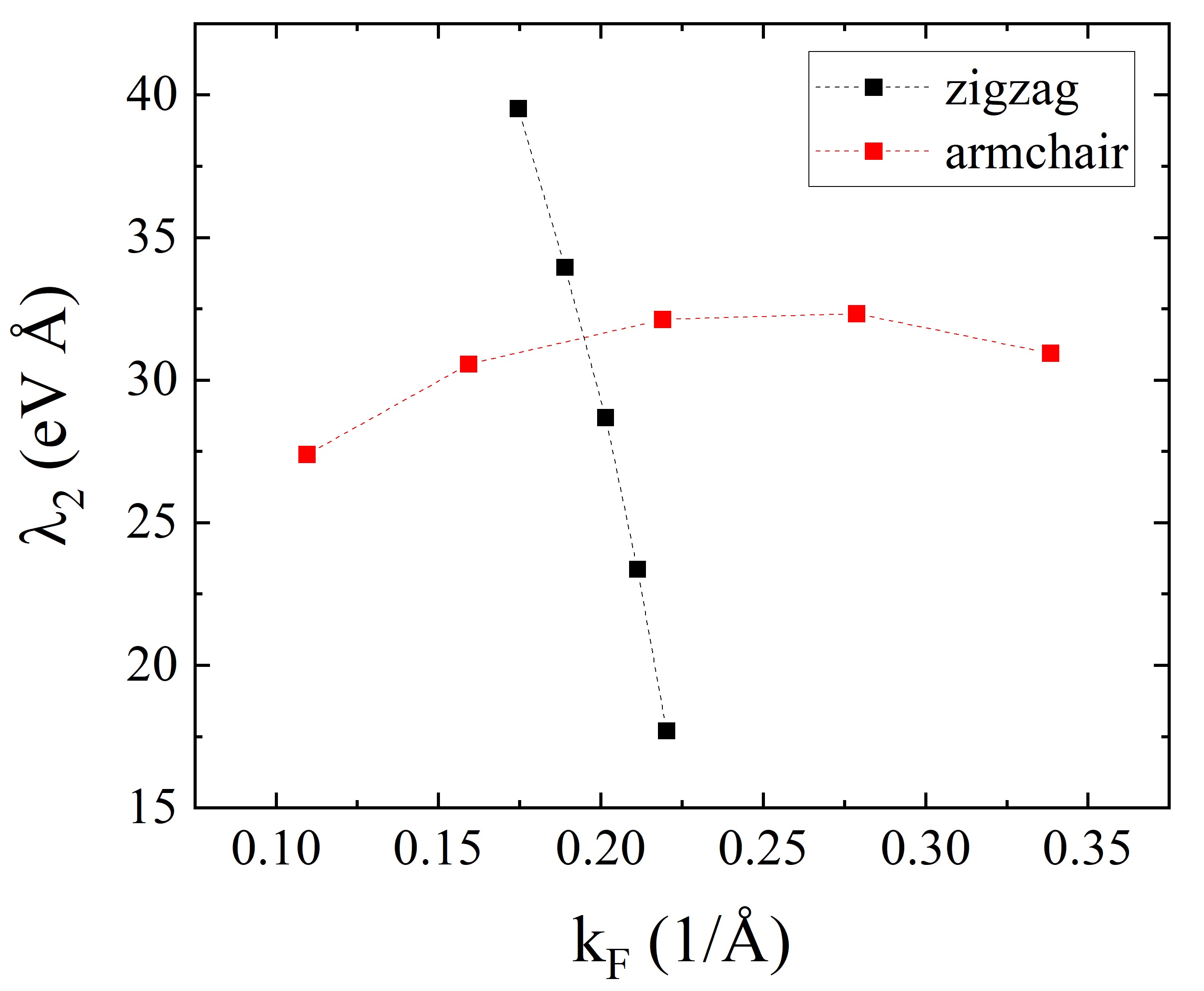}
	\caption{Coupling constant $\lambda_{2}$ dependence on the Fermi wavevector for two types of edge termination. }
	\label{figA11}
\end{figure}
It is also instructive to analyze the dependence of the coupling constant $\lambda_{2}$  on the Fermi wavevector $k_{F}$. Analyzing $\lambda_{2}$ that gave $K$ values shown in Fig. \ref{fig5}, one can note in Fig. \ref{figA11} that the qualitative behaviour of the coupling constant depends on the edge termination. For zigzag $\lambda_{2}$ changes significantly across the energy/wavevector window  analyzed, trend that can be approximately captured by the linear function $\lambda_{2} = [122 -472\cdot k_{F} (1/\textrm{\AA})]$ (eV$\cdot$\AA). For the armchair edge, the change of the coupling constant with the wavevector is significantly smaller and can be considered as constant (32 eV$\cdot$\AA).

\bibliographystyle{apsrev4-2}
\bibliography{ver5_WTe2_23Jun2022}

\begin{thebibliography}{121}%
\makeatletter
\providecommand \@ifxundefined [1]{%
 \@ifx{#1\undefined}
}%
\providecommand \@ifnum [1]{%
 \ifnum #1\expandafter \@firstoftwo
 \else \expandafter \@secondoftwo
 \fi
}%
\providecommand \@ifx [1]{%
 \ifx #1\expandafter \@firstoftwo
 \else \expandafter \@secondoftwo
 \fi
}%
\providecommand \natexlab [1]{#1}%
\providecommand \enquote  [1]{``#1''}%
\providecommand \bibnamefont  [1]{#1}%
\providecommand \bibfnamefont [1]{#1}%
\providecommand \citenamefont [1]{#1}%
\providecommand \href@noop [0]{\@secondoftwo}%
\providecommand \href [0]{\begingroup \@sanitize@url \@href}%
\providecommand \@href[1]{\@@startlink{#1}\@@href}%
\providecommand \@@href[1]{\endgroup#1\@@endlink}%
\providecommand \@sanitize@url [0]{\catcode `\\12\catcode `\$12\catcode
  `\&12\catcode `\#12\catcode `\^12\catcode `\_12\catcode `\%12\relax}%
\providecommand \@@startlink[1]{}%
\providecommand \@@endlink[0]{}%
\providecommand \url  [0]{\begingroup\@sanitize@url \@url }%
\providecommand \@url [1]{\endgroup\@href {#1}{\urlprefix }}%
\providecommand \urlprefix  [0]{URL }%
\providecommand \Eprint [0]{\href }%
\providecommand \doibase [0]{https://doi.org/}%
\providecommand \selectlanguage [0]{\@gobble}%
\providecommand \bibinfo  [0]{\@secondoftwo}%
\providecommand \bibfield  [0]{\@secondoftwo}%
\providecommand \translation [1]{[#1]}%
\providecommand \BibitemOpen [0]{}%
\providecommand \bibitemStop [0]{}%
\providecommand \bibitemNoStop [0]{.\EOS\space}%
\providecommand \EOS [0]{\spacefactor3000\relax}%
\providecommand \BibitemShut  [1]{\csname bibitem#1\endcsname}%
\let\auto@bib@innerbib\@empty
\bibitem [{\citenamefont {Hasan}\ and\ \citenamefont
  {Kane}(2010)}]{Hasan_Kane_2010}%
  \BibitemOpen
  \bibfield  {author} {\bibinfo {author} {\bibfnamefont {M.~Z.}\ \bibnamefont
  {Hasan}}\ and\ \bibinfo {author} {\bibfnamefont {C.~L.}\ \bibnamefont
  {Kane}},\ }\href {https://doi.org/10.1103/RevModPhys.82.3045} {\bibfield
  {journal} {\bibinfo  {journal} {Rev. Mod. Phys.}\ }\textbf {\bibinfo {volume}
  {82}},\ \bibinfo {pages} {3045} (\bibinfo {year} {2010})}\BibitemShut
  {NoStop}%
\bibitem [{\citenamefont {Qi}\ and\ \citenamefont
  {Zhang}(2011)}]{Qi_Zhang_2011}%
  \BibitemOpen
  \bibfield  {author} {\bibinfo {author} {\bibfnamefont {X.-L.}\ \bibnamefont
  {Qi}}\ and\ \bibinfo {author} {\bibfnamefont {S.-C.}\ \bibnamefont {Zhang}},\
  }\href {https://doi.org/10.1103/RevModPhys.83.1057} {\bibfield  {journal}
  {\bibinfo  {journal} {Rev. Mod. Phys.}\ }\textbf {\bibinfo {volume} {83}},\
  \bibinfo {pages} {1057} (\bibinfo {year} {2011})}\BibitemShut {NoStop}%
\bibitem [{\citenamefont {Alicea}(2012)}]{Alicea_2012}%
  \BibitemOpen
  \bibfield  {author} {\bibinfo {author} {\bibfnamefont {J.}~\bibnamefont
  {Alicea}},\ }\href {https://doi.org/10.1088/0034-4885/75/7/076501} {\bibfield
   {journal} {\bibinfo  {journal} {Reports on Progress in Physics}\ }\textbf
  {\bibinfo {volume} {75}},\ \bibinfo {pages} {076501} (\bibinfo {year}
  {2012})}\BibitemShut {NoStop}%
\bibitem [{\citenamefont {Ren}\ \emph {et~al.}(2016)\citenamefont {Ren},
  \citenamefont {Qiao},\ and\ \citenamefont {Niu}}]{Ren_Niu_2016}%
  \BibitemOpen
  \bibfield  {author} {\bibinfo {author} {\bibfnamefont {Y.}~\bibnamefont
  {Ren}}, \bibinfo {author} {\bibfnamefont {Z.}~\bibnamefont {Qiao}},\ and\
  \bibinfo {author} {\bibfnamefont {Q.}~\bibnamefont {Niu}},\ }\href
  {https://doi.org/10.1088/0034-4885/79/6/066501} {\bibfield  {journal}
  {\bibinfo  {journal} {Reports on Progress in Physics}\ }\textbf {\bibinfo
  {volume} {79}},\ \bibinfo {pages} {066501} (\bibinfo {year}
  {2016})}\BibitemShut {NoStop}%
\bibitem [{\citenamefont {Culcer}\ \emph {et~al.}(2020)\citenamefont {Culcer},
  \citenamefont {Keser}, \citenamefont {Li},\ and\ \citenamefont
  {Tkachov}}]{Culcer_Tkachov_2020}%
  \BibitemOpen
  \bibfield  {author} {\bibinfo {author} {\bibfnamefont {D.}~\bibnamefont
  {Culcer}}, \bibinfo {author} {\bibfnamefont {A.~C.}\ \bibnamefont {Keser}},
  \bibinfo {author} {\bibfnamefont {Y.}~\bibnamefont {Li}},\ and\ \bibinfo
  {author} {\bibfnamefont {G.}~\bibnamefont {Tkachov}},\ }\href
  {https://doi.org/10.1088/2053-1583/ab6ff7} {\bibfield  {journal} {\bibinfo
  {journal} {2D Materials}\ }\textbf {\bibinfo {volume} {7}},\ \bibinfo {pages}
  {022007} (\bibinfo {year} {2020})}\BibitemShut {NoStop}%
\bibitem [{\citenamefont {Kane}\ and\ \citenamefont
  {Mele}(2005{\natexlab{a}})}]{Kane_Mele_2005a}%
  \BibitemOpen
  \bibfield  {author} {\bibinfo {author} {\bibfnamefont {C.~L.}\ \bibnamefont
  {Kane}}\ and\ \bibinfo {author} {\bibfnamefont {E.~J.}\ \bibnamefont
  {Mele}},\ }\href {https://doi.org/10.1103/PhysRevLett.95.226801} {\bibfield
  {journal} {\bibinfo  {journal} {Phys. Rev. Lett.}\ }\textbf {\bibinfo
  {volume} {95}},\ \bibinfo {pages} {226801} (\bibinfo {year}
  {2005}{\natexlab{a}})}\BibitemShut {NoStop}%
\bibitem [{\citenamefont {Kane}\ and\ \citenamefont
  {Mele}(2005{\natexlab{b}})}]{Kane_Mele_2005b}%
  \BibitemOpen
  \bibfield  {author} {\bibinfo {author} {\bibfnamefont {C.~L.}\ \bibnamefont
  {Kane}}\ and\ \bibinfo {author} {\bibfnamefont {E.~J.}\ \bibnamefont
  {Mele}},\ }\href {https://doi.org/10.1103/PhysRevLett.95.146802} {\bibfield
  {journal} {\bibinfo  {journal} {Phys. Rev. Lett.}\ }\textbf {\bibinfo
  {volume} {95}},\ \bibinfo {pages} {146802} (\bibinfo {year}
  {2005}{\natexlab{b}})}\BibitemShut {NoStop}%
\bibitem [{\citenamefont {Yao}\ \emph {et~al.}(2007)\citenamefont {Yao},
  \citenamefont {Ye}, \citenamefont {Qi}, \citenamefont {Zhang},\ and\
  \citenamefont {Fang}}]{Yao_Fang_2007}%
  \BibitemOpen
  \bibfield  {author} {\bibinfo {author} {\bibfnamefont {Y.}~\bibnamefont
  {Yao}}, \bibinfo {author} {\bibfnamefont {F.}~\bibnamefont {Ye}}, \bibinfo
  {author} {\bibfnamefont {X.-L.}\ \bibnamefont {Qi}}, \bibinfo {author}
  {\bibfnamefont {S.-C.}\ \bibnamefont {Zhang}},\ and\ \bibinfo {author}
  {\bibfnamefont {Z.}~\bibnamefont {Fang}},\ }\href
  {https://doi.org/10.1103/PhysRevB.75.041401} {\bibfield  {journal} {\bibinfo
  {journal} {Phys. Rev. B}\ }\textbf {\bibinfo {volume} {75}},\ \bibinfo
  {pages} {041401} (\bibinfo {year} {2007})}\BibitemShut {NoStop}%
\bibitem [{\citenamefont {Li}\ \emph {et~al.}(2018)\citenamefont {Li},
  \citenamefont {Hanke}, \citenamefont {Hankiewicz}, \citenamefont {Reis},
  \citenamefont {Sch\"afer}, \citenamefont {Claessen}, \citenamefont {Wu},\
  and\ \citenamefont {Thomale}}]{Li_Thomale_2018}%
  \BibitemOpen
  \bibfield  {author} {\bibinfo {author} {\bibfnamefont {G.}~\bibnamefont
  {Li}}, \bibinfo {author} {\bibfnamefont {W.}~\bibnamefont {Hanke}}, \bibinfo
  {author} {\bibfnamefont {E.~M.}\ \bibnamefont {Hankiewicz}}, \bibinfo
  {author} {\bibfnamefont {F.}~\bibnamefont {Reis}}, \bibinfo {author}
  {\bibfnamefont {J.}~\bibnamefont {Sch\"afer}}, \bibinfo {author}
  {\bibfnamefont {R.}~\bibnamefont {Claessen}}, \bibinfo {author}
  {\bibfnamefont {C.}~\bibnamefont {Wu}},\ and\ \bibinfo {author}
  {\bibfnamefont {R.}~\bibnamefont {Thomale}},\ }\href
  {https://doi.org/10.1103/PhysRevB.98.165146} {\bibfield  {journal} {\bibinfo
  {journal} {Phys. Rev. B}\ }\textbf {\bibinfo {volume} {98}},\ \bibinfo
  {pages} {165146} (\bibinfo {year} {2018})}\BibitemShut {NoStop}%
\bibitem [{\citenamefont {Reis}\ \emph {et~al.}(2017)\citenamefont {Reis},
  \citenamefont {Li}, \citenamefont {Dudy}, \citenamefont {Bauernfeind},
  \citenamefont {Glass}, \citenamefont {Hanke}, \citenamefont {Thomale},
  \citenamefont {Schäfer},\ and\ \citenamefont
  {Claessen}}]{Reis_Claessen_2017}%
  \BibitemOpen
  \bibfield  {author} {\bibinfo {author} {\bibfnamefont {F.}~\bibnamefont
  {Reis}}, \bibinfo {author} {\bibfnamefont {G.}~\bibnamefont {Li}}, \bibinfo
  {author} {\bibfnamefont {L.}~\bibnamefont {Dudy}}, \bibinfo {author}
  {\bibfnamefont {M.}~\bibnamefont {Bauernfeind}}, \bibinfo {author}
  {\bibfnamefont {S.}~\bibnamefont {Glass}}, \bibinfo {author} {\bibfnamefont
  {W.}~\bibnamefont {Hanke}}, \bibinfo {author} {\bibfnamefont
  {R.}~\bibnamefont {Thomale}}, \bibinfo {author} {\bibfnamefont
  {J.}~\bibnamefont {Schäfer}},\ and\ \bibinfo {author} {\bibfnamefont
  {R.}~\bibnamefont {Claessen}},\ }\href
  {https://doi.org/10.1126/science.aai8142} {\bibfield  {journal} {\bibinfo
  {journal} {Science}\ }\textbf {\bibinfo {volume} {357}},\ \bibinfo {pages}
  {287} (\bibinfo {year} {2017})}\BibitemShut {NoStop}%
\bibitem [{\citenamefont {St{\"u}hler}\ \emph {et~al.}(2020)\citenamefont
  {St{\"u}hler}, \citenamefont {Reis}, \citenamefont {M{\"u}ller},
  \citenamefont {Helbig}, \citenamefont {Schwemmer}, \citenamefont {Thomale},
  \citenamefont {Sch{\"a}fer},\ and\ \citenamefont
  {Claessen}}]{Stuhler_Claessen_2020}%
  \BibitemOpen
  \bibfield  {author} {\bibinfo {author} {\bibfnamefont {R.}~\bibnamefont
  {St{\"u}hler}}, \bibinfo {author} {\bibfnamefont {F.}~\bibnamefont {Reis}},
  \bibinfo {author} {\bibfnamefont {T.}~\bibnamefont {M{\"u}ller}}, \bibinfo
  {author} {\bibfnamefont {T.}~\bibnamefont {Helbig}}, \bibinfo {author}
  {\bibfnamefont {T.}~\bibnamefont {Schwemmer}}, \bibinfo {author}
  {\bibfnamefont {R.}~\bibnamefont {Thomale}}, \bibinfo {author} {\bibfnamefont
  {J.}~\bibnamefont {Sch{\"a}fer}},\ and\ \bibinfo {author} {\bibfnamefont
  {R.}~\bibnamefont {Claessen}},\ }\href
  {https://doi.org/10.1038/s41567-019-0697-z} {\bibfield  {journal} {\bibinfo
  {journal} {Nature Physics}\ }\textbf {\bibinfo {volume} {16}},\ \bibinfo
  {pages} {47} (\bibinfo {year} {2020})}\BibitemShut {NoStop}%
\bibitem [{\citenamefont {Deng}\ \emph {et~al.}(2018)\citenamefont {Deng},
  \citenamefont {Xia}, \citenamefont {Ma}, \citenamefont {Chen}, \citenamefont
  {Shan}, \citenamefont {Zhai}, \citenamefont {Li}, \citenamefont {Zhao},
  \citenamefont {Xu}, \citenamefont {Duan}, \citenamefont {Zhang},
  \citenamefont {Wang},\ and\ \citenamefont {Hou}}]{Deng_Hou_2018}%
  \BibitemOpen
  \bibfield  {author} {\bibinfo {author} {\bibfnamefont {J.}~\bibnamefont
  {Deng}}, \bibinfo {author} {\bibfnamefont {B.}~\bibnamefont {Xia}}, \bibinfo
  {author} {\bibfnamefont {X.}~\bibnamefont {Ma}}, \bibinfo {author}
  {\bibfnamefont {H.}~\bibnamefont {Chen}}, \bibinfo {author} {\bibfnamefont
  {H.}~\bibnamefont {Shan}}, \bibinfo {author} {\bibfnamefont {X.}~\bibnamefont
  {Zhai}}, \bibinfo {author} {\bibfnamefont {B.}~\bibnamefont {Li}}, \bibinfo
  {author} {\bibfnamefont {A.}~\bibnamefont {Zhao}}, \bibinfo {author}
  {\bibfnamefont {Y.}~\bibnamefont {Xu}}, \bibinfo {author} {\bibfnamefont
  {W.}~\bibnamefont {Duan}}, \bibinfo {author} {\bibfnamefont {S.-C.}\
  \bibnamefont {Zhang}}, \bibinfo {author} {\bibfnamefont {B.}~\bibnamefont
  {Wang}},\ and\ \bibinfo {author} {\bibfnamefont {J.~G.}\ \bibnamefont
  {Hou}},\ }\href {https://doi.org/10.1038/s41563-018-0203-5} {\bibfield
  {journal} {\bibinfo  {journal} {Nature Materials}\ }\textbf {\bibinfo
  {volume} {17}},\ \bibinfo {pages} {1081} (\bibinfo {year}
  {2018})}\BibitemShut {NoStop}%
\bibitem [{\citenamefont {Marrazzo}\ \emph {et~al.}(2018)\citenamefont
  {Marrazzo}, \citenamefont {Gibertini}, \citenamefont {Campi}, \citenamefont
  {Mounet},\ and\ \citenamefont {Marzari}}]{Marrazzo_Marzari_2018}%
  \BibitemOpen
  \bibfield  {author} {\bibinfo {author} {\bibfnamefont {A.}~\bibnamefont
  {Marrazzo}}, \bibinfo {author} {\bibfnamefont {M.}~\bibnamefont {Gibertini}},
  \bibinfo {author} {\bibfnamefont {D.}~\bibnamefont {Campi}}, \bibinfo
  {author} {\bibfnamefont {N.}~\bibnamefont {Mounet}},\ and\ \bibinfo {author}
  {\bibfnamefont {N.}~\bibnamefont {Marzari}},\ }\href
  {https://doi.org/10.1103/PhysRevLett.120.117701} {\bibfield  {journal}
  {\bibinfo  {journal} {Phys. Rev. Lett.}\ }\textbf {\bibinfo {volume} {120}},\
  \bibinfo {pages} {117701} (\bibinfo {year} {2018})}\BibitemShut {NoStop}%
\bibitem [{\citenamefont {Kandrai}\ \emph {et~al.}(2020)\citenamefont
  {Kandrai}, \citenamefont {Vancsó}, \citenamefont {Kukucska}, \citenamefont
  {Koltai}, \citenamefont {Baranka}, \citenamefont {Ákos Hoffmann},
  \citenamefont {Áron Pekker}, \citenamefont {Kamarás}, \citenamefont
  {Horváth}, \citenamefont {Vymazalová}, \citenamefont {Tapasztó},\ and\
  \citenamefont {Nemes-Incze}}]{Kandrai_Nemes-Incze_2020}%
  \BibitemOpen
  \bibfield  {author} {\bibinfo {author} {\bibfnamefont {K.}~\bibnamefont
  {Kandrai}}, \bibinfo {author} {\bibfnamefont {P.}~\bibnamefont {Vancsó}},
  \bibinfo {author} {\bibfnamefont {G.}~\bibnamefont {Kukucska}}, \bibinfo
  {author} {\bibfnamefont {J.}~\bibnamefont {Koltai}}, \bibinfo {author}
  {\bibfnamefont {G.}~\bibnamefont {Baranka}}, \bibinfo {author} {\bibnamefont
  {Ákos Hoffmann}}, \bibinfo {author} {\bibnamefont {Áron Pekker}}, \bibinfo
  {author} {\bibfnamefont {K.}~\bibnamefont {Kamarás}}, \bibinfo {author}
  {\bibfnamefont {Z.~E.}\ \bibnamefont {Horváth}}, \bibinfo {author}
  {\bibfnamefont {A.}~\bibnamefont {Vymazalová}}, \bibinfo {author}
  {\bibfnamefont {L.}~\bibnamefont {Tapasztó}},\ and\ \bibinfo {author}
  {\bibfnamefont {P.}~\bibnamefont {Nemes-Incze}},\ }\href
  {https://doi.org/10.1021/acs.nanolett.0c01499} {\bibfield  {journal}
  {\bibinfo  {journal} {Nano Letters}\ }\textbf {\bibinfo {volume} {20}},\
  \bibinfo {pages} {5207} (\bibinfo {year} {2020})}\BibitemShut {NoStop}%
\bibitem [{\citenamefont {Wu}\ \emph {et~al.}(2019)\citenamefont {Wu},
  \citenamefont {Fink}, \citenamefont {Hanke}, \citenamefont {Thomale},\ and\
  \citenamefont {Di~Sante}}]{Wu_DiSante_2019}%
  \BibitemOpen
  \bibfield  {author} {\bibinfo {author} {\bibfnamefont {X.}~\bibnamefont
  {Wu}}, \bibinfo {author} {\bibfnamefont {M.}~\bibnamefont {Fink}}, \bibinfo
  {author} {\bibfnamefont {W.}~\bibnamefont {Hanke}}, \bibinfo {author}
  {\bibfnamefont {R.}~\bibnamefont {Thomale}},\ and\ \bibinfo {author}
  {\bibfnamefont {D.}~\bibnamefont {Di~Sante}},\ }\href
  {https://doi.org/10.1103/PhysRevB.100.041117} {\bibfield  {journal} {\bibinfo
   {journal} {Phys. Rev. B}\ }\textbf {\bibinfo {volume} {100}},\ \bibinfo
  {pages} {041117} (\bibinfo {year} {2019})}\BibitemShut {NoStop}%
\bibitem [{\citenamefont {K{\"o}nig}\ \emph {et~al.}(2007)\citenamefont
  {K{\"o}nig}, \citenamefont {Wiedmann}, \citenamefont {Br{\"u}ne},
  \citenamefont {Roth}, \citenamefont {Buhmann}, \citenamefont {Molenkamp},
  \citenamefont {Qi},\ and\ \citenamefont {Zhang}}]{Konig_Zhang_2007}%
  \BibitemOpen
  \bibfield  {author} {\bibinfo {author} {\bibfnamefont {M.}~\bibnamefont
  {K{\"o}nig}}, \bibinfo {author} {\bibfnamefont {S.}~\bibnamefont {Wiedmann}},
  \bibinfo {author} {\bibfnamefont {C.}~\bibnamefont {Br{\"u}ne}}, \bibinfo
  {author} {\bibfnamefont {A.}~\bibnamefont {Roth}}, \bibinfo {author}
  {\bibfnamefont {H.}~\bibnamefont {Buhmann}}, \bibinfo {author} {\bibfnamefont
  {L.~W.}\ \bibnamefont {Molenkamp}}, \bibinfo {author} {\bibfnamefont {X.-L.}\
  \bibnamefont {Qi}},\ and\ \bibinfo {author} {\bibfnamefont {S.-C.}\
  \bibnamefont {Zhang}},\ }\href {https://doi.org/10.1126/science.1148047}
  {\bibfield  {journal} {\bibinfo  {journal} {Science}\ }\textbf {\bibinfo
  {volume} {318}},\ \bibinfo {pages} {766} (\bibinfo {year}
  {2007})}\BibitemShut {NoStop}%
\bibitem [{\citenamefont {Roth}\ \emph {et~al.}(2009)\citenamefont {Roth},
  \citenamefont {Brüne}, \citenamefont {Buhmann}, \citenamefont {Molenkamp},
  \citenamefont {Maciejko}, \citenamefont {Qi},\ and\ \citenamefont
  {Zhang}}]{Roth_Zhang_2009}%
  \BibitemOpen
  \bibfield  {author} {\bibinfo {author} {\bibfnamefont {A.}~\bibnamefont
  {Roth}}, \bibinfo {author} {\bibfnamefont {C.}~\bibnamefont {Brüne}},
  \bibinfo {author} {\bibfnamefont {H.}~\bibnamefont {Buhmann}}, \bibinfo
  {author} {\bibfnamefont {L.~W.}\ \bibnamefont {Molenkamp}}, \bibinfo {author}
  {\bibfnamefont {J.}~\bibnamefont {Maciejko}}, \bibinfo {author}
  {\bibfnamefont {X.-L.}\ \bibnamefont {Qi}},\ and\ \bibinfo {author}
  {\bibfnamefont {S.-C.}\ \bibnamefont {Zhang}},\ }\href
  {https://doi.org/10.1126/science.1174736} {\bibfield  {journal} {\bibinfo
  {journal} {Science}\ }\textbf {\bibinfo {volume} {325}},\ \bibinfo {pages}
  {294} (\bibinfo {year} {2009})}\BibitemShut {NoStop}%
\bibitem [{\citenamefont {Knez}\ \emph {et~al.}(2011)\citenamefont {Knez},
  \citenamefont {Du},\ and\ \citenamefont {Sullivan}}]{Knez_Sullivan_2011}%
  \BibitemOpen
  \bibfield  {author} {\bibinfo {author} {\bibfnamefont {I.}~\bibnamefont
  {Knez}}, \bibinfo {author} {\bibfnamefont {R.-R.}\ \bibnamefont {Du}},\ and\
  \bibinfo {author} {\bibfnamefont {G.}~\bibnamefont {Sullivan}},\ }\href
  {https://doi.org/10.1103/PhysRevLett.107.136603} {\bibfield  {journal}
  {\bibinfo  {journal} {Phys. Rev. Lett.}\ }\textbf {\bibinfo {volume} {107}},\
  \bibinfo {pages} {136603} (\bibinfo {year} {2011})}\BibitemShut {NoStop}%
\bibitem [{\citenamefont {Spanton}\ \emph {et~al.}(2014)\citenamefont
  {Spanton}, \citenamefont {Nowack}, \citenamefont {Du}, \citenamefont
  {Sullivan}, \citenamefont {Du},\ and\ \citenamefont
  {Moler}}]{Spanton_Moler_2014}%
  \BibitemOpen
  \bibfield  {author} {\bibinfo {author} {\bibfnamefont {E.~M.}\ \bibnamefont
  {Spanton}}, \bibinfo {author} {\bibfnamefont {K.~C.}\ \bibnamefont {Nowack}},
  \bibinfo {author} {\bibfnamefont {L.}~\bibnamefont {Du}}, \bibinfo {author}
  {\bibfnamefont {G.}~\bibnamefont {Sullivan}}, \bibinfo {author}
  {\bibfnamefont {R.-R.}\ \bibnamefont {Du}},\ and\ \bibinfo {author}
  {\bibfnamefont {K.~A.}\ \bibnamefont {Moler}},\ }\href
  {https://doi.org/10.1103/PhysRevLett.113.026804} {\bibfield  {journal}
  {\bibinfo  {journal} {Phys. Rev. Lett.}\ }\textbf {\bibinfo {volume} {113}},\
  \bibinfo {pages} {026804} (\bibinfo {year} {2014})}\BibitemShut {NoStop}%
\bibitem [{\citenamefont {Pribiag}\ \emph {et~al.}(2015)\citenamefont
  {Pribiag}, \citenamefont {Beukman}, \citenamefont {Qu}, \citenamefont
  {Cassidy}, \citenamefont {Charpentier}, \citenamefont {Wegscheider},\ and\
  \citenamefont {Kouwenhoven}}]{Pribiag_Kouwenhoven_2015}%
  \BibitemOpen
  \bibfield  {author} {\bibinfo {author} {\bibfnamefont {V.~S.}\ \bibnamefont
  {Pribiag}}, \bibinfo {author} {\bibfnamefont {A.~J.~A.}\ \bibnamefont
  {Beukman}}, \bibinfo {author} {\bibfnamefont {F.}~\bibnamefont {Qu}},
  \bibinfo {author} {\bibfnamefont {M.~C.}\ \bibnamefont {Cassidy}}, \bibinfo
  {author} {\bibfnamefont {C.}~\bibnamefont {Charpentier}}, \bibinfo {author}
  {\bibfnamefont {W.}~\bibnamefont {Wegscheider}},\ and\ \bibinfo {author}
  {\bibfnamefont {L.~P.}\ \bibnamefont {Kouwenhoven}},\ }\href
  {https://doi.org/10.1038/nnano.2015.86} {\bibfield  {journal} {\bibinfo
  {journal} {Nature Nanotechnology}\ }\textbf {\bibinfo {volume} {10}},\
  \bibinfo {pages} {593} (\bibinfo {year} {2015})}\BibitemShut {NoStop}%
\bibitem [{\citenamefont {Du}\ \emph {et~al.}(2015)\citenamefont {Du},
  \citenamefont {Knez}, \citenamefont {Sullivan},\ and\ \citenamefont
  {Du}}]{Du_Du_2015}%
  \BibitemOpen
  \bibfield  {author} {\bibinfo {author} {\bibfnamefont {L.}~\bibnamefont
  {Du}}, \bibinfo {author} {\bibfnamefont {I.}~\bibnamefont {Knez}}, \bibinfo
  {author} {\bibfnamefont {G.}~\bibnamefont {Sullivan}},\ and\ \bibinfo
  {author} {\bibfnamefont {R.-R.}\ \bibnamefont {Du}},\ }\href
  {https://doi.org/10.1103/PhysRevLett.114.096802} {\bibfield  {journal}
  {\bibinfo  {journal} {Phys. Rev. Lett.}\ }\textbf {\bibinfo {volume} {114}},\
  \bibinfo {pages} {096802} (\bibinfo {year} {2015})}\BibitemShut {NoStop}%
\bibitem [{\citenamefont {Li}\ \emph {et~al.}(2015)\citenamefont {Li},
  \citenamefont {Wang}, \citenamefont {Fu}, \citenamefont {Du}, \citenamefont
  {Schreiber}, \citenamefont {Mu}, \citenamefont {Liu}, \citenamefont
  {Sullivan}, \citenamefont {Cs\'athy}, \citenamefont {Lin},\ and\
  \citenamefont {Du}}]{Li_Du_2015}%
  \BibitemOpen
  \bibfield  {author} {\bibinfo {author} {\bibfnamefont {T.}~\bibnamefont
  {Li}}, \bibinfo {author} {\bibfnamefont {P.}~\bibnamefont {Wang}}, \bibinfo
  {author} {\bibfnamefont {H.}~\bibnamefont {Fu}}, \bibinfo {author}
  {\bibfnamefont {L.}~\bibnamefont {Du}}, \bibinfo {author} {\bibfnamefont
  {K.~A.}\ \bibnamefont {Schreiber}}, \bibinfo {author} {\bibfnamefont
  {X.}~\bibnamefont {Mu}}, \bibinfo {author} {\bibfnamefont {X.}~\bibnamefont
  {Liu}}, \bibinfo {author} {\bibfnamefont {G.}~\bibnamefont {Sullivan}},
  \bibinfo {author} {\bibfnamefont {G.~A.}\ \bibnamefont {Cs\'athy}}, \bibinfo
  {author} {\bibfnamefont {X.}~\bibnamefont {Lin}},\ and\ \bibinfo {author}
  {\bibfnamefont {R.-R.}\ \bibnamefont {Du}},\ }\href
  {https://doi.org/10.1103/PhysRevLett.115.136804} {\bibfield  {journal}
  {\bibinfo  {journal} {Phys. Rev. Lett.}\ }\textbf {\bibinfo {volume} {115}},\
  \bibinfo {pages} {136804} (\bibinfo {year} {2015})}\BibitemShut {NoStop}%
\bibitem [{\citenamefont {Du}\ \emph {et~al.}(2017{\natexlab{a}})\citenamefont
  {Du}, \citenamefont {Li}, \citenamefont {Lou}, \citenamefont {Wu},
  \citenamefont {Liu}, \citenamefont {Han}, \citenamefont {Zhang},
  \citenamefont {Sullivan}, \citenamefont {Ikhlassi}, \citenamefont {Chang},\
  and\ \citenamefont {Du}}]{Du_Du_2017}%
  \BibitemOpen
  \bibfield  {author} {\bibinfo {author} {\bibfnamefont {L.}~\bibnamefont
  {Du}}, \bibinfo {author} {\bibfnamefont {T.}~\bibnamefont {Li}}, \bibinfo
  {author} {\bibfnamefont {W.}~\bibnamefont {Lou}}, \bibinfo {author}
  {\bibfnamefont {X.}~\bibnamefont {Wu}}, \bibinfo {author} {\bibfnamefont
  {X.}~\bibnamefont {Liu}}, \bibinfo {author} {\bibfnamefont {Z.}~\bibnamefont
  {Han}}, \bibinfo {author} {\bibfnamefont {C.}~\bibnamefont {Zhang}}, \bibinfo
  {author} {\bibfnamefont {G.}~\bibnamefont {Sullivan}}, \bibinfo {author}
  {\bibfnamefont {A.}~\bibnamefont {Ikhlassi}}, \bibinfo {author}
  {\bibfnamefont {K.}~\bibnamefont {Chang}},\ and\ \bibinfo {author}
  {\bibfnamefont {R.-R.}\ \bibnamefont {Du}},\ }\href
  {https://doi.org/10.1103/PhysRevLett.119.056803} {\bibfield  {journal}
  {\bibinfo  {journal} {Phys. Rev. Lett.}\ }\textbf {\bibinfo {volume} {119}},\
  \bibinfo {pages} {056803} (\bibinfo {year} {2017}{\natexlab{a}})}\BibitemShut
  {NoStop}%
\bibitem [{\citenamefont {Du}\ \emph {et~al.}(2017{\natexlab{b}})\citenamefont
  {Du}, \citenamefont {Li}, \citenamefont {Lou}, \citenamefont {Sullivan},
  \citenamefont {Chang}, \citenamefont {Kono},\ and\ \citenamefont
  {Du}}]{Du_Du_2017b}%
  \BibitemOpen
  \bibfield  {author} {\bibinfo {author} {\bibfnamefont {L.}~\bibnamefont
  {Du}}, \bibinfo {author} {\bibfnamefont {X.}~\bibnamefont {Li}}, \bibinfo
  {author} {\bibfnamefont {W.}~\bibnamefont {Lou}}, \bibinfo {author}
  {\bibfnamefont {G.}~\bibnamefont {Sullivan}}, \bibinfo {author}
  {\bibfnamefont {K.}~\bibnamefont {Chang}}, \bibinfo {author} {\bibfnamefont
  {J.}~\bibnamefont {Kono}},\ and\ \bibinfo {author} {\bibfnamefont {R.-R.}\
  \bibnamefont {Du}},\ }\href {https://doi.org/10.1038/s41467-017-01988-1}
  {\bibfield  {journal} {\bibinfo  {journal} {Nature Communications}\ }\textbf
  {\bibinfo {volume} {8}},\ \bibinfo {pages} {1971} (\bibinfo {year}
  {2017}{\natexlab{b}})}\BibitemShut {NoStop}%
\bibitem [{\citenamefont {Bendias}\ \emph {et~al.}(2018)\citenamefont
  {Bendias}, \citenamefont {Shamim}, \citenamefont {Herrmann}, \citenamefont
  {Budewitz}, \citenamefont {Shekhar}, \citenamefont {Leubner}, \citenamefont
  {Kleinlein}, \citenamefont {Bocquillon}, \citenamefont {Buhmann},\ and\
  \citenamefont {Molenkamp}}]{Bendias_Molenkamp_2018}%
  \BibitemOpen
  \bibfield  {author} {\bibinfo {author} {\bibfnamefont {K.}~\bibnamefont
  {Bendias}}, \bibinfo {author} {\bibfnamefont {S.}~\bibnamefont {Shamim}},
  \bibinfo {author} {\bibfnamefont {O.}~\bibnamefont {Herrmann}}, \bibinfo
  {author} {\bibfnamefont {A.}~\bibnamefont {Budewitz}}, \bibinfo {author}
  {\bibfnamefont {P.}~\bibnamefont {Shekhar}}, \bibinfo {author} {\bibfnamefont
  {P.}~\bibnamefont {Leubner}}, \bibinfo {author} {\bibfnamefont
  {J.}~\bibnamefont {Kleinlein}}, \bibinfo {author} {\bibfnamefont
  {E.}~\bibnamefont {Bocquillon}}, \bibinfo {author} {\bibfnamefont
  {H.}~\bibnamefont {Buhmann}},\ and\ \bibinfo {author} {\bibfnamefont {L.~W.}\
  \bibnamefont {Molenkamp}},\ }\href
  {https://doi.org/10.1021/acs.nanolett.8b01405} {\bibfield  {journal}
  {\bibinfo  {journal} {Nano Letters}\ }\textbf {\bibinfo {volume} {18}},\
  \bibinfo {pages} {4831} (\bibinfo {year} {2018})}\BibitemShut {NoStop}%
\bibitem [{\citenamefont {Lunczer}\ \emph {et~al.}(2019)\citenamefont
  {Lunczer}, \citenamefont {Leubner}, \citenamefont {Endres}, \citenamefont
  {M\"uller}, \citenamefont {Br\"une}, \citenamefont {Buhmann},\ and\
  \citenamefont {Molenkamp}}]{Lunczer_Molenkamp_2019}%
  \BibitemOpen
  \bibfield  {author} {\bibinfo {author} {\bibfnamefont {L.}~\bibnamefont
  {Lunczer}}, \bibinfo {author} {\bibfnamefont {P.}~\bibnamefont {Leubner}},
  \bibinfo {author} {\bibfnamefont {M.}~\bibnamefont {Endres}}, \bibinfo
  {author} {\bibfnamefont {V.~L.}\ \bibnamefont {M\"uller}}, \bibinfo {author}
  {\bibfnamefont {C.}~\bibnamefont {Br\"une}}, \bibinfo {author} {\bibfnamefont
  {H.}~\bibnamefont {Buhmann}},\ and\ \bibinfo {author} {\bibfnamefont {L.~W.}\
  \bibnamefont {Molenkamp}},\ }\href
  {https://doi.org/10.1103/PhysRevLett.123.047701} {\bibfield  {journal}
  {\bibinfo  {journal} {Phys. Rev. Lett.}\ }\textbf {\bibinfo {volume} {123}},\
  \bibinfo {pages} {047701} (\bibinfo {year} {2019})}\BibitemShut {NoStop}%
\bibitem [{\citenamefont {Xiao}\ \emph {et~al.}(2019)\citenamefont {Xiao},
  \citenamefont {Liu}, \citenamefont {Samarth},\ and\ \citenamefont
  {Hu}}]{Xiao_Hu_2019}%
  \BibitemOpen
  \bibfield  {author} {\bibinfo {author} {\bibfnamefont {D.}~\bibnamefont
  {Xiao}}, \bibinfo {author} {\bibfnamefont {C.-X.}\ \bibnamefont {Liu}},
  \bibinfo {author} {\bibfnamefont {N.}~\bibnamefont {Samarth}},\ and\ \bibinfo
  {author} {\bibfnamefont {L.-H.}\ \bibnamefont {Hu}},\ }\href
  {https://doi.org/10.1103/PhysRevLett.122.186802} {\bibfield  {journal}
  {\bibinfo  {journal} {Phys. Rev. Lett.}\ }\textbf {\bibinfo {volume} {122}},\
  \bibinfo {pages} {186802} (\bibinfo {year} {2019})}\BibitemShut {NoStop}%
\bibitem [{\citenamefont {Han}\ \emph {et~al.}(2019)\citenamefont {Han},
  \citenamefont {Li}, \citenamefont {Zhang}, \citenamefont {Sullivan},\ and\
  \citenamefont {Du}}]{Han_Du_2019}%
  \BibitemOpen
  \bibfield  {author} {\bibinfo {author} {\bibfnamefont {Z.}~\bibnamefont
  {Han}}, \bibinfo {author} {\bibfnamefont {T.}~\bibnamefont {Li}}, \bibinfo
  {author} {\bibfnamefont {L.}~\bibnamefont {Zhang}}, \bibinfo {author}
  {\bibfnamefont {G.}~\bibnamefont {Sullivan}},\ and\ \bibinfo {author}
  {\bibfnamefont {R.-R.}\ \bibnamefont {Du}},\ }\href
  {https://doi.org/10.1103/PhysRevLett.123.126803} {\bibfield  {journal}
  {\bibinfo  {journal} {Phys. Rev. Lett.}\ }\textbf {\bibinfo {volume} {123}},\
  \bibinfo {pages} {126803} (\bibinfo {year} {2019})}\BibitemShut {NoStop}%
\bibitem [{\citenamefont {Piatrusha}\ \emph {et~al.}(2019)\citenamefont
  {Piatrusha}, \citenamefont {Tikhonov}, \citenamefont {Kvon}, \citenamefont
  {Mikhailov}, \citenamefont {Dvoretsky},\ and\ \citenamefont
  {Khrapai}}]{Piatrusha_Khrapai_2019}%
  \BibitemOpen
  \bibfield  {author} {\bibinfo {author} {\bibfnamefont {S.~U.}\ \bibnamefont
  {Piatrusha}}, \bibinfo {author} {\bibfnamefont {E.~S.}\ \bibnamefont
  {Tikhonov}}, \bibinfo {author} {\bibfnamefont {Z.~D.}\ \bibnamefont {Kvon}},
  \bibinfo {author} {\bibfnamefont {N.~N.}\ \bibnamefont {Mikhailov}}, \bibinfo
  {author} {\bibfnamefont {S.~A.}\ \bibnamefont {Dvoretsky}},\ and\ \bibinfo
  {author} {\bibfnamefont {V.~S.}\ \bibnamefont {Khrapai}},\ }\href
  {https://doi.org/10.1103/PhysRevLett.123.056801} {\bibfield  {journal}
  {\bibinfo  {journal} {Phys. Rev. Lett.}\ }\textbf {\bibinfo {volume} {123}},\
  \bibinfo {pages} {056801} (\bibinfo {year} {2019})}\BibitemShut {NoStop}%
\bibitem [{\citenamefont {Strunz}\ \emph {et~al.}(2020)\citenamefont {Strunz},
  \citenamefont {Wiedenmann}, \citenamefont {Fleckenstein}, \citenamefont
  {Lunczer}, \citenamefont {Beugeling}, \citenamefont {M{\"u}ller},
  \citenamefont {Shekhar}, \citenamefont {Ziani}, \citenamefont {Shamim},
  \citenamefont {Kleinlein}, \citenamefont {Buhmann}, \citenamefont
  {Trauzettel},\ and\ \citenamefont {Molenkamp}}]{Strunz_Molenkamp_2020}%
  \BibitemOpen
  \bibfield  {author} {\bibinfo {author} {\bibfnamefont {J.}~\bibnamefont
  {Strunz}}, \bibinfo {author} {\bibfnamefont {J.}~\bibnamefont {Wiedenmann}},
  \bibinfo {author} {\bibfnamefont {C.}~\bibnamefont {Fleckenstein}}, \bibinfo
  {author} {\bibfnamefont {L.}~\bibnamefont {Lunczer}}, \bibinfo {author}
  {\bibfnamefont {W.}~\bibnamefont {Beugeling}}, \bibinfo {author}
  {\bibfnamefont {V.~L.}\ \bibnamefont {M{\"u}ller}}, \bibinfo {author}
  {\bibfnamefont {P.}~\bibnamefont {Shekhar}}, \bibinfo {author} {\bibfnamefont
  {N.~T.}\ \bibnamefont {Ziani}}, \bibinfo {author} {\bibfnamefont
  {S.}~\bibnamefont {Shamim}}, \bibinfo {author} {\bibfnamefont
  {J.}~\bibnamefont {Kleinlein}}, \bibinfo {author} {\bibfnamefont
  {H.}~\bibnamefont {Buhmann}}, \bibinfo {author} {\bibfnamefont
  {B.}~\bibnamefont {Trauzettel}},\ and\ \bibinfo {author} {\bibfnamefont
  {L.~W.}\ \bibnamefont {Molenkamp}},\ }\href
  {https://doi.org/10.1038/s41567-019-0692-4} {\bibfield  {journal} {\bibinfo
  {journal} {Nature Physics}\ }\textbf {\bibinfo {volume} {16}},\ \bibinfo
  {pages} {83} (\bibinfo {year} {2020})}\BibitemShut {NoStop}%
\bibitem [{\citenamefont {Shamim}\ \emph {et~al.}(2020)\citenamefont {Shamim},
  \citenamefont {Beugeling}, \citenamefont {Böttcher}, \citenamefont
  {Shekhar}, \citenamefont {Budewitz}, \citenamefont {Leubner}, \citenamefont
  {Lunczer}, \citenamefont {Hankiewicz}, \citenamefont {Buhmann},\ and\
  \citenamefont {Molenkamp}}]{Shamim_Molenkamp_2020}%
  \BibitemOpen
  \bibfield  {author} {\bibinfo {author} {\bibfnamefont {S.}~\bibnamefont
  {Shamim}}, \bibinfo {author} {\bibfnamefont {W.}~\bibnamefont {Beugeling}},
  \bibinfo {author} {\bibfnamefont {J.}~\bibnamefont {Böttcher}}, \bibinfo
  {author} {\bibfnamefont {P.}~\bibnamefont {Shekhar}}, \bibinfo {author}
  {\bibfnamefont {A.}~\bibnamefont {Budewitz}}, \bibinfo {author}
  {\bibfnamefont {P.}~\bibnamefont {Leubner}}, \bibinfo {author} {\bibfnamefont
  {L.}~\bibnamefont {Lunczer}}, \bibinfo {author} {\bibfnamefont {E.~M.}\
  \bibnamefont {Hankiewicz}}, \bibinfo {author} {\bibfnamefont
  {H.}~\bibnamefont {Buhmann}},\ and\ \bibinfo {author} {\bibfnamefont {L.~W.}\
  \bibnamefont {Molenkamp}},\ }\href {https://doi.org/10.1126/sciadv.aba4625}
  {\bibfield  {journal} {\bibinfo  {journal} {Science Advances}\ }\textbf
  {\bibinfo {volume} {6}},\ \bibinfo {pages} {eaba4625} (\bibinfo {year}
  {2020})}\BibitemShut {NoStop}%
\bibitem [{\citenamefont {Dartiailh}\ \emph {et~al.}(2020)\citenamefont
  {Dartiailh}, \citenamefont {Hartinger}, \citenamefont {Gourmelon},
  \citenamefont {Bendias}, \citenamefont {Bartolomei}, \citenamefont {Kamata},
  \citenamefont {Berroir}, \citenamefont {F\`eve}, \citenamefont
  {Pla\ifmmode~\mbox{\c{c}}\else \c{c}\fi{}ais}, \citenamefont {Lunczer},
  \citenamefont {Schlereth}, \citenamefont {Buhmann}, \citenamefont
  {Molenkamp},\ and\ \citenamefont {Bocquillon}}]{Dartiailh_Bocquillon_2020}%
  \BibitemOpen
  \bibfield  {author} {\bibinfo {author} {\bibfnamefont {M.~C.}\ \bibnamefont
  {Dartiailh}}, \bibinfo {author} {\bibfnamefont {S.}~\bibnamefont
  {Hartinger}}, \bibinfo {author} {\bibfnamefont {A.}~\bibnamefont
  {Gourmelon}}, \bibinfo {author} {\bibfnamefont {K.}~\bibnamefont {Bendias}},
  \bibinfo {author} {\bibfnamefont {H.}~\bibnamefont {Bartolomei}}, \bibinfo
  {author} {\bibfnamefont {H.}~\bibnamefont {Kamata}}, \bibinfo {author}
  {\bibfnamefont {J.-M.}\ \bibnamefont {Berroir}}, \bibinfo {author}
  {\bibfnamefont {G.}~\bibnamefont {F\`eve}}, \bibinfo {author} {\bibfnamefont
  {B.}~\bibnamefont {Pla\ifmmode~\mbox{\c{c}}\else \c{c}\fi{}ais}}, \bibinfo
  {author} {\bibfnamefont {L.}~\bibnamefont {Lunczer}}, \bibinfo {author}
  {\bibfnamefont {R.}~\bibnamefont {Schlereth}}, \bibinfo {author}
  {\bibfnamefont {H.}~\bibnamefont {Buhmann}}, \bibinfo {author} {\bibfnamefont
  {L.~W.}\ \bibnamefont {Molenkamp}},\ and\ \bibinfo {author} {\bibfnamefont
  {E.}~\bibnamefont {Bocquillon}},\ }\href
  {https://doi.org/10.1103/PhysRevLett.124.076802} {\bibfield  {journal}
  {\bibinfo  {journal} {Phys. Rev. Lett.}\ }\textbf {\bibinfo {volume} {124}},\
  \bibinfo {pages} {076802} (\bibinfo {year} {2020})}\BibitemShut {NoStop}%
\bibitem [{\citenamefont {Shamim}\ \emph {et~al.}(2021)\citenamefont {Shamim},
  \citenamefont {Beugeling}, \citenamefont {Shekhar}, \citenamefont {Bendias},
  \citenamefont {Lunczer}, \citenamefont {Kleinlein}, \citenamefont {Buhmann},\
  and\ \citenamefont {Molenkamp}}]{Shamim_Molenkamp_2021}%
  \BibitemOpen
  \bibfield  {author} {\bibinfo {author} {\bibfnamefont {S.}~\bibnamefont
  {Shamim}}, \bibinfo {author} {\bibfnamefont {W.}~\bibnamefont {Beugeling}},
  \bibinfo {author} {\bibfnamefont {P.}~\bibnamefont {Shekhar}}, \bibinfo
  {author} {\bibfnamefont {K.}~\bibnamefont {Bendias}}, \bibinfo {author}
  {\bibfnamefont {L.}~\bibnamefont {Lunczer}}, \bibinfo {author} {\bibfnamefont
  {J.}~\bibnamefont {Kleinlein}}, \bibinfo {author} {\bibfnamefont
  {H.}~\bibnamefont {Buhmann}},\ and\ \bibinfo {author} {\bibfnamefont {L.~W.}\
  \bibnamefont {Molenkamp}},\ }\href
  {https://doi.org/10.1038/s41467-021-23262-1} {\bibfield  {journal} {\bibinfo
  {journal} {Nature Communications}\ }\textbf {\bibinfo {volume} {12}},\
  \bibinfo {pages} {3193} (\bibinfo {year} {2021})}\BibitemShut {NoStop}%
\bibitem [{\citenamefont {V\"ayrynen}\ \emph {et~al.}(2013)\citenamefont
  {V\"ayrynen}, \citenamefont {Goldstein},\ and\ \citenamefont
  {Glazman}}]{Vayrynen_Glazman_2013}%
  \BibitemOpen
  \bibfield  {author} {\bibinfo {author} {\bibfnamefont {J.~I.}\ \bibnamefont
  {V\"ayrynen}}, \bibinfo {author} {\bibfnamefont {M.}~\bibnamefont
  {Goldstein}},\ and\ \bibinfo {author} {\bibfnamefont {L.~I.}\ \bibnamefont
  {Glazman}},\ }\href {https://doi.org/10.1103/PhysRevLett.110.216402}
  {\bibfield  {journal} {\bibinfo  {journal} {Phys. Rev. Lett.}\ }\textbf
  {\bibinfo {volume} {110}},\ \bibinfo {pages} {216402} (\bibinfo {year}
  {2013})}\BibitemShut {NoStop}%
\bibitem [{\citenamefont {V\"ayrynen}\ \emph {et~al.}(2014)\citenamefont
  {V\"ayrynen}, \citenamefont {Goldstein}, \citenamefont {Gefen},\ and\
  \citenamefont {Glazman}}]{Vayrynen_Glazman_2014}%
  \BibitemOpen
  \bibfield  {author} {\bibinfo {author} {\bibfnamefont {J.~I.}\ \bibnamefont
  {V\"ayrynen}}, \bibinfo {author} {\bibfnamefont {M.}~\bibnamefont
  {Goldstein}}, \bibinfo {author} {\bibfnamefont {Y.}~\bibnamefont {Gefen}},\
  and\ \bibinfo {author} {\bibfnamefont {L.~I.}\ \bibnamefont {Glazman}},\
  }\href {https://doi.org/10.1103/PhysRevB.90.115309} {\bibfield  {journal}
  {\bibinfo  {journal} {Phys. Rev. B}\ }\textbf {\bibinfo {volume} {90}},\
  \bibinfo {pages} {115309} (\bibinfo {year} {2014})}\BibitemShut {NoStop}%
\bibitem [{\citenamefont {Novelli}\ \emph {et~al.}(2019)\citenamefont
  {Novelli}, \citenamefont {Taddei}, \citenamefont {Geim},\ and\ \citenamefont
  {Polini}}]{Novelli_Polini_2019}%
  \BibitemOpen
  \bibfield  {author} {\bibinfo {author} {\bibfnamefont {P.}~\bibnamefont
  {Novelli}}, \bibinfo {author} {\bibfnamefont {F.}~\bibnamefont {Taddei}},
  \bibinfo {author} {\bibfnamefont {A.~K.}\ \bibnamefont {Geim}},\ and\
  \bibinfo {author} {\bibfnamefont {M.}~\bibnamefont {Polini}},\ }\href
  {https://doi.org/10.1103/PhysRevLett.122.016601} {\bibfield  {journal}
  {\bibinfo  {journal} {Phys. Rev. Lett.}\ }\textbf {\bibinfo {volume} {122}},\
  \bibinfo {pages} {016601} (\bibinfo {year} {2019})}\BibitemShut {NoStop}%
\bibitem [{\citenamefont {Maciejko}\ \emph {et~al.}(2009)\citenamefont
  {Maciejko}, \citenamefont {Liu}, \citenamefont {Oreg}, \citenamefont {Qi},
  \citenamefont {Wu},\ and\ \citenamefont {Zhang}}]{Maciejko_Zhang_2009}%
  \BibitemOpen
  \bibfield  {author} {\bibinfo {author} {\bibfnamefont {J.}~\bibnamefont
  {Maciejko}}, \bibinfo {author} {\bibfnamefont {C.}~\bibnamefont {Liu}},
  \bibinfo {author} {\bibfnamefont {Y.}~\bibnamefont {Oreg}}, \bibinfo {author}
  {\bibfnamefont {X.-L.}\ \bibnamefont {Qi}}, \bibinfo {author} {\bibfnamefont
  {C.}~\bibnamefont {Wu}},\ and\ \bibinfo {author} {\bibfnamefont {S.-C.}\
  \bibnamefont {Zhang}},\ }\href
  {https://doi.org/10.1103/PhysRevLett.102.256803} {\bibfield  {journal}
  {\bibinfo  {journal} {Phys. Rev. Lett.}\ }\textbf {\bibinfo {volume} {102}},\
  \bibinfo {pages} {256803} (\bibinfo {year} {2009})}\BibitemShut {NoStop}%
\bibitem [{\citenamefont {Tanaka}\ \emph {et~al.}(2011)\citenamefont {Tanaka},
  \citenamefont {Furusaki},\ and\ \citenamefont
  {Matveev}}]{Tanaka_Matveev_2011}%
  \BibitemOpen
  \bibfield  {author} {\bibinfo {author} {\bibfnamefont {Y.}~\bibnamefont
  {Tanaka}}, \bibinfo {author} {\bibfnamefont {A.}~\bibnamefont {Furusaki}},\
  and\ \bibinfo {author} {\bibfnamefont {K.~A.}\ \bibnamefont {Matveev}},\
  }\href {https://doi.org/10.1103/PhysRevLett.106.236402} {\bibfield  {journal}
  {\bibinfo  {journal} {Phys. Rev. Lett.}\ }\textbf {\bibinfo {volume} {106}},\
  \bibinfo {pages} {236402} (\bibinfo {year} {2011})}\BibitemShut {NoStop}%
\bibitem [{\citenamefont {Altshuler}\ \emph {et~al.}(2013)\citenamefont
  {Altshuler}, \citenamefont {Aleiner},\ and\ \citenamefont
  {Yudson}}]{Altshuler_Yudson_2013}%
  \BibitemOpen
  \bibfield  {author} {\bibinfo {author} {\bibfnamefont {B.~L.}\ \bibnamefont
  {Altshuler}}, \bibinfo {author} {\bibfnamefont {I.~L.}\ \bibnamefont
  {Aleiner}},\ and\ \bibinfo {author} {\bibfnamefont {V.~I.}\ \bibnamefont
  {Yudson}},\ }\href {https://doi.org/10.1103/PhysRevLett.111.086401}
  {\bibfield  {journal} {\bibinfo  {journal} {Phys. Rev. Lett.}\ }\textbf
  {\bibinfo {volume} {111}},\ \bibinfo {pages} {086401} (\bibinfo {year}
  {2013})}\BibitemShut {NoStop}%
\bibitem [{\citenamefont {V\"ayrynen}\ \emph {et~al.}(2018)\citenamefont
  {V\"ayrynen}, \citenamefont {Pikulin},\ and\ \citenamefont
  {Alicea}}]{Vayrynen_Alicea_2018}%
  \BibitemOpen
  \bibfield  {author} {\bibinfo {author} {\bibfnamefont {J.~I.}\ \bibnamefont
  {V\"ayrynen}}, \bibinfo {author} {\bibfnamefont {D.~I.}\ \bibnamefont
  {Pikulin}},\ and\ \bibinfo {author} {\bibfnamefont {J.}~\bibnamefont
  {Alicea}},\ }\href {https://doi.org/10.1103/PhysRevLett.121.106601}
  {\bibfield  {journal} {\bibinfo  {journal} {Phys. Rev. Lett.}\ }\textbf
  {\bibinfo {volume} {121}},\ \bibinfo {pages} {106601} (\bibinfo {year}
  {2018})}\BibitemShut {NoStop}%
\bibitem [{\citenamefont {Wu}\ \emph {et~al.}(2006)\citenamefont {Wu},
  \citenamefont {Bernevig},\ and\ \citenamefont {Zhang}}]{Wu_Zhang_2006}%
  \BibitemOpen
  \bibfield  {author} {\bibinfo {author} {\bibfnamefont {C.}~\bibnamefont
  {Wu}}, \bibinfo {author} {\bibfnamefont {B.~A.}\ \bibnamefont {Bernevig}},\
  and\ \bibinfo {author} {\bibfnamefont {S.-C.}\ \bibnamefont {Zhang}},\ }\href
  {https://doi.org/10.1103/PhysRevLett.96.106401} {\bibfield  {journal}
  {\bibinfo  {journal} {Phys. Rev. Lett.}\ }\textbf {\bibinfo {volume} {96}},\
  \bibinfo {pages} {106401} (\bibinfo {year} {2006})}\BibitemShut {NoStop}%
\bibitem [{\citenamefont {Xu}\ and\ \citenamefont
  {Moore}(2006)}]{Xu_Moore_2006}%
  \BibitemOpen
  \bibfield  {author} {\bibinfo {author} {\bibfnamefont {C.}~\bibnamefont
  {Xu}}\ and\ \bibinfo {author} {\bibfnamefont {J.~E.}\ \bibnamefont {Moore}},\
  }\href {https://doi.org/10.1103/PhysRevB.73.045322} {\bibfield  {journal}
  {\bibinfo  {journal} {Phys. Rev. B}\ }\textbf {\bibinfo {volume} {73}},\
  \bibinfo {pages} {045322} (\bibinfo {year} {2006})}\BibitemShut {NoStop}%
\bibitem [{\citenamefont {Del~Maestro}\ \emph {et~al.}(2013)\citenamefont
  {Del~Maestro}, \citenamefont {Hyart},\ and\ \citenamefont
  {Rosenow}}]{Maestro_Rosenow_2013}%
  \BibitemOpen
  \bibfield  {author} {\bibinfo {author} {\bibfnamefont {A.}~\bibnamefont
  {Del~Maestro}}, \bibinfo {author} {\bibfnamefont {T.}~\bibnamefont {Hyart}},\
  and\ \bibinfo {author} {\bibfnamefont {B.}~\bibnamefont {Rosenow}},\ }\href
  {https://doi.org/10.1103/PhysRevB.87.165440} {\bibfield  {journal} {\bibinfo
  {journal} {Phys. Rev. B}\ }\textbf {\bibinfo {volume} {87}},\ \bibinfo
  {pages} {165440} (\bibinfo {year} {2013})}\BibitemShut {NoStop}%
\bibitem [{\citenamefont {Hsu}\ \emph {et~al.}(2017)\citenamefont {Hsu},
  \citenamefont {Stano}, \citenamefont {Klinovaja},\ and\ \citenamefont
  {Loss}}]{Hsu_Loss_2017}%
  \BibitemOpen
  \bibfield  {author} {\bibinfo {author} {\bibfnamefont {C.-H.}\ \bibnamefont
  {Hsu}}, \bibinfo {author} {\bibfnamefont {P.}~\bibnamefont {Stano}}, \bibinfo
  {author} {\bibfnamefont {J.}~\bibnamefont {Klinovaja}},\ and\ \bibinfo
  {author} {\bibfnamefont {D.}~\bibnamefont {Loss}},\ }\href
  {https://doi.org/10.1103/PhysRevB.96.081405} {\bibfield  {journal} {\bibinfo
  {journal} {Phys. Rev. B}\ }\textbf {\bibinfo {volume} {96}},\ \bibinfo
  {pages} {081405} (\bibinfo {year} {2017})}\BibitemShut {NoStop}%
\bibitem [{\citenamefont {Hsu}\ \emph {et~al.}(2018)\citenamefont {Hsu},
  \citenamefont {Stano}, \citenamefont {Klinovaja},\ and\ \citenamefont
  {Loss}}]{Hsu_Loss_2018}%
  \BibitemOpen
  \bibfield  {author} {\bibinfo {author} {\bibfnamefont {C.-H.}\ \bibnamefont
  {Hsu}}, \bibinfo {author} {\bibfnamefont {P.}~\bibnamefont {Stano}}, \bibinfo
  {author} {\bibfnamefont {J.}~\bibnamefont {Klinovaja}},\ and\ \bibinfo
  {author} {\bibfnamefont {D.}~\bibnamefont {Loss}},\ }\href
  {https://doi.org/10.1103/PhysRevB.97.125432} {\bibfield  {journal} {\bibinfo
  {journal} {Phys. Rev. B}\ }\textbf {\bibinfo {volume} {97}},\ \bibinfo
  {pages} {125432} (\bibinfo {year} {2018})}\BibitemShut {NoStop}%
\bibitem [{\citenamefont {Schmidt}\ \emph {et~al.}(2012)\citenamefont
  {Schmidt}, \citenamefont {Rachel}, \citenamefont {von Oppen},\ and\
  \citenamefont {Glazman}}]{Schmidt_Glazman_2012}%
  \BibitemOpen
  \bibfield  {author} {\bibinfo {author} {\bibfnamefont {T.~L.}\ \bibnamefont
  {Schmidt}}, \bibinfo {author} {\bibfnamefont {S.}~\bibnamefont {Rachel}},
  \bibinfo {author} {\bibfnamefont {F.}~\bibnamefont {von Oppen}},\ and\
  \bibinfo {author} {\bibfnamefont {L.~I.}\ \bibnamefont {Glazman}},\ }\href
  {https://doi.org/10.1103/PhysRevLett.108.156402} {\bibfield  {journal}
  {\bibinfo  {journal} {Phys. Rev. Lett.}\ }\textbf {\bibinfo {volume} {108}},\
  \bibinfo {pages} {156402} (\bibinfo {year} {2012})}\BibitemShut {NoStop}%
\bibitem [{\citenamefont {Chou}\ \emph {et~al.}(2015)\citenamefont {Chou},
  \citenamefont {Levchenko},\ and\ \citenamefont {Foster}}]{Chou_Foster_2015}%
  \BibitemOpen
  \bibfield  {author} {\bibinfo {author} {\bibfnamefont {Y.-Z.}\ \bibnamefont
  {Chou}}, \bibinfo {author} {\bibfnamefont {A.}~\bibnamefont {Levchenko}},\
  and\ \bibinfo {author} {\bibfnamefont {M.~S.}\ \bibnamefont {Foster}},\
  }\href {https://doi.org/10.1103/PhysRevLett.115.186404} {\bibfield  {journal}
  {\bibinfo  {journal} {Phys. Rev. Lett.}\ }\textbf {\bibinfo {volume} {115}},\
  \bibinfo {pages} {186404} (\bibinfo {year} {2015})}\BibitemShut {NoStop}%
\bibitem [{\citenamefont {Schleder}\ \emph {et~al.}(2021)\citenamefont
  {Schleder}, \citenamefont {Focassio},\ and\ \citenamefont
  {Fazzio}}]{Schleder_Fazzio_2021}%
  \BibitemOpen
  \bibfield  {author} {\bibinfo {author} {\bibfnamefont {G.~R.}\ \bibnamefont
  {Schleder}}, \bibinfo {author} {\bibfnamefont {B.}~\bibnamefont {Focassio}},\
  and\ \bibinfo {author} {\bibfnamefont {A.}~\bibnamefont {Fazzio}},\ }\href
  {https://doi.org/10.1063/5.0055035} {\bibfield  {journal} {\bibinfo
  {journal} {Applied Physics Reviews}\ }\textbf {\bibinfo {volume} {8}},\
  \bibinfo {pages} {031409} (\bibinfo {year} {2021})}\BibitemShut {NoStop}%
\bibitem [{\citenamefont {Di~Sante}\ \emph {et~al.}(2017)\citenamefont
  {Di~Sante}, \citenamefont {Das}, \citenamefont {Bigi}, \citenamefont
  {Erg\"onenc}, \citenamefont {G\"urtler}, \citenamefont {Krieger},
  \citenamefont {Schmitt}, \citenamefont {Ali}, \citenamefont {Rossi},
  \citenamefont {Thomale}, \citenamefont {Franchini}, \citenamefont {Picozzi},
  \citenamefont {Fujii}, \citenamefont {Strocov}, \citenamefont {Sangiovanni},
  \citenamefont {Vobornik}, \citenamefont {Cava},\ and\ \citenamefont
  {Panaccione}}]{DiSante_Panaccione_2017}%
  \BibitemOpen
  \bibfield  {author} {\bibinfo {author} {\bibfnamefont {D.}~\bibnamefont
  {Di~Sante}}, \bibinfo {author} {\bibfnamefont {P.~K.}\ \bibnamefont {Das}},
  \bibinfo {author} {\bibfnamefont {C.}~\bibnamefont {Bigi}}, \bibinfo {author}
  {\bibfnamefont {Z.}~\bibnamefont {Erg\"onenc}}, \bibinfo {author}
  {\bibfnamefont {N.}~\bibnamefont {G\"urtler}}, \bibinfo {author}
  {\bibfnamefont {J.~A.}\ \bibnamefont {Krieger}}, \bibinfo {author}
  {\bibfnamefont {T.}~\bibnamefont {Schmitt}}, \bibinfo {author} {\bibfnamefont
  {M.~N.}\ \bibnamefont {Ali}}, \bibinfo {author} {\bibfnamefont
  {G.}~\bibnamefont {Rossi}}, \bibinfo {author} {\bibfnamefont
  {R.}~\bibnamefont {Thomale}}, \bibinfo {author} {\bibfnamefont
  {C.}~\bibnamefont {Franchini}}, \bibinfo {author} {\bibfnamefont
  {S.}~\bibnamefont {Picozzi}}, \bibinfo {author} {\bibfnamefont
  {J.}~\bibnamefont {Fujii}}, \bibinfo {author} {\bibfnamefont {V.~N.}\
  \bibnamefont {Strocov}}, \bibinfo {author} {\bibfnamefont {G.}~\bibnamefont
  {Sangiovanni}}, \bibinfo {author} {\bibfnamefont {I.}~\bibnamefont
  {Vobornik}}, \bibinfo {author} {\bibfnamefont {R.~J.}\ \bibnamefont {Cava}},\
  and\ \bibinfo {author} {\bibfnamefont {G.}~\bibnamefont {Panaccione}},\
  }\href {https://doi.org/10.1103/PhysRevLett.119.026403} {\bibfield  {journal}
  {\bibinfo  {journal} {Phys. Rev. Lett.}\ }\textbf {\bibinfo {volume} {119}},\
  \bibinfo {pages} {026403} (\bibinfo {year} {2017})}\BibitemShut {NoStop}%
\bibitem [{\citenamefont {Das}\ \emph {et~al.}(2019)\citenamefont {Das},
  \citenamefont {Sante}, \citenamefont {Cilento}, \citenamefont {Bigi},
  \citenamefont {Kopic}, \citenamefont {Soranzio}, \citenamefont {Sterzi},
  \citenamefont {Krieger}, \citenamefont {Vobornik}, \citenamefont {Fujii},
  \citenamefont {Okuda}, \citenamefont {Strocov}, \citenamefont {Breese},
  \citenamefont {Parmigiani}, \citenamefont {Rossi}, \citenamefont {Picozzi},
  \citenamefont {Thomale}, \citenamefont {Sangiovanni}, \citenamefont {Cava},\
  and\ \citenamefont {Panaccione}}]{Das_Panaccione_2019}%
  \BibitemOpen
  \bibfield  {author} {\bibinfo {author} {\bibfnamefont {P.~K.}\ \bibnamefont
  {Das}}, \bibinfo {author} {\bibfnamefont {D.~D.}\ \bibnamefont {Sante}},
  \bibinfo {author} {\bibfnamefont {F.}~\bibnamefont {Cilento}}, \bibinfo
  {author} {\bibfnamefont {C.}~\bibnamefont {Bigi}}, \bibinfo {author}
  {\bibfnamefont {D.}~\bibnamefont {Kopic}}, \bibinfo {author} {\bibfnamefont
  {D.}~\bibnamefont {Soranzio}}, \bibinfo {author} {\bibfnamefont
  {A.}~\bibnamefont {Sterzi}}, \bibinfo {author} {\bibfnamefont {J.~A.}\
  \bibnamefont {Krieger}}, \bibinfo {author} {\bibfnamefont {I.}~\bibnamefont
  {Vobornik}}, \bibinfo {author} {\bibfnamefont {J.}~\bibnamefont {Fujii}},
  \bibinfo {author} {\bibfnamefont {T.}~\bibnamefont {Okuda}}, \bibinfo
  {author} {\bibfnamefont {V.~N.}\ \bibnamefont {Strocov}}, \bibinfo {author}
  {\bibfnamefont {M.~B.~H.}\ \bibnamefont {Breese}}, \bibinfo {author}
  {\bibfnamefont {F.}~\bibnamefont {Parmigiani}}, \bibinfo {author}
  {\bibfnamefont {G.}~\bibnamefont {Rossi}}, \bibinfo {author} {\bibfnamefont
  {S.}~\bibnamefont {Picozzi}}, \bibinfo {author} {\bibfnamefont
  {R.}~\bibnamefont {Thomale}}, \bibinfo {author} {\bibfnamefont
  {G.}~\bibnamefont {Sangiovanni}}, \bibinfo {author} {\bibfnamefont {R.~J.}\
  \bibnamefont {Cava}},\ and\ \bibinfo {author} {\bibfnamefont
  {G.}~\bibnamefont {Panaccione}},\ }\href
  {https://doi.org/10.1088/2516-1075/ab0835} {\bibfield  {journal} {\bibinfo
  {journal} {Electronic Structure}\ }\textbf {\bibinfo {volume} {1}},\ \bibinfo
  {pages} {014003} (\bibinfo {year} {2019})}\BibitemShut {NoStop}%
\bibitem [{\citenamefont {Qian}\ \emph {et~al.}(2014)\citenamefont {Qian},
  \citenamefont {Liu}, \citenamefont {Fu},\ and\ \citenamefont
  {Li}}]{Qian_Li_2014}%
  \BibitemOpen
  \bibfield  {author} {\bibinfo {author} {\bibfnamefont {X.}~\bibnamefont
  {Qian}}, \bibinfo {author} {\bibfnamefont {J.}~\bibnamefont {Liu}}, \bibinfo
  {author} {\bibfnamefont {L.}~\bibnamefont {Fu}},\ and\ \bibinfo {author}
  {\bibfnamefont {J.}~\bibnamefont {Li}},\ }\href
  {https://doi.org/10.1126/science.1256815} {\bibfield  {journal} {\bibinfo
  {journal} {Science}\ }\textbf {\bibinfo {volume} {346}},\ \bibinfo {pages}
  {1344} (\bibinfo {year} {2014})}\BibitemShut {NoStop}%
\bibitem [{\citenamefont {Fei}\ \emph {et~al.}(2017)\citenamefont {Fei},
  \citenamefont {Palomaki}, \citenamefont {Wu}, \citenamefont {Zhao},
  \citenamefont {Cai}, \citenamefont {Sun}, \citenamefont {Nguyen},
  \citenamefont {Finney}, \citenamefont {Xu},\ and\ \citenamefont
  {Cobden}}]{Fei_Cobden_2017}%
  \BibitemOpen
  \bibfield  {author} {\bibinfo {author} {\bibfnamefont {Z.}~\bibnamefont
  {Fei}}, \bibinfo {author} {\bibfnamefont {T.}~\bibnamefont {Palomaki}},
  \bibinfo {author} {\bibfnamefont {S.}~\bibnamefont {Wu}}, \bibinfo {author}
  {\bibfnamefont {W.}~\bibnamefont {Zhao}}, \bibinfo {author} {\bibfnamefont
  {X.}~\bibnamefont {Cai}}, \bibinfo {author} {\bibfnamefont {B.}~\bibnamefont
  {Sun}}, \bibinfo {author} {\bibfnamefont {P.}~\bibnamefont {Nguyen}},
  \bibinfo {author} {\bibfnamefont {J.}~\bibnamefont {Finney}}, \bibinfo
  {author} {\bibfnamefont {X.}~\bibnamefont {Xu}},\ and\ \bibinfo {author}
  {\bibfnamefont {D.~H.}\ \bibnamefont {Cobden}},\ }\href
  {https://doi.org/10.1038/nphys4091} {\bibfield  {journal} {\bibinfo
  {journal} {Nature Physics}\ }\textbf {\bibinfo {volume} {13}},\ \bibinfo
  {pages} {677} (\bibinfo {year} {2017})}\BibitemShut {NoStop}%
\bibitem [{\citenamefont {Tang}\ \emph {et~al.}(2017)\citenamefont {Tang},
  \citenamefont {Zhang}, \citenamefont {Wong}, \citenamefont {Pedramrazi},
  \citenamefont {Tsai}, \citenamefont {Jia}, \citenamefont {Moritz},
  \citenamefont {Claassen}, \citenamefont {Ryu}, \citenamefont {Kahn},
  \citenamefont {Jiang}, \citenamefont {Yan}, \citenamefont {Hashimoto},
  \citenamefont {Lu}, \citenamefont {Moore}, \citenamefont {Hwang},
  \citenamefont {Hwang}, \citenamefont {Hussain}, \citenamefont {Chen},
  \citenamefont {Ugeda}, \citenamefont {Liu}, \citenamefont {Xie},
  \citenamefont {Devereaux}, \citenamefont {Crommie}, \citenamefont {Mo},\ and\
  \citenamefont {Shen}}]{Tang_Shen_2017}%
  \BibitemOpen
  \bibfield  {author} {\bibinfo {author} {\bibfnamefont {S.}~\bibnamefont
  {Tang}}, \bibinfo {author} {\bibfnamefont {C.}~\bibnamefont {Zhang}},
  \bibinfo {author} {\bibfnamefont {D.}~\bibnamefont {Wong}}, \bibinfo {author}
  {\bibfnamefont {Z.}~\bibnamefont {Pedramrazi}}, \bibinfo {author}
  {\bibfnamefont {H.-Z.}\ \bibnamefont {Tsai}}, \bibinfo {author}
  {\bibfnamefont {C.}~\bibnamefont {Jia}}, \bibinfo {author} {\bibfnamefont
  {B.}~\bibnamefont {Moritz}}, \bibinfo {author} {\bibfnamefont
  {M.}~\bibnamefont {Claassen}}, \bibinfo {author} {\bibfnamefont
  {H.}~\bibnamefont {Ryu}}, \bibinfo {author} {\bibfnamefont {S.}~\bibnamefont
  {Kahn}}, \bibinfo {author} {\bibfnamefont {J.}~\bibnamefont {Jiang}},
  \bibinfo {author} {\bibfnamefont {H.}~\bibnamefont {Yan}}, \bibinfo {author}
  {\bibfnamefont {M.}~\bibnamefont {Hashimoto}}, \bibinfo {author}
  {\bibfnamefont {D.}~\bibnamefont {Lu}}, \bibinfo {author} {\bibfnamefont
  {R.~G.}\ \bibnamefont {Moore}}, \bibinfo {author} {\bibfnamefont {C.-C.}\
  \bibnamefont {Hwang}}, \bibinfo {author} {\bibfnamefont {C.}~\bibnamefont
  {Hwang}}, \bibinfo {author} {\bibfnamefont {Z.}~\bibnamefont {Hussain}},
  \bibinfo {author} {\bibfnamefont {Y.}~\bibnamefont {Chen}}, \bibinfo {author}
  {\bibfnamefont {M.~M.}\ \bibnamefont {Ugeda}}, \bibinfo {author}
  {\bibfnamefont {Z.}~\bibnamefont {Liu}}, \bibinfo {author} {\bibfnamefont
  {X.}~\bibnamefont {Xie}}, \bibinfo {author} {\bibfnamefont {T.~P.}\
  \bibnamefont {Devereaux}}, \bibinfo {author} {\bibfnamefont {M.~F.}\
  \bibnamefont {Crommie}}, \bibinfo {author} {\bibfnamefont {S.-K.}\
  \bibnamefont {Mo}},\ and\ \bibinfo {author} {\bibfnamefont {Z.-X.}\
  \bibnamefont {Shen}},\ }\href {https://doi.org/10.1038/nphys4174} {\bibfield
  {journal} {\bibinfo  {journal} {Nature Physics}\ }\textbf {\bibinfo {volume}
  {13}},\ \bibinfo {pages} {683} (\bibinfo {year} {2017})}\BibitemShut
  {NoStop}%
\bibitem [{\citenamefont {Wu}\ \emph {et~al.}(2018)\citenamefont {Wu},
  \citenamefont {Fatemi}, \citenamefont {Gibson}, \citenamefont {Watanabe},
  \citenamefont {Taniguchi}, \citenamefont {Cava},\ and\ \citenamefont
  {Jarillo-Herrero}}]{Wu_Jarillo-Herrero_2018}%
  \BibitemOpen
  \bibfield  {author} {\bibinfo {author} {\bibfnamefont {S.}~\bibnamefont
  {Wu}}, \bibinfo {author} {\bibfnamefont {V.}~\bibnamefont {Fatemi}}, \bibinfo
  {author} {\bibfnamefont {Q.~D.}\ \bibnamefont {Gibson}}, \bibinfo {author}
  {\bibfnamefont {K.}~\bibnamefont {Watanabe}}, \bibinfo {author}
  {\bibfnamefont {T.}~\bibnamefont {Taniguchi}}, \bibinfo {author}
  {\bibfnamefont {R.~J.}\ \bibnamefont {Cava}},\ and\ \bibinfo {author}
  {\bibfnamefont {P.}~\bibnamefont {Jarillo-Herrero}},\ }\href
  {https://doi.org/10.1126/science.aan6003} {\bibfield  {journal} {\bibinfo
  {journal} {Science}\ }\textbf {\bibinfo {volume} {359}},\ \bibinfo {pages}
  {76} (\bibinfo {year} {2018})}\BibitemShut {NoStop}%
\bibitem [{\citenamefont {Li}\ \emph {et~al.}(2020)\citenamefont {Li},
  \citenamefont {Song},\ and\ \citenamefont {Tang}}]{Li_Tang_2020}%
  \BibitemOpen
  \bibfield  {author} {\bibinfo {author} {\bibfnamefont {Z.}~\bibnamefont
  {Li}}, \bibinfo {author} {\bibfnamefont {Y.}~\bibnamefont {Song}},\ and\
  \bibinfo {author} {\bibfnamefont {S.}~\bibnamefont {Tang}},\ }\href
  {https://doi.org/10.1088/1361-648x/ab8660} {\bibfield  {journal} {\bibinfo
  {journal} {Journal of Physics: Condensed Matter}\ }\textbf {\bibinfo {volume}
  {32}},\ \bibinfo {pages} {333001} (\bibinfo {year} {2020})}\BibitemShut
  {NoStop}%
\bibitem [{\citenamefont {Jia}\ \emph {et~al.}(2017)\citenamefont {Jia},
  \citenamefont {Song}, \citenamefont {Li}, \citenamefont {Ran}, \citenamefont
  {Lu}, \citenamefont {Zheng}, \citenamefont {Zhu}, \citenamefont {Shi},
  \citenamefont {Sun}, \citenamefont {Wen}, \citenamefont {Xing},\ and\
  \citenamefont {Li}}]{Jia_Li_2017}%
  \BibitemOpen
  \bibfield  {author} {\bibinfo {author} {\bibfnamefont {Z.-Y.}\ \bibnamefont
  {Jia}}, \bibinfo {author} {\bibfnamefont {Y.-H.}\ \bibnamefont {Song}},
  \bibinfo {author} {\bibfnamefont {X.-B.}\ \bibnamefont {Li}}, \bibinfo
  {author} {\bibfnamefont {K.}~\bibnamefont {Ran}}, \bibinfo {author}
  {\bibfnamefont {P.}~\bibnamefont {Lu}}, \bibinfo {author} {\bibfnamefont
  {H.-J.}\ \bibnamefont {Zheng}}, \bibinfo {author} {\bibfnamefont {X.-Y.}\
  \bibnamefont {Zhu}}, \bibinfo {author} {\bibfnamefont {Z.-Q.}\ \bibnamefont
  {Shi}}, \bibinfo {author} {\bibfnamefont {J.}~\bibnamefont {Sun}}, \bibinfo
  {author} {\bibfnamefont {J.}~\bibnamefont {Wen}}, \bibinfo {author}
  {\bibfnamefont {D.}~\bibnamefont {Xing}},\ and\ \bibinfo {author}
  {\bibfnamefont {S.-C.}\ \bibnamefont {Li}},\ }\href
  {https://doi.org/10.1103/PhysRevB.96.041108} {\bibfield  {journal} {\bibinfo
  {journal} {Phys. Rev. B}\ }\textbf {\bibinfo {volume} {96}},\ \bibinfo
  {pages} {041108} (\bibinfo {year} {2017})}\BibitemShut {NoStop}%
\bibitem [{\citenamefont {Peng}\ \emph {et~al.}(2017)\citenamefont {Peng},
  \citenamefont {Yuan}, \citenamefont {Li}, \citenamefont {Yang}, \citenamefont
  {Xian}, \citenamefont {Yi}, \citenamefont {Shi},\ and\ \citenamefont
  {Fu}}]{Peng_Fu_2017}%
  \BibitemOpen
  \bibfield  {author} {\bibinfo {author} {\bibfnamefont {L.}~\bibnamefont
  {Peng}}, \bibinfo {author} {\bibfnamefont {Y.}~\bibnamefont {Yuan}}, \bibinfo
  {author} {\bibfnamefont {G.}~\bibnamefont {Li}}, \bibinfo {author}
  {\bibfnamefont {X.}~\bibnamefont {Yang}}, \bibinfo {author} {\bibfnamefont
  {J.-J.}\ \bibnamefont {Xian}}, \bibinfo {author} {\bibfnamefont {C.-J.}\
  \bibnamefont {Yi}}, \bibinfo {author} {\bibfnamefont {Y.-G.}\ \bibnamefont
  {Shi}},\ and\ \bibinfo {author} {\bibfnamefont {Y.-S.}\ \bibnamefont {Fu}},\
  }\href {https://doi.org/10.1038/s41467-017-00745-8} {\bibfield  {journal}
  {\bibinfo  {journal} {Nature Communications}\ }\textbf {\bibinfo {volume}
  {8}},\ \bibinfo {pages} {659} (\bibinfo {year} {2017})}\BibitemShut {NoStop}%
\bibitem [{\citenamefont {Maximenko}\ \emph {et~al.}(2022)\citenamefont
  {Maximenko}, \citenamefont {Chang}, \citenamefont {Chen}, \citenamefont
  {Hirsbrunner}, \citenamefont {Swiech}, \citenamefont {Hughes}, \citenamefont
  {Wagner},\ and\ \citenamefont {Madhavan}}]{Maximenko_Madhavan_2022}%
  \BibitemOpen
  \bibfield  {author} {\bibinfo {author} {\bibfnamefont {Y.}~\bibnamefont
  {Maximenko}}, \bibinfo {author} {\bibfnamefont {Y.}~\bibnamefont {Chang}},
  \bibinfo {author} {\bibfnamefont {G.}~\bibnamefont {Chen}}, \bibinfo {author}
  {\bibfnamefont {M.~R.}\ \bibnamefont {Hirsbrunner}}, \bibinfo {author}
  {\bibfnamefont {W.}~\bibnamefont {Swiech}}, \bibinfo {author} {\bibfnamefont
  {T.~L.}\ \bibnamefont {Hughes}}, \bibinfo {author} {\bibfnamefont {L.~K.}\
  \bibnamefont {Wagner}},\ and\ \bibinfo {author} {\bibfnamefont
  {V.}~\bibnamefont {Madhavan}},\ }\href
  {https://doi.org/10.1038/s41535-022-00433-x} {\bibfield  {journal} {\bibinfo
  {journal} {npj Quantum Materials}\ }\textbf {\bibinfo {volume} {7}},\
  \bibinfo {pages} {29} (\bibinfo {year} {2022})}\BibitemShut {NoStop}%
\bibitem [{\citenamefont {Song}\ \emph {et~al.}(2018)\citenamefont {Song},
  \citenamefont {Jia}, \citenamefont {Zhang}, \citenamefont {Zhu},
  \citenamefont {Shi}, \citenamefont {Wang}, \citenamefont {Zhu}, \citenamefont
  {Yuan}, \citenamefont {Zhang}, \citenamefont {Xing},\ and\ \citenamefont
  {Li}}]{Song_Li_2018}%
  \BibitemOpen
  \bibfield  {author} {\bibinfo {author} {\bibfnamefont {Y.-H.}\ \bibnamefont
  {Song}}, \bibinfo {author} {\bibfnamefont {Z.-Y.}\ \bibnamefont {Jia}},
  \bibinfo {author} {\bibfnamefont {D.}~\bibnamefont {Zhang}}, \bibinfo
  {author} {\bibfnamefont {X.-Y.}\ \bibnamefont {Zhu}}, \bibinfo {author}
  {\bibfnamefont {Z.-Q.}\ \bibnamefont {Shi}}, \bibinfo {author} {\bibfnamefont
  {H.}~\bibnamefont {Wang}}, \bibinfo {author} {\bibfnamefont {L.}~\bibnamefont
  {Zhu}}, \bibinfo {author} {\bibfnamefont {Q.-Q.}\ \bibnamefont {Yuan}},
  \bibinfo {author} {\bibfnamefont {H.}~\bibnamefont {Zhang}}, \bibinfo
  {author} {\bibfnamefont {D.-Y.}\ \bibnamefont {Xing}},\ and\ \bibinfo
  {author} {\bibfnamefont {S.-C.}\ \bibnamefont {Li}},\ }\href
  {https://doi.org/10.1038/s41467-018-06635-x} {\bibfield  {journal} {\bibinfo
  {journal} {Nature Communications}\ }\textbf {\bibinfo {volume} {9}},\
  \bibinfo {pages} {4071} (\bibinfo {year} {2018})}\BibitemShut {NoStop}%
\bibitem [{\citenamefont {L{\"u}pke}\ \emph {et~al.}(2020)\citenamefont
  {L{\"u}pke}, \citenamefont {Waters}, \citenamefont {de~la Barrera},
  \citenamefont {Widom}, \citenamefont {Mandrus}, \citenamefont {Yan},
  \citenamefont {Feenstra},\ and\ \citenamefont {Hunt}}]{Lupke_Hunt_2020}%
  \BibitemOpen
  \bibfield  {author} {\bibinfo {author} {\bibfnamefont {F.}~\bibnamefont
  {L{\"u}pke}}, \bibinfo {author} {\bibfnamefont {D.}~\bibnamefont {Waters}},
  \bibinfo {author} {\bibfnamefont {S.~C.}\ \bibnamefont {de~la Barrera}},
  \bibinfo {author} {\bibfnamefont {M.}~\bibnamefont {Widom}}, \bibinfo
  {author} {\bibfnamefont {D.~G.}\ \bibnamefont {Mandrus}}, \bibinfo {author}
  {\bibfnamefont {J.}~\bibnamefont {Yan}}, \bibinfo {author} {\bibfnamefont
  {R.~M.}\ \bibnamefont {Feenstra}},\ and\ \bibinfo {author} {\bibfnamefont
  {B.~M.}\ \bibnamefont {Hunt}},\ }\href
  {https://doi.org/10.1038/s41567-020-0816-x} {\bibfield  {journal} {\bibinfo
  {journal} {Nature Physics}\ }\textbf {\bibinfo {volume} {16}},\ \bibinfo
  {pages} {526} (\bibinfo {year} {2020})}\BibitemShut {NoStop}%
\bibitem [{\citenamefont {Tao}\ \emph {et~al.}(2022)\citenamefont {Tao},
  \citenamefont {Tong}, \citenamefont {Das}, \citenamefont {Ho}, \citenamefont
  {Sato}, \citenamefont {Haze}, \citenamefont {Jia}, \citenamefont {Que},
  \citenamefont {Bussolotti}, \citenamefont {Goh}, \citenamefont {Wang},
  \citenamefont {Lin}, \citenamefont {Bansil}, \citenamefont {Mukherjee},
  \citenamefont {Hasegawa},\ and\ \citenamefont {Weber}}]{Tao_Weber_2022}%
  \BibitemOpen
  \bibfield  {author} {\bibinfo {author} {\bibfnamefont {W.}~\bibnamefont
  {Tao}}, \bibinfo {author} {\bibfnamefont {Z.~J.}\ \bibnamefont {Tong}},
  \bibinfo {author} {\bibfnamefont {A.}~\bibnamefont {Das}}, \bibinfo {author}
  {\bibfnamefont {D.-Q.}\ \bibnamefont {Ho}}, \bibinfo {author} {\bibfnamefont
  {Y.}~\bibnamefont {Sato}}, \bibinfo {author} {\bibfnamefont {M.}~\bibnamefont
  {Haze}}, \bibinfo {author} {\bibfnamefont {J.}~\bibnamefont {Jia}}, \bibinfo
  {author} {\bibfnamefont {Y.}~\bibnamefont {Que}}, \bibinfo {author}
  {\bibfnamefont {F.}~\bibnamefont {Bussolotti}}, \bibinfo {author}
  {\bibfnamefont {K.~E.~J.}\ \bibnamefont {Goh}}, \bibinfo {author}
  {\bibfnamefont {B.}~\bibnamefont {Wang}}, \bibinfo {author} {\bibfnamefont
  {H.}~\bibnamefont {Lin}}, \bibinfo {author} {\bibfnamefont {A.}~\bibnamefont
  {Bansil}}, \bibinfo {author} {\bibfnamefont {S.}~\bibnamefont {Mukherjee}},
  \bibinfo {author} {\bibfnamefont {Y.}~\bibnamefont {Hasegawa}},\ and\
  \bibinfo {author} {\bibfnamefont {B.}~\bibnamefont {Weber}},\ }\href
  {https://doi.org/10.1103/PhysRevB.105.094512} {\bibfield  {journal} {\bibinfo
   {journal} {Phys. Rev. B}\ }\textbf {\bibinfo {volume} {105}},\ \bibinfo
  {pages} {094512} (\bibinfo {year} {2022})}\BibitemShut {NoStop}%
\bibitem [{\citenamefont {Zhao}\ \emph
  {et~al.}(2020{\natexlab{a}})\citenamefont {Zhao}, \citenamefont {Hu},
  \citenamefont {Qin}, \citenamefont {Xia}, \citenamefont {Liu}, \citenamefont
  {Wang}, \citenamefont {Guan}, \citenamefont {Li}, \citenamefont {Zheng},
  \citenamefont {Liu},\ and\ \citenamefont {Jia}}]{Zhao_Jia_2020}%
  \BibitemOpen
  \bibfield  {author} {\bibinfo {author} {\bibfnamefont {C.}~\bibnamefont
  {Zhao}}, \bibinfo {author} {\bibfnamefont {M.}~\bibnamefont {Hu}}, \bibinfo
  {author} {\bibfnamefont {J.}~\bibnamefont {Qin}}, \bibinfo {author}
  {\bibfnamefont {B.}~\bibnamefont {Xia}}, \bibinfo {author} {\bibfnamefont
  {C.}~\bibnamefont {Liu}}, \bibinfo {author} {\bibfnamefont {S.}~\bibnamefont
  {Wang}}, \bibinfo {author} {\bibfnamefont {D.}~\bibnamefont {Guan}}, \bibinfo
  {author} {\bibfnamefont {Y.}~\bibnamefont {Li}}, \bibinfo {author}
  {\bibfnamefont {H.}~\bibnamefont {Zheng}}, \bibinfo {author} {\bibfnamefont
  {J.}~\bibnamefont {Liu}},\ and\ \bibinfo {author} {\bibfnamefont
  {J.}~\bibnamefont {Jia}},\ }\href
  {https://doi.org/10.1103/PhysRevLett.125.046801} {\bibfield  {journal}
  {\bibinfo  {journal} {Phys. Rev. Lett.}\ }\textbf {\bibinfo {volume} {125}},\
  \bibinfo {pages} {046801} (\bibinfo {year} {2020}{\natexlab{a}})}\BibitemShut
  {NoStop}%
\bibitem [{\citenamefont {Zhao}\ \emph
  {et~al.}(2020{\natexlab{b}})\citenamefont {Zhao}, \citenamefont {Fei},
  \citenamefont {Song}, \citenamefont {Choi}, \citenamefont {Palomaki},
  \citenamefont {Sun}, \citenamefont {Malinowski}, \citenamefont {McGuire},
  \citenamefont {Chu}, \citenamefont {Xu},\ and\ \citenamefont
  {Cobden}}]{Zhao_Cobden_2020}%
  \BibitemOpen
  \bibfield  {author} {\bibinfo {author} {\bibfnamefont {W.}~\bibnamefont
  {Zhao}}, \bibinfo {author} {\bibfnamefont {Z.}~\bibnamefont {Fei}}, \bibinfo
  {author} {\bibfnamefont {T.}~\bibnamefont {Song}}, \bibinfo {author}
  {\bibfnamefont {H.~K.}\ \bibnamefont {Choi}}, \bibinfo {author}
  {\bibfnamefont {T.}~\bibnamefont {Palomaki}}, \bibinfo {author}
  {\bibfnamefont {B.}~\bibnamefont {Sun}}, \bibinfo {author} {\bibfnamefont
  {P.}~\bibnamefont {Malinowski}}, \bibinfo {author} {\bibfnamefont {M.~A.}\
  \bibnamefont {McGuire}}, \bibinfo {author} {\bibfnamefont {J.-H.}\
  \bibnamefont {Chu}}, \bibinfo {author} {\bibfnamefont {X.}~\bibnamefont
  {Xu}},\ and\ \bibinfo {author} {\bibfnamefont {D.~H.}\ \bibnamefont
  {Cobden}},\ }\href {https://doi.org/10.1038/s41563-020-0620-0} {\bibfield
  {journal} {\bibinfo  {journal} {Nature Materials}\ }\textbf {\bibinfo
  {volume} {19}},\ \bibinfo {pages} {503} (\bibinfo {year}
  {2020}{\natexlab{b}})}\BibitemShut {NoStop}%
\bibitem [{\citenamefont {Fatemi}\ \emph {et~al.}(2018)\citenamefont {Fatemi},
  \citenamefont {Wu}, \citenamefont {Cao}, \citenamefont {Bretheau},
  \citenamefont {Gibson}, \citenamefont {Watanabe}, \citenamefont {Taniguchi},
  \citenamefont {Cava},\ and\ \citenamefont
  {Jarillo-Herrero}}]{Fatemi_Jarillo-Herrero_2018}%
  \BibitemOpen
  \bibfield  {author} {\bibinfo {author} {\bibfnamefont {V.}~\bibnamefont
  {Fatemi}}, \bibinfo {author} {\bibfnamefont {S.}~\bibnamefont {Wu}}, \bibinfo
  {author} {\bibfnamefont {Y.}~\bibnamefont {Cao}}, \bibinfo {author}
  {\bibfnamefont {L.}~\bibnamefont {Bretheau}}, \bibinfo {author}
  {\bibfnamefont {Q.~D.}\ \bibnamefont {Gibson}}, \bibinfo {author}
  {\bibfnamefont {K.}~\bibnamefont {Watanabe}}, \bibinfo {author}
  {\bibfnamefont {T.}~\bibnamefont {Taniguchi}}, \bibinfo {author}
  {\bibfnamefont {R.~J.}\ \bibnamefont {Cava}},\ and\ \bibinfo {author}
  {\bibfnamefont {P.}~\bibnamefont {Jarillo-Herrero}},\ }\href
  {https://doi.org/10.1126/science.aar4642} {\bibfield  {journal} {\bibinfo
  {journal} {Science}\ }\textbf {\bibinfo {volume} {362}},\ \bibinfo {pages}
  {926} (\bibinfo {year} {2018})}\BibitemShut {NoStop}%
\bibitem [{\citenamefont {Sajadi}\ \emph {et~al.}(2018)\citenamefont {Sajadi},
  \citenamefont {Palomaki}, \citenamefont {Fei}, \citenamefont {Zhao},
  \citenamefont {Bement}, \citenamefont {Olsen}, \citenamefont {Luescher},
  \citenamefont {Xu}, \citenamefont {Folk},\ and\ \citenamefont
  {Cobden}}]{Sajadi_Cobden_2018}%
  \BibitemOpen
  \bibfield  {author} {\bibinfo {author} {\bibfnamefont {E.}~\bibnamefont
  {Sajadi}}, \bibinfo {author} {\bibfnamefont {T.}~\bibnamefont {Palomaki}},
  \bibinfo {author} {\bibfnamefont {Z.}~\bibnamefont {Fei}}, \bibinfo {author}
  {\bibfnamefont {W.}~\bibnamefont {Zhao}}, \bibinfo {author} {\bibfnamefont
  {P.}~\bibnamefont {Bement}}, \bibinfo {author} {\bibfnamefont
  {C.}~\bibnamefont {Olsen}}, \bibinfo {author} {\bibfnamefont
  {S.}~\bibnamefont {Luescher}}, \bibinfo {author} {\bibfnamefont
  {X.}~\bibnamefont {Xu}}, \bibinfo {author} {\bibfnamefont {J.~A.}\
  \bibnamefont {Folk}},\ and\ \bibinfo {author} {\bibfnamefont {D.~H.}\
  \bibnamefont {Cobden}},\ }\href {https://doi.org/10.1126/science.aar4426}
  {\bibfield  {journal} {\bibinfo  {journal} {Science}\ }\textbf {\bibinfo
  {volume} {362}},\ \bibinfo {pages} {922} (\bibinfo {year}
  {2018})}\BibitemShut {NoStop}%
\bibitem [{\citenamefont {Jia}\ \emph {et~al.}(2022)\citenamefont {Jia},
  \citenamefont {Wang}, \citenamefont {Chiu}, \citenamefont {Song},
  \citenamefont {Yu}, \citenamefont {J{\"a}ck}, \citenamefont {Lei},
  \citenamefont {Klemenz}, \citenamefont {Cevallos}, \citenamefont {Onyszczak},
  \citenamefont {Fishchenko}, \citenamefont {Liu}, \citenamefont {Farahi},
  \citenamefont {Xie}, \citenamefont {Xu}, \citenamefont {Watanabe},
  \citenamefont {Taniguchi}, \citenamefont {Bernevig}, \citenamefont {Cava},
  \citenamefont {Schoop}, \citenamefont {Yazdani},\ and\ \citenamefont
  {Wu}}]{Jia_Wu_2022}%
  \BibitemOpen
  \bibfield  {author} {\bibinfo {author} {\bibfnamefont {Y.}~\bibnamefont
  {Jia}}, \bibinfo {author} {\bibfnamefont {P.}~\bibnamefont {Wang}}, \bibinfo
  {author} {\bibfnamefont {C.-L.}\ \bibnamefont {Chiu}}, \bibinfo {author}
  {\bibfnamefont {Z.}~\bibnamefont {Song}}, \bibinfo {author} {\bibfnamefont
  {G.}~\bibnamefont {Yu}}, \bibinfo {author} {\bibfnamefont {B.}~\bibnamefont
  {J{\"a}ck}}, \bibinfo {author} {\bibfnamefont {S.}~\bibnamefont {Lei}},
  \bibinfo {author} {\bibfnamefont {S.}~\bibnamefont {Klemenz}}, \bibinfo
  {author} {\bibfnamefont {F.~A.}\ \bibnamefont {Cevallos}}, \bibinfo {author}
  {\bibfnamefont {M.}~\bibnamefont {Onyszczak}}, \bibinfo {author}
  {\bibfnamefont {N.}~\bibnamefont {Fishchenko}}, \bibinfo {author}
  {\bibfnamefont {X.}~\bibnamefont {Liu}}, \bibinfo {author} {\bibfnamefont
  {G.}~\bibnamefont {Farahi}}, \bibinfo {author} {\bibfnamefont
  {F.}~\bibnamefont {Xie}}, \bibinfo {author} {\bibfnamefont {Y.}~\bibnamefont
  {Xu}}, \bibinfo {author} {\bibfnamefont {K.}~\bibnamefont {Watanabe}},
  \bibinfo {author} {\bibfnamefont {T.}~\bibnamefont {Taniguchi}}, \bibinfo
  {author} {\bibfnamefont {B.~A.}\ \bibnamefont {Bernevig}}, \bibinfo {author}
  {\bibfnamefont {R.~J.}\ \bibnamefont {Cava}}, \bibinfo {author}
  {\bibfnamefont {L.~M.}\ \bibnamefont {Schoop}}, \bibinfo {author}
  {\bibfnamefont {A.}~\bibnamefont {Yazdani}},\ and\ \bibinfo {author}
  {\bibfnamefont {S.}~\bibnamefont {Wu}},\ }\href
  {https://doi.org/10.1038/s41567-021-01422-w} {\bibfield  {journal} {\bibinfo
  {journal} {Nature Physics}\ }\textbf {\bibinfo {volume} {18}},\ \bibinfo
  {pages} {87} (\bibinfo {year} {2022})}\BibitemShut {NoStop}%
\bibitem [{\citenamefont {Sun}\ \emph {et~al.}(2022)\citenamefont {Sun},
  \citenamefont {Zhao}, \citenamefont {Palomaki}, \citenamefont {Fei},
  \citenamefont {Runburg}, \citenamefont {Malinowski}, \citenamefont {Huang},
  \citenamefont {Cenker}, \citenamefont {Cui}, \citenamefont {Chu},
  \citenamefont {Xu}, \citenamefont {Ataei}, \citenamefont {Varsano},
  \citenamefont {Palummo}, \citenamefont {Molinari}, \citenamefont {Rontani},\
  and\ \citenamefont {Cobden}}]{Sun_Cobden_2022}%
  \BibitemOpen
  \bibfield  {author} {\bibinfo {author} {\bibfnamefont {B.}~\bibnamefont
  {Sun}}, \bibinfo {author} {\bibfnamefont {W.}~\bibnamefont {Zhao}}, \bibinfo
  {author} {\bibfnamefont {T.}~\bibnamefont {Palomaki}}, \bibinfo {author}
  {\bibfnamefont {Z.}~\bibnamefont {Fei}}, \bibinfo {author} {\bibfnamefont
  {E.}~\bibnamefont {Runburg}}, \bibinfo {author} {\bibfnamefont
  {P.}~\bibnamefont {Malinowski}}, \bibinfo {author} {\bibfnamefont
  {X.}~\bibnamefont {Huang}}, \bibinfo {author} {\bibfnamefont
  {J.}~\bibnamefont {Cenker}}, \bibinfo {author} {\bibfnamefont {Y.-T.}\
  \bibnamefont {Cui}}, \bibinfo {author} {\bibfnamefont {J.-H.}\ \bibnamefont
  {Chu}}, \bibinfo {author} {\bibfnamefont {X.}~\bibnamefont {Xu}}, \bibinfo
  {author} {\bibfnamefont {S.~S.}\ \bibnamefont {Ataei}}, \bibinfo {author}
  {\bibfnamefont {D.}~\bibnamefont {Varsano}}, \bibinfo {author} {\bibfnamefont
  {M.}~\bibnamefont {Palummo}}, \bibinfo {author} {\bibfnamefont
  {E.}~\bibnamefont {Molinari}}, \bibinfo {author} {\bibfnamefont
  {M.}~\bibnamefont {Rontani}},\ and\ \bibinfo {author} {\bibfnamefont {D.~H.}\
  \bibnamefont {Cobden}},\ }\href {https://doi.org/10.1038/s41567-021-01427-5}
  {\bibfield  {journal} {\bibinfo  {journal} {Nature Physics}\ }\textbf
  {\bibinfo {volume} {18}},\ \bibinfo {pages} {94} (\bibinfo {year}
  {2022})}\BibitemShut {NoStop}%
\bibitem [{\citenamefont {Wang}\ \emph {et~al.}(2022)\citenamefont {Wang},
  \citenamefont {Yu}, \citenamefont {Kwan}, \citenamefont {Jia}, \citenamefont
  {Lei}, \citenamefont {Klemenz}, \citenamefont {Cevallos}, \citenamefont
  {Singha}, \citenamefont {Devakul}, \citenamefont {Watanabe}, \citenamefont
  {Taniguchi}, \citenamefont {Sondhi}, \citenamefont {Cava}, \citenamefont
  {Schoop}, \citenamefont {Parameswaran},\ and\ \citenamefont
  {Wu}}]{Wang_Wu_2022}%
  \BibitemOpen
  \bibfield  {author} {\bibinfo {author} {\bibfnamefont {P.}~\bibnamefont
  {Wang}}, \bibinfo {author} {\bibfnamefont {G.}~\bibnamefont {Yu}}, \bibinfo
  {author} {\bibfnamefont {Y.~H.}\ \bibnamefont {Kwan}}, \bibinfo {author}
  {\bibfnamefont {Y.}~\bibnamefont {Jia}}, \bibinfo {author} {\bibfnamefont
  {S.}~\bibnamefont {Lei}}, \bibinfo {author} {\bibfnamefont {S.}~\bibnamefont
  {Klemenz}}, \bibinfo {author} {\bibfnamefont {F.~A.}\ \bibnamefont
  {Cevallos}}, \bibinfo {author} {\bibfnamefont {R.}~\bibnamefont {Singha}},
  \bibinfo {author} {\bibfnamefont {T.}~\bibnamefont {Devakul}}, \bibinfo
  {author} {\bibfnamefont {K.}~\bibnamefont {Watanabe}}, \bibinfo {author}
  {\bibfnamefont {T.}~\bibnamefont {Taniguchi}}, \bibinfo {author}
  {\bibfnamefont {S.~L.}\ \bibnamefont {Sondhi}}, \bibinfo {author}
  {\bibfnamefont {R.~J.}\ \bibnamefont {Cava}}, \bibinfo {author}
  {\bibfnamefont {L.~M.}\ \bibnamefont {Schoop}}, \bibinfo {author}
  {\bibfnamefont {S.~A.}\ \bibnamefont {Parameswaran}},\ and\ \bibinfo {author}
  {\bibfnamefont {S.}~\bibnamefont {Wu}},\ }\href
  {https://doi.org/10.1038/s41586-022-04514-6} {\bibfield  {journal} {\bibinfo
  {journal} {Nature}\ }\textbf {\bibinfo {volume} {605}},\ \bibinfo {pages}
  {57} (\bibinfo {year} {2022})}\BibitemShut {NoStop}%
\bibitem [{\citenamefont {Lv}\ \emph {et~al.}(2015)\citenamefont {Lv},
  \citenamefont {Lu}, \citenamefont {Shao}, \citenamefont {Liu}, \citenamefont
  {Tan},\ and\ \citenamefont {Sun}}]{Lv_Sun_2015}%
  \BibitemOpen
  \bibfield  {author} {\bibinfo {author} {\bibfnamefont {H.~Y.}\ \bibnamefont
  {Lv}}, \bibinfo {author} {\bibfnamefont {W.~J.}\ \bibnamefont {Lu}}, \bibinfo
  {author} {\bibfnamefont {D.~F.}\ \bibnamefont {Shao}}, \bibinfo {author}
  {\bibfnamefont {Y.}~\bibnamefont {Liu}}, \bibinfo {author} {\bibfnamefont
  {S.~G.}\ \bibnamefont {Tan}},\ and\ \bibinfo {author} {\bibfnamefont {Y.~P.}\
  \bibnamefont {Sun}},\ }\href {https://doi.org/10.1209/0295-5075/110/37004}
  {\bibfield  {journal} {\bibinfo  {journal} {{EPL} (Europhysics Letters)}\
  }\textbf {\bibinfo {volume} {110}},\ \bibinfo {pages} {37004} (\bibinfo
  {year} {2015})}\BibitemShut {NoStop}%
\bibitem [{\citenamefont {Zheng}\ \emph {et~al.}(2016)\citenamefont {Zheng},
  \citenamefont {Cai}, \citenamefont {Ge}, \citenamefont {Zhang}, \citenamefont
  {Liu}, \citenamefont {Lu}, \citenamefont {Zhang}, \citenamefont {Qiu},
  \citenamefont {Taniguchi}, \citenamefont {Watanabe}, \citenamefont {Jia},
  \citenamefont {Qi}, \citenamefont {Chen}, \citenamefont {Sun},\ and\
  \citenamefont {Feng}}]{Zheng_Feng_2016}%
  \BibitemOpen
  \bibfield  {author} {\bibinfo {author} {\bibfnamefont {F.}~\bibnamefont
  {Zheng}}, \bibinfo {author} {\bibfnamefont {C.}~\bibnamefont {Cai}}, \bibinfo
  {author} {\bibfnamefont {S.}~\bibnamefont {Ge}}, \bibinfo {author}
  {\bibfnamefont {X.}~\bibnamefont {Zhang}}, \bibinfo {author} {\bibfnamefont
  {X.}~\bibnamefont {Liu}}, \bibinfo {author} {\bibfnamefont {H.}~\bibnamefont
  {Lu}}, \bibinfo {author} {\bibfnamefont {Y.}~\bibnamefont {Zhang}}, \bibinfo
  {author} {\bibfnamefont {J.}~\bibnamefont {Qiu}}, \bibinfo {author}
  {\bibfnamefont {T.}~\bibnamefont {Taniguchi}}, \bibinfo {author}
  {\bibfnamefont {K.}~\bibnamefont {Watanabe}}, \bibinfo {author}
  {\bibfnamefont {S.}~\bibnamefont {Jia}}, \bibinfo {author} {\bibfnamefont
  {J.}~\bibnamefont {Qi}}, \bibinfo {author} {\bibfnamefont {J.-H.}\
  \bibnamefont {Chen}}, \bibinfo {author} {\bibfnamefont {D.}~\bibnamefont
  {Sun}},\ and\ \bibinfo {author} {\bibfnamefont {J.}~\bibnamefont {Feng}},\
  }\href {https://doi.org/https://doi.org/10.1002/adma.201600100} {\bibfield
  {journal} {\bibinfo  {journal} {Advanced Materials}\ }\textbf {\bibinfo
  {volume} {28}},\ \bibinfo {pages} {4845} (\bibinfo {year}
  {2016})}\BibitemShut {NoStop}%
\bibitem [{\citenamefont {Xiang}\ \emph {et~al.}(2016)\citenamefont {Xiang},
  \citenamefont {Xu}, \citenamefont {Liu}, \citenamefont {Xia}, \citenamefont
  {Lu}, \citenamefont {Yin},\ and\ \citenamefont {Liu}}]{Xiang_Liu_2016}%
  \BibitemOpen
  \bibfield  {author} {\bibinfo {author} {\bibfnamefont {H.}~\bibnamefont
  {Xiang}}, \bibinfo {author} {\bibfnamefont {B.}~\bibnamefont {Xu}}, \bibinfo
  {author} {\bibfnamefont {J.}~\bibnamefont {Liu}}, \bibinfo {author}
  {\bibfnamefont {Y.}~\bibnamefont {Xia}}, \bibinfo {author} {\bibfnamefont
  {H.}~\bibnamefont {Lu}}, \bibinfo {author} {\bibfnamefont {J.}~\bibnamefont
  {Yin}},\ and\ \bibinfo {author} {\bibfnamefont {Z.}~\bibnamefont {Liu}},\
  }\href {https://doi.org/10.1063/1.4962662} {\bibfield  {journal} {\bibinfo
  {journal} {AIP Advances}\ }\textbf {\bibinfo {volume} {6}},\ \bibinfo {pages}
  {095005} (\bibinfo {year} {2016})}\BibitemShut {NoStop}%
\bibitem [{\citenamefont {Lin}\ and\ \citenamefont {Ni}(2017)}]{Lin_Ni_2017}%
  \BibitemOpen
  \bibfield  {author} {\bibinfo {author} {\bibfnamefont {X.}~\bibnamefont
  {Lin}}\ and\ \bibinfo {author} {\bibfnamefont {J.}~\bibnamefont {Ni}},\
  }\href {https://doi.org/10.1103/PhysRevB.95.245436} {\bibfield  {journal}
  {\bibinfo  {journal} {Phys. Rev. B}\ }\textbf {\bibinfo {volume} {95}},\
  \bibinfo {pages} {245436} (\bibinfo {year} {2017})}\BibitemShut {NoStop}%
\bibitem [{\citenamefont {Hu}\ \emph {et~al.}(2018)\citenamefont {Hu},
  \citenamefont {Kang}, \citenamefont {Yang}, \citenamefont {Huang},\ and\
  \citenamefont {Liu}}]{Hu_Liu_2018}%
  \BibitemOpen
  \bibfield  {author} {\bibinfo {author} {\bibfnamefont {L.}~\bibnamefont
  {Hu}}, \bibinfo {author} {\bibfnamefont {L.}~\bibnamefont {Kang}}, \bibinfo
  {author} {\bibfnamefont {J.}~\bibnamefont {Yang}}, \bibinfo {author}
  {\bibfnamefont {B.}~\bibnamefont {Huang}},\ and\ \bibinfo {author}
  {\bibfnamefont {F.}~\bibnamefont {Liu}},\ }\href
  {https://doi.org/10.1039/C8NR04391D} {\bibfield  {journal} {\bibinfo
  {journal} {Nanoscale}\ }\textbf {\bibinfo {volume} {10}},\ \bibinfo {pages}
  {22231} (\bibinfo {year} {2018})}\BibitemShut {NoStop}%
\bibitem [{\citenamefont {Jelver}\ \emph {et~al.}(2019)\citenamefont {Jelver},
  \citenamefont {Stradi}, \citenamefont {Stokbro}, \citenamefont {Olsen},\ and\
  \citenamefont {Jacobsen}}]{Jelver_Jacobsen_2019}%
  \BibitemOpen
  \bibfield  {author} {\bibinfo {author} {\bibfnamefont {L.}~\bibnamefont
  {Jelver}}, \bibinfo {author} {\bibfnamefont {D.}~\bibnamefont {Stradi}},
  \bibinfo {author} {\bibfnamefont {K.}~\bibnamefont {Stokbro}}, \bibinfo
  {author} {\bibfnamefont {T.}~\bibnamefont {Olsen}},\ and\ \bibinfo {author}
  {\bibfnamefont {K.~W.}\ \bibnamefont {Jacobsen}},\ }\href
  {https://doi.org/10.1103/PhysRevB.99.155420} {\bibfield  {journal} {\bibinfo
  {journal} {Phys. Rev. B}\ }\textbf {\bibinfo {volume} {99}},\ \bibinfo
  {pages} {155420} (\bibinfo {year} {2019})}\BibitemShut {NoStop}%
\bibitem [{\citenamefont {Ok}\ \emph {et~al.}(2019)\citenamefont {Ok},
  \citenamefont {Muechler}, \citenamefont {Di~Sante}, \citenamefont
  {Sangiovanni}, \citenamefont {Thomale},\ and\ \citenamefont
  {Neupert}}]{Ok_Neupert_2019}%
  \BibitemOpen
  \bibfield  {author} {\bibinfo {author} {\bibfnamefont {S.}~\bibnamefont
  {Ok}}, \bibinfo {author} {\bibfnamefont {L.}~\bibnamefont {Muechler}},
  \bibinfo {author} {\bibfnamefont {D.}~\bibnamefont {Di~Sante}}, \bibinfo
  {author} {\bibfnamefont {G.}~\bibnamefont {Sangiovanni}}, \bibinfo {author}
  {\bibfnamefont {R.}~\bibnamefont {Thomale}},\ and\ \bibinfo {author}
  {\bibfnamefont {T.}~\bibnamefont {Neupert}},\ }\href
  {https://doi.org/10.1103/PhysRevB.99.121105} {\bibfield  {journal} {\bibinfo
  {journal} {Phys. Rev. B}\ }\textbf {\bibinfo {volume} {99}},\ \bibinfo
  {pages} {121105} (\bibinfo {year} {2019})}\BibitemShut {NoStop}%
\bibitem [{\citenamefont {Lau}\ \emph {et~al.}(2019)\citenamefont {Lau},
  \citenamefont {Ray}, \citenamefont {Varjas},\ and\ \citenamefont
  {Akhmerov}}]{Lau_Akhmerov_2019}%
  \BibitemOpen
  \bibfield  {author} {\bibinfo {author} {\bibfnamefont {A.}~\bibnamefont
  {Lau}}, \bibinfo {author} {\bibfnamefont {R.}~\bibnamefont {Ray}}, \bibinfo
  {author} {\bibfnamefont {D.}~\bibnamefont {Varjas}},\ and\ \bibinfo {author}
  {\bibfnamefont {A.~R.}\ \bibnamefont {Akhmerov}},\ }\href
  {https://doi.org/10.1103/PhysRevMaterials.3.054206} {\bibfield  {journal}
  {\bibinfo  {journal} {Phys. Rev. Materials}\ }\textbf {\bibinfo {volume}
  {3}},\ \bibinfo {pages} {054206} (\bibinfo {year} {2019})}\BibitemShut
  {NoStop}%
\bibitem [{\citenamefont {Muechler}\ \emph {et~al.}(2020)\citenamefont
  {Muechler}, \citenamefont {Hu}, \citenamefont {Lin}, \citenamefont {Yang},\
  and\ \citenamefont {Car}}]{Muechler_Car_2020}%
  \BibitemOpen
  \bibfield  {author} {\bibinfo {author} {\bibfnamefont {L.}~\bibnamefont
  {Muechler}}, \bibinfo {author} {\bibfnamefont {W.}~\bibnamefont {Hu}},
  \bibinfo {author} {\bibfnamefont {L.}~\bibnamefont {Lin}}, \bibinfo {author}
  {\bibfnamefont {C.}~\bibnamefont {Yang}},\ and\ \bibinfo {author}
  {\bibfnamefont {R.}~\bibnamefont {Car}},\ }\href
  {https://doi.org/10.1103/PhysRevB.102.041103} {\bibfield  {journal} {\bibinfo
   {journal} {Phys. Rev. B}\ }\textbf {\bibinfo {volume} {102}},\ \bibinfo
  {pages} {041103} (\bibinfo {year} {2020})}\BibitemShut {NoStop}%
\bibitem [{\citenamefont {Zhang}\ and\ \citenamefont
  {Li}(2020)}]{Zhang_Li_2020}%
  \BibitemOpen
  \bibfield  {author} {\bibinfo {author} {\bibfnamefont {H.}~\bibnamefont
  {Zhang}}\ and\ \bibinfo {author} {\bibfnamefont {Z.}~\bibnamefont {Li}},\
  }\href {https://doi.org/10.1088/1361-648x/ab9051} {\bibfield  {journal}
  {\bibinfo  {journal} {Journal of Physics: Condensed Matter}\ }\textbf
  {\bibinfo {volume} {32}},\ \bibinfo {pages} {365303} (\bibinfo {year}
  {2020})}\BibitemShut {NoStop}%
\bibitem [{\citenamefont {Yang}\ \emph {et~al.}(2020)\citenamefont {Yang},
  \citenamefont {Mo}, \citenamefont {Fu}, \citenamefont {Yang}, \citenamefont
  {Zheng}, \citenamefont {Wang}, \citenamefont {Liu}, \citenamefont {Hao},\
  and\ \citenamefont {Zhang}}]{Yang_Zhang_2020}%
  \BibitemOpen
  \bibfield  {author} {\bibinfo {author} {\bibfnamefont {W.}~\bibnamefont
  {Yang}}, \bibinfo {author} {\bibfnamefont {C.-J.}\ \bibnamefont {Mo}},
  \bibinfo {author} {\bibfnamefont {S.-B.}\ \bibnamefont {Fu}}, \bibinfo
  {author} {\bibfnamefont {Y.}~\bibnamefont {Yang}}, \bibinfo {author}
  {\bibfnamefont {F.-W.}\ \bibnamefont {Zheng}}, \bibinfo {author}
  {\bibfnamefont {X.-H.}\ \bibnamefont {Wang}}, \bibinfo {author}
  {\bibfnamefont {Y.-A.}\ \bibnamefont {Liu}}, \bibinfo {author} {\bibfnamefont
  {N.}~\bibnamefont {Hao}},\ and\ \bibinfo {author} {\bibfnamefont
  {P.}~\bibnamefont {Zhang}},\ }\href
  {https://doi.org/10.1103/PhysRevLett.125.237006} {\bibfield  {journal}
  {\bibinfo  {journal} {Phys. Rev. Lett.}\ }\textbf {\bibinfo {volume} {125}},\
  \bibinfo {pages} {237006} (\bibinfo {year} {2020})}\BibitemShut {NoStop}%
\bibitem [{\citenamefont {Lu}\ \emph {et~al.}(2021)\citenamefont {Lu},
  \citenamefont {Prange},\ and\ \citenamefont {Sushko}}]{Lu_Sushko_2021}%
  \BibitemOpen
  \bibfield  {author} {\bibinfo {author} {\bibfnamefont {Z.}~\bibnamefont
  {Lu}}, \bibinfo {author} {\bibfnamefont {M.~P.}\ \bibnamefont {Prange}},\
  and\ \bibinfo {author} {\bibfnamefont {P.~V.}\ \bibnamefont {Sushko}},\
  }\href {https://doi.org/10.1021/acs.jpclett.1c01617} {\bibfield  {journal}
  {\bibinfo  {journal} {The Journal of Physical Chemistry Letters}\ }\textbf
  {\bibinfo {volume} {12}},\ \bibinfo {pages} {6596} (\bibinfo {year}
  {2021})},\ \bibinfo {note} {pMID: 34251220}\BibitemShut {NoStop}%
\bibitem [{\citenamefont {Muechler}\ \emph
  {et~al.}(2016{\natexlab{a}})\citenamefont {Muechler}, \citenamefont
  {Alexandradinata}, \citenamefont {Neupert},\ and\ \citenamefont
  {Car}}]{Meuchler_Carr_2016}%
  \BibitemOpen
  \bibfield  {author} {\bibinfo {author} {\bibfnamefont {L.}~\bibnamefont
  {Muechler}}, \bibinfo {author} {\bibfnamefont {A.}~\bibnamefont
  {Alexandradinata}}, \bibinfo {author} {\bibfnamefont {T.}~\bibnamefont
  {Neupert}},\ and\ \bibinfo {author} {\bibfnamefont {R.}~\bibnamefont {Car}},\
  }\href {https://doi.org/10.1103/PhysRevX.6.041069} {\bibfield  {journal}
  {\bibinfo  {journal} {Phys. Rev. X}\ }\textbf {\bibinfo {volume} {6}},\
  \bibinfo {pages} {041069} (\bibinfo {year} {2016}{\natexlab{a}})}\BibitemShut
  {NoStop}%
\bibitem [{\citenamefont {Hsu}\ \emph {et~al.}(2020)\citenamefont {Hsu},
  \citenamefont {Cole}, \citenamefont {Zhang},\ and\ \citenamefont
  {Sau}}]{Hsu_Sau_2020}%
  \BibitemOpen
  \bibfield  {author} {\bibinfo {author} {\bibfnamefont {Y.-T.}\ \bibnamefont
  {Hsu}}, \bibinfo {author} {\bibfnamefont {W.~S.}\ \bibnamefont {Cole}},
  \bibinfo {author} {\bibfnamefont {R.-X.}\ \bibnamefont {Zhang}},\ and\
  \bibinfo {author} {\bibfnamefont {J.~D.}\ \bibnamefont {Sau}},\ }\href
  {https://doi.org/10.1103/PhysRevLett.125.097001} {\bibfield  {journal}
  {\bibinfo  {journal} {Phys. Rev. Lett.}\ }\textbf {\bibinfo {volume} {125}},\
  \bibinfo {pages} {097001} (\bibinfo {year} {2020})}\BibitemShut {NoStop}%
\bibitem [{\citenamefont {Copenhaver}\ and\ \citenamefont
  {V\"ayrynen}(2022)}]{Copenhaver_Vayrynen_2021}%
  \BibitemOpen
  \bibfield  {author} {\bibinfo {author} {\bibfnamefont {J.}~\bibnamefont
  {Copenhaver}}\ and\ \bibinfo {author} {\bibfnamefont {J.~I.}\ \bibnamefont
  {V\"ayrynen}},\ }\href {https://doi.org/10.1103/PhysRevB.105.115402}
  {\bibfield  {journal} {\bibinfo  {journal} {Phys. Rev. B}\ }\textbf {\bibinfo
  {volume} {105}},\ \bibinfo {pages} {115402} (\bibinfo {year}
  {2022})}\BibitemShut {NoStop}%
\bibitem [{\citenamefont {Hu}\ \emph {et~al.}(2021)\citenamefont {Hu},
  \citenamefont {Ma}, \citenamefont {Wan},\ and\ \citenamefont
  {Liu}}]{Hu_Liu_2021}%
  \BibitemOpen
  \bibfield  {author} {\bibinfo {author} {\bibfnamefont {M.}~\bibnamefont
  {Hu}}, \bibinfo {author} {\bibfnamefont {G.}~\bibnamefont {Ma}}, \bibinfo
  {author} {\bibfnamefont {C.~Y.}\ \bibnamefont {Wan}},\ and\ \bibinfo {author}
  {\bibfnamefont {J.}~\bibnamefont {Liu}},\ }\href
  {https://doi.org/10.1103/PhysRevB.104.035156} {\bibfield  {journal} {\bibinfo
   {journal} {Phys. Rev. B}\ }\textbf {\bibinfo {volume} {104}},\ \bibinfo
  {pages} {035156} (\bibinfo {year} {2021})}\BibitemShut {NoStop}%
\bibitem [{\citenamefont {Xu}\ \emph {et~al.}(2018)\citenamefont {Xu},
  \citenamefont {Ma}, \citenamefont {Shen}, \citenamefont {Fatemi},
  \citenamefont {Wu}, \citenamefont {Chang}, \citenamefont {Chang},
  \citenamefont {Valdivia}, \citenamefont {Chan}, \citenamefont {Gibson},
  \citenamefont {Zhou}, \citenamefont {Liu}, \citenamefont {Watanabe},
  \citenamefont {Taniguchi}, \citenamefont {Lin}, \citenamefont {Cava},
  \citenamefont {Fu}, \citenamefont {Gedik},\ and\ \citenamefont
  {Jarillo-Herrero}}]{Xu_Jarillo-Herrero_2018}%
  \BibitemOpen
  \bibfield  {author} {\bibinfo {author} {\bibfnamefont {S.-Y.}\ \bibnamefont
  {Xu}}, \bibinfo {author} {\bibfnamefont {Q.}~\bibnamefont {Ma}}, \bibinfo
  {author} {\bibfnamefont {H.}~\bibnamefont {Shen}}, \bibinfo {author}
  {\bibfnamefont {V.}~\bibnamefont {Fatemi}}, \bibinfo {author} {\bibfnamefont
  {S.}~\bibnamefont {Wu}}, \bibinfo {author} {\bibfnamefont {T.-R.}\
  \bibnamefont {Chang}}, \bibinfo {author} {\bibfnamefont {G.}~\bibnamefont
  {Chang}}, \bibinfo {author} {\bibfnamefont {A.~M.~M.}\ \bibnamefont
  {Valdivia}}, \bibinfo {author} {\bibfnamefont {C.-K.}\ \bibnamefont {Chan}},
  \bibinfo {author} {\bibfnamefont {Q.~D.}\ \bibnamefont {Gibson}}, \bibinfo
  {author} {\bibfnamefont {J.}~\bibnamefont {Zhou}}, \bibinfo {author}
  {\bibfnamefont {Z.}~\bibnamefont {Liu}}, \bibinfo {author} {\bibfnamefont
  {K.}~\bibnamefont {Watanabe}}, \bibinfo {author} {\bibfnamefont
  {T.}~\bibnamefont {Taniguchi}}, \bibinfo {author} {\bibfnamefont
  {H.}~\bibnamefont {Lin}}, \bibinfo {author} {\bibfnamefont {R.~J.}\
  \bibnamefont {Cava}}, \bibinfo {author} {\bibfnamefont {L.}~\bibnamefont
  {Fu}}, \bibinfo {author} {\bibfnamefont {N.}~\bibnamefont {Gedik}},\ and\
  \bibinfo {author} {\bibfnamefont {P.}~\bibnamefont {Jarillo-Herrero}},\
  }\href {https://doi.org/10.1038/s41567-018-0189-6} {\bibfield  {journal}
  {\bibinfo  {journal} {Nature Physics}\ }\textbf {\bibinfo {volume} {14}},\
  \bibinfo {pages} {900} (\bibinfo {year} {2018})}\BibitemShut {NoStop}%
\bibitem [{\citenamefont {Shi}\ and\ \citenamefont
  {Song}(2019)}]{Shi_Song_2019}%
  \BibitemOpen
  \bibfield  {author} {\bibinfo {author} {\bibfnamefont {L.-k.}\ \bibnamefont
  {Shi}}\ and\ \bibinfo {author} {\bibfnamefont {J.~C.~W.}\ \bibnamefont
  {Song}},\ }\href {https://doi.org/10.1103/PhysRevB.99.035403} {\bibfield
  {journal} {\bibinfo  {journal} {Phys. Rev. B}\ }\textbf {\bibinfo {volume}
  {99}},\ \bibinfo {pages} {035403} (\bibinfo {year} {2019})}\BibitemShut
  {NoStop}%
\bibitem [{\citenamefont {Xie}\ \emph {et~al.}(2020)\citenamefont {Xie},
  \citenamefont {Zhou},\ and\ \citenamefont {Law}}]{Xie_Law_2020}%
  \BibitemOpen
  \bibfield  {author} {\bibinfo {author} {\bibfnamefont {Y.-M.}\ \bibnamefont
  {Xie}}, \bibinfo {author} {\bibfnamefont {B.~T.}\ \bibnamefont {Zhou}},\ and\
  \bibinfo {author} {\bibfnamefont {K.~T.}\ \bibnamefont {Law}},\ }\href
  {https://doi.org/10.1103/PhysRevLett.125.107001} {\bibfield  {journal}
  {\bibinfo  {journal} {Phys. Rev. Lett.}\ }\textbf {\bibinfo {volume} {125}},\
  \bibinfo {pages} {107001} (\bibinfo {year} {2020})}\BibitemShut {NoStop}%
\bibitem [{\citenamefont {Garcia}\ \emph {et~al.}(2020)\citenamefont {Garcia},
  \citenamefont {Vila}, \citenamefont {Hsu}, \citenamefont {Waintal},
  \citenamefont {Pereira},\ and\ \citenamefont {Roche}}]{Garcia_Roche_2020}%
  \BibitemOpen
  \bibfield  {author} {\bibinfo {author} {\bibfnamefont {J.~H.}\ \bibnamefont
  {Garcia}}, \bibinfo {author} {\bibfnamefont {M.}~\bibnamefont {Vila}},
  \bibinfo {author} {\bibfnamefont {C.-H.}\ \bibnamefont {Hsu}}, \bibinfo
  {author} {\bibfnamefont {X.}~\bibnamefont {Waintal}}, \bibinfo {author}
  {\bibfnamefont {V.~M.}\ \bibnamefont {Pereira}},\ and\ \bibinfo {author}
  {\bibfnamefont {S.}~\bibnamefont {Roche}},\ }\href
  {https://doi.org/10.1103/PhysRevLett.125.256603} {\bibfield  {journal}
  {\bibinfo  {journal} {Phys. Rev. Lett.}\ }\textbf {\bibinfo {volume} {125}},\
  \bibinfo {pages} {256603} (\bibinfo {year} {2020})}\BibitemShut {NoStop}%
\bibitem [{\citenamefont {Nandy}\ and\ \citenamefont
  {Pesin}(2022)}]{Nandy_Pesin_2021}%
  \BibitemOpen
  \bibfield  {author} {\bibinfo {author} {\bibfnamefont {S.}~\bibnamefont
  {Nandy}}\ and\ \bibinfo {author} {\bibfnamefont {D.~A.}\ \bibnamefont
  {Pesin}},\ }\href {https://doi.org/10.21468/SciPostPhys.12.4.120} {\bibfield
  {journal} {\bibinfo  {journal} {SciPost Phys.}\ }\textbf {\bibinfo {volume}
  {12}},\ \bibinfo {pages} {120} (\bibinfo {year} {2022})}\BibitemShut
  {NoStop}%
\bibitem [{\citenamefont {Choe}\ \emph {et~al.}(2016)\citenamefont {Choe},
  \citenamefont {Sung},\ and\ \citenamefont {Chang}}]{Choe_Chang_2016}%
  \BibitemOpen
  \bibfield  {author} {\bibinfo {author} {\bibfnamefont {D.-H.}\ \bibnamefont
  {Choe}}, \bibinfo {author} {\bibfnamefont {H.-J.}\ \bibnamefont {Sung}},\
  and\ \bibinfo {author} {\bibfnamefont {K.~J.}\ \bibnamefont {Chang}},\ }\href
  {https://doi.org/10.1103/PhysRevB.93.125109} {\bibfield  {journal} {\bibinfo
  {journal} {Phys. Rev. B}\ }\textbf {\bibinfo {volume} {93}},\ \bibinfo
  {pages} {125109} (\bibinfo {year} {2016})}\BibitemShut {NoStop}%
\bibitem [{\citenamefont {Arora}\ \emph {et~al.}(2020)\citenamefont {Arora},
  \citenamefont {Shi},\ and\ \citenamefont {Song}}]{Arora_Song_2020}%
  \BibitemOpen
  \bibfield  {author} {\bibinfo {author} {\bibfnamefont {A.}~\bibnamefont
  {Arora}}, \bibinfo {author} {\bibfnamefont {L.-k.}\ \bibnamefont {Shi}},\
  and\ \bibinfo {author} {\bibfnamefont {J.~C.~W.}\ \bibnamefont {Song}},\
  }\href {https://doi.org/10.1103/PhysRevB.102.161402} {\bibfield  {journal}
  {\bibinfo  {journal} {Phys. Rev. B}\ }\textbf {\bibinfo {volume} {102}},\
  \bibinfo {pages} {161402} (\bibinfo {year} {2020})}\BibitemShut {NoStop}%
\bibitem [{\citenamefont {Lee}\ and\ \citenamefont {Son}(2021)}]{Lee_Son_2021}%
  \BibitemOpen
  \bibfield  {author} {\bibinfo {author} {\bibfnamefont {J.-H.}\ \bibnamefont
  {Lee}}\ and\ \bibinfo {author} {\bibfnamefont {Y.-W.}\ \bibnamefont {Son}},\
  }\href {https://doi.org/10.1039/D1CP02214H} {\bibfield  {journal} {\bibinfo
  {journal} {Phys. Chem. Chem. Phys.}\ }\textbf {\bibinfo {volume} {23}},\
  \bibinfo {pages} {17279} (\bibinfo {year} {2021})}\BibitemShut {NoStop}%
\bibitem [{\citenamefont {Crépel}\ and\ \citenamefont
  {Fu}(2022)}]{Crepel_Fu_2021}%
  \BibitemOpen
  \bibfield  {author} {\bibinfo {author} {\bibfnamefont {V.}~\bibnamefont
  {Crépel}}\ and\ \bibinfo {author} {\bibfnamefont {L.}~\bibnamefont {Fu}},\
  }\href {https://doi.org/10.1073/pnas.2117735119} {\bibfield  {journal}
  {\bibinfo  {journal} {Proceedings of the National Academy of Sciences}\
  }\textbf {\bibinfo {volume} {119}},\ \bibinfo {pages} {e2117735119} (\bibinfo
  {year} {2022})}\BibitemShut {NoStop}%
\bibitem [{\citenamefont {Varsano}\ \emph {et~al.}(2020)\citenamefont
  {Varsano}, \citenamefont {Palummo}, \citenamefont {Molinari},\ and\
  \citenamefont {Rontani}}]{Varsano_Rontani_2020}%
  \BibitemOpen
  \bibfield  {author} {\bibinfo {author} {\bibfnamefont {D.}~\bibnamefont
  {Varsano}}, \bibinfo {author} {\bibfnamefont {M.}~\bibnamefont {Palummo}},
  \bibinfo {author} {\bibfnamefont {E.}~\bibnamefont {Molinari}},\ and\
  \bibinfo {author} {\bibfnamefont {M.}~\bibnamefont {Rontani}},\ }\href
  {https://doi.org/10.1038/s41565-020-0650-4} {\bibfield  {journal} {\bibinfo
  {journal} {Nature Nanotechnology}\ }\textbf {\bibinfo {volume} {15}},\
  \bibinfo {pages} {367} (\bibinfo {year} {2020})}\BibitemShut {NoStop}%
\bibitem [{\citenamefont {Lee}(2021)}]{Lee_2021}%
  \BibitemOpen
  \bibfield  {author} {\bibinfo {author} {\bibfnamefont {P.~A.}\ \bibnamefont
  {Lee}},\ }\href {https://doi.org/10.1103/PhysRevB.103.L041101} {\bibfield
  {journal} {\bibinfo  {journal} {Phys. Rev. B}\ }\textbf {\bibinfo {volume}
  {103}},\ \bibinfo {pages} {L041101} (\bibinfo {year} {2021})}\BibitemShut
  {NoStop}%
\bibitem [{\citenamefont {He}\ and\ \citenamefont {Lee}(2021)}]{He_Lee_2021}%
  \BibitemOpen
  \bibfield  {author} {\bibinfo {author} {\bibfnamefont {W.-Y.}\ \bibnamefont
  {He}}\ and\ \bibinfo {author} {\bibfnamefont {P.~A.}\ \bibnamefont {Lee}},\
  }\href {https://doi.org/10.1103/PhysRevB.104.L041110} {\bibfield  {journal}
  {\bibinfo  {journal} {Phys. Rev. B}\ }\textbf {\bibinfo {volume} {104}},\
  \bibinfo {pages} {L041110} (\bibinfo {year} {2021})}\BibitemShut {NoStop}%
\bibitem [{\citenamefont {Kwan}\ \emph {et~al.}(2021)\citenamefont {Kwan},
  \citenamefont {Devakul}, \citenamefont {Sondhi},\ and\ \citenamefont
  {Parameswaran}}]{Kwan_Parameswaran_2021}%
  \BibitemOpen
  \bibfield  {author} {\bibinfo {author} {\bibfnamefont {Y.~H.}\ \bibnamefont
  {Kwan}}, \bibinfo {author} {\bibfnamefont {T.}~\bibnamefont {Devakul}},
  \bibinfo {author} {\bibfnamefont {S.~L.}\ \bibnamefont {Sondhi}},\ and\
  \bibinfo {author} {\bibfnamefont {S.~A.}\ \bibnamefont {Parameswaran}},\
  }\href {https://doi.org/10.1103/PhysRevB.104.125133} {\bibfield  {journal}
  {\bibinfo  {journal} {Phys. Rev. B}\ }\textbf {\bibinfo {volume} {104}},\
  \bibinfo {pages} {125133} (\bibinfo {year} {2021})}\BibitemShut {NoStop}%
\bibitem [{\citenamefont {Muechler}\ \emph
  {et~al.}(2016{\natexlab{b}})\citenamefont {Muechler}, \citenamefont
  {Alexandradinata}, \citenamefont {Neupert},\ and\ \citenamefont
  {Car}}]{Muechler_Car_2016}%
  \BibitemOpen
  \bibfield  {author} {\bibinfo {author} {\bibfnamefont {L.}~\bibnamefont
  {Muechler}}, \bibinfo {author} {\bibfnamefont {A.}~\bibnamefont
  {Alexandradinata}}, \bibinfo {author} {\bibfnamefont {T.}~\bibnamefont
  {Neupert}},\ and\ \bibinfo {author} {\bibfnamefont {R.}~\bibnamefont {Car}},\
  }\href {https://doi.org/10.1103/PhysRevX.6.041069} {\bibfield  {journal}
  {\bibinfo  {journal} {Phys. Rev. X}\ }\textbf {\bibinfo {volume} {6}},\
  \bibinfo {pages} {041069} (\bibinfo {year} {2016}{\natexlab{b}})}\BibitemShut
  {NoStop}%
\bibitem [{\citenamefont {Bernevig}\ \emph {et~al.}(2006)\citenamefont
  {Bernevig}, \citenamefont {Hughes},\ and\ \citenamefont
  {Zhang}}]{Bernevig_Zhang_2006}%
  \BibitemOpen
  \bibfield  {author} {\bibinfo {author} {\bibfnamefont {B.~A.}\ \bibnamefont
  {Bernevig}}, \bibinfo {author} {\bibfnamefont {T.~L.}\ \bibnamefont
  {Hughes}},\ and\ \bibinfo {author} {\bibfnamefont {S.-C.}\ \bibnamefont
  {Zhang}},\ }\href {https://doi.org/10.1126/science.1133734} {\bibfield
  {journal} {\bibinfo  {journal} {Science}\ }\textbf {\bibinfo {volume}
  {314}},\ \bibinfo {pages} {1757} (\bibinfo {year} {2006})}\BibitemShut
  {NoStop}%
\bibitem [{\citenamefont {Landauer}(1987)}]{Landauer_1987}%
  \BibitemOpen
  \bibfield  {author} {\bibinfo {author} {\bibfnamefont {R.}~\bibnamefont
  {Landauer}},\ }\href {https://doi.org/10.1007/BF01304229} {\bibfield
  {journal} {\bibinfo  {journal} {Zeitschrift f{\"u}r Physik B Condensed
  Matter}\ }\textbf {\bibinfo {volume} {68}},\ \bibinfo {pages} {217} (\bibinfo
  {year} {1987})}\BibitemShut {NoStop}%
\bibitem [{\citenamefont {Datta}(1997)}]{Datta_1997}%
  \BibitemOpen
  \bibfield  {author} {\bibinfo {author} {\bibfnamefont {S.}~\bibnamefont
  {Datta}},\ }\href@noop {} {\emph {\bibinfo {title} {Electronic transport in
  mesoscopic systems}}}\ (\bibinfo  {publisher} {Cambridge University Press},\
  \bibinfo {year} {1997})\BibitemShut {NoStop}%
\bibitem [{\citenamefont {Thouless}\ and\ \citenamefont
  {Kirkpatrick}(1981)}]{Thouless_Kirkpatrick_1981}%
  \BibitemOpen
  \bibfield  {author} {\bibinfo {author} {\bibfnamefont {D.~J.}\ \bibnamefont
  {Thouless}}\ and\ \bibinfo {author} {\bibfnamefont {S.}~\bibnamefont
  {Kirkpatrick}},\ }\href {https://doi.org/10.1088/0022-3719/14/3/007}
  {\bibfield  {journal} {\bibinfo  {journal} {Journal of Physics C: Solid State
  Physics}\ }\textbf {\bibinfo {volume} {14}},\ \bibinfo {pages} {235}
  (\bibinfo {year} {1981})}\BibitemShut {NoStop}%
\bibitem [{\citenamefont {Caroli}\ \emph {et~al.}(1971)\citenamefont {Caroli},
  \citenamefont {Combescot}, \citenamefont {Nozieres},\ and\ \citenamefont
  {Saint-James}}]{Caroli_Saint-James_1971}%
  \BibitemOpen
  \bibfield  {author} {\bibinfo {author} {\bibfnamefont {C.}~\bibnamefont
  {Caroli}}, \bibinfo {author} {\bibfnamefont {R.}~\bibnamefont {Combescot}},
  \bibinfo {author} {\bibfnamefont {P.}~\bibnamefont {Nozieres}},\ and\
  \bibinfo {author} {\bibfnamefont {D.}~\bibnamefont {Saint-James}},\ }\href
  {https://doi.org/10.1088/0022-3719/4/8/018} {\bibfield  {journal} {\bibinfo
  {journal} {Journal of Physics C: Solid State Physics}\ }\textbf {\bibinfo
  {volume} {4}},\ \bibinfo {pages} {916} (\bibinfo {year} {1971})}\BibitemShut
  {NoStop}%
\bibitem [{\citenamefont {Lee}\ and\ \citenamefont
  {Fisher}(1981)}]{Lee_Fisher_1981}%
  \BibitemOpen
  \bibfield  {author} {\bibinfo {author} {\bibfnamefont {P.~A.}\ \bibnamefont
  {Lee}}\ and\ \bibinfo {author} {\bibfnamefont {D.~S.}\ \bibnamefont
  {Fisher}},\ }\href {https://doi.org/10.1103/PhysRevLett.47.882} {\bibfield
  {journal} {\bibinfo  {journal} {Phys. Rev. Lett.}\ }\textbf {\bibinfo
  {volume} {47}},\ \bibinfo {pages} {882} (\bibinfo {year} {1981})}\BibitemShut
  {NoStop}%
\bibitem [{\citenamefont {Li}\ \emph {et~al.}(2009)\citenamefont {Li},
  \citenamefont {Chu}, \citenamefont {Jain},\ and\ \citenamefont
  {Shen}}]{Li_Shen_2009}%
  \BibitemOpen
  \bibfield  {author} {\bibinfo {author} {\bibfnamefont {J.}~\bibnamefont
  {Li}}, \bibinfo {author} {\bibfnamefont {R.-L.}\ \bibnamefont {Chu}},
  \bibinfo {author} {\bibfnamefont {J.~K.}\ \bibnamefont {Jain}},\ and\
  \bibinfo {author} {\bibfnamefont {S.-Q.}\ \bibnamefont {Shen}},\ }\href
  {https://doi.org/10.1103/PhysRevLett.102.136806} {\bibfield  {journal}
  {\bibinfo  {journal} {Phys. Rev. Lett.}\ }\textbf {\bibinfo {volume} {102}},\
  \bibinfo {pages} {136806} (\bibinfo {year} {2009})}\BibitemShut {NoStop}%
\bibitem [{\citenamefont {Groth}\ \emph {et~al.}(2009)\citenamefont {Groth},
  \citenamefont {Wimmer}, \citenamefont {Akhmerov}, \citenamefont
  {Tworzyd\l{}o},\ and\ \citenamefont {Beenakker}}]{Groth_Beenakker_2009}%
  \BibitemOpen
  \bibfield  {author} {\bibinfo {author} {\bibfnamefont {C.~W.}\ \bibnamefont
  {Groth}}, \bibinfo {author} {\bibfnamefont {M.}~\bibnamefont {Wimmer}},
  \bibinfo {author} {\bibfnamefont {A.~R.}\ \bibnamefont {Akhmerov}}, \bibinfo
  {author} {\bibfnamefont {J.}~\bibnamefont {Tworzyd\l{}o}},\ and\ \bibinfo
  {author} {\bibfnamefont {C.~W.~J.}\ \bibnamefont {Beenakker}},\ }\href
  {https://doi.org/10.1103/PhysRevLett.103.196805} {\bibfield  {journal}
  {\bibinfo  {journal} {Phys. Rev. Lett.}\ }\textbf {\bibinfo {volume} {103}},\
  \bibinfo {pages} {196805} (\bibinfo {year} {2009})}\BibitemShut {NoStop}%
\bibitem [{\citenamefont {Zhou}\ \emph {et~al.}(2008)\citenamefont {Zhou},
  \citenamefont {Lu}, \citenamefont {Chu}, \citenamefont {Shen},\ and\
  \citenamefont {Niu}}]{Zhou_Niu_2008}%
  \BibitemOpen
  \bibfield  {author} {\bibinfo {author} {\bibfnamefont {B.}~\bibnamefont
  {Zhou}}, \bibinfo {author} {\bibfnamefont {H.-Z.}\ \bibnamefont {Lu}},
  \bibinfo {author} {\bibfnamefont {R.-L.}\ \bibnamefont {Chu}}, \bibinfo
  {author} {\bibfnamefont {S.-Q.}\ \bibnamefont {Shen}},\ and\ \bibinfo
  {author} {\bibfnamefont {Q.}~\bibnamefont {Niu}},\ }\href
  {https://doi.org/10.1103/PhysRevLett.101.246807} {\bibfield  {journal}
  {\bibinfo  {journal} {Phys. Rev. Lett.}\ }\textbf {\bibinfo {volume} {101}},\
  \bibinfo {pages} {246807} (\bibinfo {year} {2008})}\BibitemShut {NoStop}%
\bibitem [{\citenamefont {Bieniek}\ \emph {et~al.}(2017)\citenamefont
  {Bieniek}, \citenamefont {Wo{\'{z}}niak},\ and\ \citenamefont
  {Potasz}}]{Bieniek_Potasz_2017}%
  \BibitemOpen
  \bibfield  {author} {\bibinfo {author} {\bibfnamefont {M.}~\bibnamefont
  {Bieniek}}, \bibinfo {author} {\bibfnamefont {T.}~\bibnamefont
  {Wo{\'{z}}niak}},\ and\ \bibinfo {author} {\bibfnamefont {P.}~\bibnamefont
  {Potasz}},\ }\href {https://doi.org/10.1088/1361-648x/aa5e79} {\bibfield
  {journal} {\bibinfo  {journal} {Journal of Physics: Condensed Matter}\
  }\textbf {\bibinfo {volume} {29}},\ \bibinfo {pages} {155501} (\bibinfo
  {year} {2017})}\BibitemShut {NoStop}%
\bibitem [{\citenamefont {Song}\ \emph {et~al.}(2021)\citenamefont {Song},
  \citenamefont {Zhang}, \citenamefont {Yang}, \citenamefont {Ji},
  \citenamefont {Sun}, \citenamefont {Liu}, \citenamefont {Wang},\ and\
  \citenamefont {Gao}}]{Song_Gao_2021}%
  \BibitemOpen
  \bibfield  {author} {\bibinfo {author} {\bibfnamefont {X.}~\bibnamefont
  {Song}}, \bibinfo {author} {\bibfnamefont {T.}~\bibnamefont {Zhang}},
  \bibinfo {author} {\bibfnamefont {H.}~\bibnamefont {Yang}}, \bibinfo {author}
  {\bibfnamefont {H.}~\bibnamefont {Ji}}, \bibinfo {author} {\bibfnamefont
  {J.}~\bibnamefont {Sun}}, \bibinfo {author} {\bibfnamefont {L.}~\bibnamefont
  {Liu}}, \bibinfo {author} {\bibfnamefont {Y.}~\bibnamefont {Wang}},\ and\
  \bibinfo {author} {\bibfnamefont {H.}~\bibnamefont {Gao}},\ }\href
  {https://doi.org/10.1007/s12274-021-3668-5} {\bibfield  {journal} {\bibinfo
  {journal} {Nano Research}\ }\textbf {\bibinfo {volume} {14}},\ \bibinfo
  {pages} {3810} (\bibinfo {year} {2021})}\BibitemShut {NoStop}%
\bibitem [{\citenamefont {M\"uller}\ \emph {et~al.}(2017)\citenamefont
  {M\"uller}, \citenamefont {Thomale}, \citenamefont {Trauzettel},
  \citenamefont {Bocquillon},\ and\ \citenamefont
  {Kashuba}}]{Muller_Kashuba_2017}%
  \BibitemOpen
  \bibfield  {author} {\bibinfo {author} {\bibfnamefont {T.}~\bibnamefont
  {M\"uller}}, \bibinfo {author} {\bibfnamefont {R.}~\bibnamefont {Thomale}},
  \bibinfo {author} {\bibfnamefont {B.}~\bibnamefont {Trauzettel}}, \bibinfo
  {author} {\bibfnamefont {E.}~\bibnamefont {Bocquillon}},\ and\ \bibinfo
  {author} {\bibfnamefont {O.}~\bibnamefont {Kashuba}},\ }\href
  {https://doi.org/10.1103/PhysRevB.95.245114} {\bibfield  {journal} {\bibinfo
  {journal} {Phys. Rev. B}\ }\textbf {\bibinfo {volume} {95}},\ \bibinfo
  {pages} {245114} (\bibinfo {year} {2017})}\BibitemShut {NoStop}%
\bibitem [{\citenamefont {Glazman}\ \emph {et~al.}(1992)\citenamefont
  {Glazman}, \citenamefont {Ruzin},\ and\ \citenamefont
  {Shklovskii}}]{Glazman_Shklovskii_1992}%
  \BibitemOpen
  \bibfield  {author} {\bibinfo {author} {\bibfnamefont {L.~I.}\ \bibnamefont
  {Glazman}}, \bibinfo {author} {\bibfnamefont {I.~M.}\ \bibnamefont {Ruzin}},\
  and\ \bibinfo {author} {\bibfnamefont {B.~I.}\ \bibnamefont {Shklovskii}},\
  }\href {https://doi.org/10.1103/PhysRevB.45.8454} {\bibfield  {journal}
  {\bibinfo  {journal} {Phys. Rev. B}\ }\textbf {\bibinfo {volume} {45}},\
  \bibinfo {pages} {8454} (\bibinfo {year} {1992})}\BibitemShut {NoStop}%
\bibitem [{\citenamefont {Teo}\ and\ \citenamefont
  {Kane}(2009)}]{Teo_Kane_2009}%
  \BibitemOpen
  \bibfield  {author} {\bibinfo {author} {\bibfnamefont {J.~C.~Y.}\
  \bibnamefont {Teo}}\ and\ \bibinfo {author} {\bibfnamefont {C.~L.}\
  \bibnamefont {Kane}},\ }\href {https://doi.org/10.1103/PhysRevB.79.235321}
  {\bibfield  {journal} {\bibinfo  {journal} {Phys. Rev. B}\ }\textbf {\bibinfo
  {volume} {79}},\ \bibinfo {pages} {235321} (\bibinfo {year}
  {2009})}\BibitemShut {NoStop}%
\bibitem [{\citenamefont {Li}\ and\ \citenamefont
  {Das~Sarma}(1991)}]{Li_DasSarma_1991}%
  \BibitemOpen
  \bibfield  {author} {\bibinfo {author} {\bibfnamefont {Q.~P.}\ \bibnamefont
  {Li}}\ and\ \bibinfo {author} {\bibfnamefont {S.}~\bibnamefont {Das~Sarma}},\
  }\href {https://doi.org/10.1103/PhysRevB.43.11768} {\bibfield  {journal}
  {\bibinfo  {journal} {Phys. Rev. B}\ }\textbf {\bibinfo {volume} {43}},\
  \bibinfo {pages} {11768} (\bibinfo {year} {1991})}\BibitemShut {NoStop}%
\bibitem [{\citenamefont {Ando}(2010)}]{Ando_2010}%
  \BibitemOpen
  \bibfield  {author} {\bibinfo {author} {\bibfnamefont {T.}~\bibnamefont
  {Ando}},\ }\href {https://doi.org/10.1143/JPSJ.79.024706} {\bibfield
  {journal} {\bibinfo  {journal} {Journal of the Physical Society of Japan}\
  }\textbf {\bibinfo {volume} {79}},\ \bibinfo {pages} {024706} (\bibinfo
  {year} {2010})}\BibitemShut {NoStop}%
\bibitem [{\citenamefont {Kainaris}\ \emph {et~al.}(2014)\citenamefont
  {Kainaris}, \citenamefont {Gornyi}, \citenamefont {Carr},\ and\ \citenamefont
  {Mirlin}}]{PhysRevB.90.075118}%
  \BibitemOpen
  \bibfield  {author} {\bibinfo {author} {\bibfnamefont {N.}~\bibnamefont
  {Kainaris}}, \bibinfo {author} {\bibfnamefont {I.~V.}\ \bibnamefont
  {Gornyi}}, \bibinfo {author} {\bibfnamefont {S.~T.}\ \bibnamefont {Carr}},\
  and\ \bibinfo {author} {\bibfnamefont {A.~D.}\ \bibnamefont {Mirlin}},\
  }\href {https://doi.org/10.1103/PhysRevB.90.075118} {\bibfield  {journal}
  {\bibinfo  {journal} {Phys. Rev. B}\ }\textbf {\bibinfo {volume} {90}},\
  \bibinfo {pages} {075118} (\bibinfo {year} {2014})}\BibitemShut {NoStop}%
\bibitem [{\citenamefont {Imambekov}\ \emph {et~al.}(2012)\citenamefont
  {Imambekov}, \citenamefont {Schmidt},\ and\ \citenamefont
  {Glazman}}]{Imambekov_Glazman_2012}%
  \BibitemOpen
  \bibfield  {author} {\bibinfo {author} {\bibfnamefont {A.}~\bibnamefont
  {Imambekov}}, \bibinfo {author} {\bibfnamefont {T.~L.}\ \bibnamefont
  {Schmidt}},\ and\ \bibinfo {author} {\bibfnamefont {L.~I.}\ \bibnamefont
  {Glazman}},\ }\href {https://doi.org/10.1103/RevModPhys.84.1253} {\bibfield
  {journal} {\bibinfo  {journal} {Rev. Mod. Phys.}\ }\textbf {\bibinfo {volume}
  {84}},\ \bibinfo {pages} {1253} (\bibinfo {year} {2012})}\BibitemShut
  {NoStop}%
\bibitem [{\citenamefont {Markhof}\ and\ \citenamefont
  {Meden}(2016)}]{Markhof_Meden_2016}%
  \BibitemOpen
  \bibfield  {author} {\bibinfo {author} {\bibfnamefont {L.}~\bibnamefont
  {Markhof}}\ and\ \bibinfo {author} {\bibfnamefont {V.}~\bibnamefont
  {Meden}},\ }\href {https://doi.org/10.1103/PhysRevB.93.085108} {\bibfield
  {journal} {\bibinfo  {journal} {Phys. Rev. B}\ }\textbf {\bibinfo {volume}
  {93}},\ \bibinfo {pages} {085108} (\bibinfo {year} {2016})}\BibitemShut
  {NoStop}%
\bibitem [{\citenamefont {Daviet}\ and\ \citenamefont
  {Dupuis}(2020)}]{Daviet_Dupuis_2020}%
  \BibitemOpen
  \bibfield  {author} {\bibinfo {author} {\bibfnamefont {R.}~\bibnamefont
  {Daviet}}\ and\ \bibinfo {author} {\bibfnamefont {N.}~\bibnamefont
  {Dupuis}},\ }\href {https://doi.org/10.1103/PhysRevLett.125.235301}
  {\bibfield  {journal} {\bibinfo  {journal} {Phys. Rev. Lett.}\ }\textbf
  {\bibinfo {volume} {125}},\ \bibinfo {pages} {235301} (\bibinfo {year}
  {2020})}\BibitemShut {NoStop}%
\bibitem [{\citenamefont {Weber}()}]{Weber_2022}%
  \BibitemOpen
  \bibfield  {author} {\bibinfo {author} {\bibfnamefont {B.}~\bibnamefont
  {Weber}},\ }\bibinfo {note} {private communication}\BibitemShut {NoStop}%
\bibitem [{\citenamefont {Solyom}(1979)}]{Solyom_1979}%
  \BibitemOpen
  \bibfield  {author} {\bibinfo {author} {\bibfnamefont {J.}~\bibnamefont
  {Solyom}},\ }\href {https://doi.org/10.1080/00018737900101375} {\bibfield
  {journal} {\bibinfo  {journal} {Advances in Physics}\ }\textbf {\bibinfo
  {volume} {28}},\ \bibinfo {pages} {201} (\bibinfo {year} {1979})}\BibitemShut
  {NoStop}%
\bibitem [{\citenamefont {Bieniek}\ \emph {et~al.}(2020)\citenamefont
  {Bieniek}, \citenamefont {Szulakowska},\ and\ \citenamefont
  {Hawrylak}}]{Bieniek_Hawrylak_2020}%
  \BibitemOpen
  \bibfield  {author} {\bibinfo {author} {\bibfnamefont {M.}~\bibnamefont
  {Bieniek}}, \bibinfo {author} {\bibfnamefont {L.}~\bibnamefont
  {Szulakowska}},\ and\ \bibinfo {author} {\bibfnamefont {P.}~\bibnamefont
  {Hawrylak}},\ }\href {https://doi.org/10.1103/PhysRevB.101.125423} {\bibfield
   {journal} {\bibinfo  {journal} {Phys. Rev. B}\ }\textbf {\bibinfo {volume}
  {101}},\ \bibinfo {pages} {125423} (\bibinfo {year} {2020})}\BibitemShut
  {NoStop}%
\end{thebibliography}%

\end{document}